\documentclass[11pt]{article}

\usepackage[utf8]{inputenc}
\usepackage[T1]{fontenc}
\usepackage{lmodern}
\usepackage[margin=1in]{geometry}
\usepackage{amsmath,amssymb,bm}
\usepackage{graphicx}
\usepackage{subcaption}
\usepackage{booktabs}
\usepackage{xcolor}
\usepackage{authblk}
\usepackage{tikz}
\usetikzlibrary{positioning,calc,arrows.meta,shapes.geometric,fit,backgrounds}
\usepackage[numbers,super,sort&compress]{natbib}
\usepackage[colorlinks=true,allcolors=blue!55!black]{hyperref}
\usepackage[colorinlistoftodos]{todonotes}

\graphicspath{{figures/}}

\definecolor{legoRed}{HTML}{C0392B}
\definecolor{legoBlue}{HTML}{2B7BBA}
\definecolor{legoGreen}{HTML}{27AE60}
\definecolor{legoYellow}{HTML}{F1C40F}
\definecolor{legoGrey}{HTML}{7F8C8D}
\definecolor{legoSlate}{HTML}{34495E}
\definecolor{legoPurple}{HTML}{8E44AD}
\definecolor{legoTeal}{HTML}{16A085}

\newcommand{\repourl}{https://github.com/climate-federation/legoESM}
\newcommand{\jaxgrad}{\texttt{jax.grad}}

\title{\vspace{-1.6cm}\textbf{\Large legoESM: a modular, differentiable, multiscale, AI-ready Earth system model built with AI agents}}

\author[1,2,10,$\ast$]{Pierre Gentine}
\author[2,3]{Dhruv Balwada}
\author[4,5]{Ayta\c{c} Pa\c{c}al}
\author[1,2]{Linnia Hawkins}
\author[6]{Alistair Adcroft}
\author[1,2]{Hang Fan}
\author[1,2]{Jianing Fang}
\author[1,2]{Aya Lahlou}
\author[1,2]{Kara D. Lamb}
\author[7]{David M. Lawrence}
\author[8]{Julien Le Sommer}
\author[6]{Joseph Mouallem}
\author[1,2]{Juan Nathaniel}
\author[9]{Duncan Watson-Parris}
\author[4,5]{Veronika Eyring}

\affil[1]{\small Department of Earth and Environmental Engineering, Columbia University, New York, NY, USA}
\affil[2]{\small Learning the Earth with Artificial Intelligence and Physics (LEAP) NSF Science and Technology Center, Columbia University, New York, NY, USA}
\affil[3]{\small Lamont--Doherty Earth Observatory, Columbia University, Palisades, NY, USA}
\affil[4]{\small University of Tübingen, connAIx Research School for Applied AI, Heilbronn, Germany}
\affil[5]{\small University of Tübingen, Tübingen AI Center, Tübingen, Germany}
\affil[6]{\small Program in Atmospheric and Oceanic Sciences, Princeton University, Princeton, NJ, USA}
\affil[7]{\small NSF National Center for Atmospheric Research, Boulder, CO, USA}
\affil[8]{\small Institut des G\'eosciences de l'Environnement (IGE), Universit\'e Grenoble Alpes, CNRS, IRD, Grenoble INP, Grenoble, France}
\affil[9]{\small Scripps Institution of Oceanography and Hal\i c\i o\u{g}lu Data Science Institute, University of California San Diego, La Jolla, CA, USA}
\affil[10]{\small \'Ecole Normale Sup\'erieure -- PSL, 45 rue d'Ulm, Paris, France}
\affil[11]{\small Celest.Science, Paris, France}
\affil[$\ast$]{\small Corresponding author: \href{mailto:pg2328@columbia.edu}{pg2328@columbia.edu}}

\date{}

\begin{document}
\maketitle
\thispagestyle{empty}

\begin{abstract}
\noindent
Earth system models (ESMs) have grown tremendously in realism, yet key uncertainties persist in the climate response to greenhouse-gas forcing, particularly due to cloud radiative feedbacks. In addition, their software architecture was not designed for accelerator hardware or modern artificial intelligence (AI). Here we present \emph{legoESM}, a composable, differentiable, multiscale ESM written in JAX. It builds on decades of community-developed parameterizations and numerical methods, recast in a unified framework by AI coding agents under a human-specified scientific contract and verified through benchmarking. Dynamical cores, physics schemes, grids, complexity levels and components are swappable like building blocks, and can use conventional physics or machine-learned emulators. A single code base spans metre-scale large-eddy simulation to global simulations and weather to climate. End-to-end differentiability enables gradient-based calibration, variational data assimilation and online training. \emph{legoESM}’s modular architecture enables systematic evaluation of diverse model variants to explore structural uncertainty and test hypotheses. \emph{legoESM} produces realistic simulations across scales, reduces land-surface temperature bias through gradient-based calibration, and scales efficiently on GPUs to kilometer-scale simulations. It offers an open, community infrastructure for hypothesis testing, research and teaching in Earth sciences and a template for multiscale physical systems.

\end{abstract}

\vspace{0.4em}

\section*{Introduction}
Earth system models (ESMs) have improved substantially over recent decades, both in their fidelity with
respect to the observed climate and in the breadth of processes they represent\cite{Eyring2016,IPCC2021,Danabasoglu2020,Boucher2020,Sellar2019,ECMWF2023IFS}.
Yet, important biases persist\cite{eyringHumanInfluenceClimate2021, Brown2026} and the uncertainty in the response of the climate system to
greenhouse-gas forcing remains large, with 
equilibrium climate sensitivity still being uncertain at the $\sim$50\% level\cite{Sherwood2020}.
Projections of rainfall extremes\cite{OGorman2015}, of ocean heat uptake, and of the strength of
the land carbon sink\cite{Arora2020,Friedlingstein2022} differ substantially between models and are
often inconsistent across different regions of the world. Further, ESMs have to a large
extent not embraced the data revolution: only a fraction of the wealth of observations and
observational products (satellite retrievals, reanalyses) is currently being used for model calibration, largely because of limitation in the capability of those models to tune parameters with current sampling methods. 

Three developments could, in principle, reduce these uncertainties. First, machine learning (ML) has been used to emulate or replace empirical parameterizations such as deep convection\cite{Gentine2018,Rasp2018,BrenowitzBretherton2018,YuvalOGorman2020}
or ocean mesoscale eddies\cite{BoltonZanna2019,ZannaBolton2020, balwada2026design, perezhogin2025generalizable}, improving key processes such
as rainfall extremes and ocean eddy fluxes, and demonstrably reducing systematic errors, such as in cloud cover,
precipitation and circulation\cite{IglesiasSuarez2024,Grundner2025,Heuer2026}. Yet, because ESMs are
typically written in Fortran or C, ML components are coupled through one-way bridges that feed a
neural-network prediction into the physics model\cite{Ott2020}. As a result, online training, that is,
learning through the model's own evolution, is generally impossible, and the resulting hybrid systems
are prone to instability\cite{BrenowitzBretherton2018} and cannot be further tuned against observations.
Second, as models have grown in complexity, many modeling centers have adopted automated methods to tune and calibrate uncertain parameters \cite{Hourdin2017,Bonnet2025,Grundner2025}. However, conventional physics-based models generally do not provide an adjoint: that is, the gradient of model outputs with respect to these parameters is not available. Calibration must therefore rely on derivative-free approaches, such as Monte Carlo or ensemble methods \cite{Cleary2021,Couvreux2021,Schneider2017}. Because these methods require many model evaluations, their computational cost grows rapidly as the number of parameters increases. Third, only a handful of models or components have been ported to modern hardware, such as graphics processing units (GPUs)\cite{Fuhrer2018,Bauer2021}, and this remains the exception rather than the norm.
This has limited the ability to go to higher resolution to better resolve key processes such as
deep convection\cite{Eyring2024,Stevens2019} or to run models on other platforms, such as the cloud, further limiting ESM accessibility.

In parallel, weather forecasting has recently witnessed a (second) revolution (after the first: data assimilation): data-driven models trained on
decades of reanalysis now rival or surpass physics-based forecasts and run at a fraction of the computational cost of physics-based models \cite{Pathak2022,Bi2023,Lam2023,Bodnar2024,Lang2024}. Yet, purely learned AI weather models tend to drift and can lose physical realism at long lead times\cite{Lehmann2026}, and on their own they do not provide the conservative, multi-component interpretable system that climate projection requires. More fundamentally,
machine-learning emulators, even if skillful and fast, mainly reproduce input--output relationships without
representing the physical mechanisms that generate them. This data-driven revolution has, moreover, so far largely been confined to the atmosphere, where decades of reanalysis provide abundant training data. Yet, the sparser ocean and land records make comparable data-driven fidelity much harder to achieve, although AI ocean emulators are now emerging\cite{Dheeshjith2025Samudra,Donno2026Neptune}.
More broadly, ESMs are not merely predictive tools but also research tools: their capacity for hypothesis
testing and mechanistic attribution is central to scientific progress. Preserving that capability
while gaining the computational and optimization advantages of differentiable, ML-enhanced methods is
a central motivation for process-based, differentiable ESMs.

In addition, and related to hypothesis testing, climate science has had a long tradition of using a hierarchy of models, spanning idealized to complex configurations \cite{Held2005,Jeevanjee2017,Maher2019}. Such a hierarchy has been fundamental to building physical understanding of the climate system. Modern ESMs can themselves span a wide range of resolutions, timescales, and configurations, yet moving systematically across levels of complexity often requires substantial changes in model configuration or even the use of different modeling systems. A more modular framework could make these transitions easier and more controlled, enabling hypotheses and counterfactuals to be tested consistently across scales and levels of complexity.

Finally, most current ESMs are written in compiled languages (Fortran, C/C++) that are less widely used across the broader scientific community than Python, which can limit adoption by researchers from other disciplines and by downstream users, leading to slow model
evolution. The recent weather AI revolution has demonstrated that it is not only model performance
that drives widespread adoption, but also ease of use \cite{donoho2024data}. 

The confluence of large-scale datasets,
GPU-accelerated computing, and differentiable programming frameworks capable of scaling
gradient-based optimization enabled the machine-learning revolution. Earth system science is now
positioned to leverage the same convergence, but doing so requires ESMs written in
modern, automatically differentiable languages that can natively use GPUs and other emerging architectures, support gradient-based optimization, and interface directly with machine learning.

As such, we argue that a modern ESM should be (i) easy to use, and ideally written in Python, the \textit{lingua franca} of modern data science, yet with high performance; (ii) hardware-agnostic, running on any of central processing units (CPUs),  graphics processing units (GPUs) and tensor processing units (TPUs) as accelerators keep evolving;
(iii) composable, so that different numerical methods, physical parameterizations or entire modules
(for example the whole atmosphere) can be swapped, combined and trained jointly for controlled
experiments and hypothesis testing; (iv) able to integrate ML seamlessly, whether by replacing a single parameterization or
an entire component (for example swapping the whole atmosphere for a neural network, as in recent AI
weather models\cite{Pathak2022,Bi2023,Lam2023,Kochkov2024,Bodnar2024,Lang2024}), which requires the code to be end-to-end
differentiable; (v) solvable on many or adaptive
grids tailored to the problem at hand; and (vi) able to run across spatial and temporal scales.

A growing ecosystem of differentiable models has begun to demonstrate the benefits of parts of these ideas,
though so far this has been for one component or one spatial or temporal scale at a time. For the atmosphere, differentiable dynamical
cores and intermediate-complexity models such as the JAX Circulation Model\cite{Davenport2026}, the
neural general circulation model NeuralGCM\cite{Kochkov2024} and SpeedyWeather.jl\cite{Klower2024}
have demonstrated the use of gradients for calibration or online learning. For the ocean, JAX and
automatic-differentiation efforts, from Veros\cite{Hafner2018,Meunier2025} to FESOM-JAX\cite{koldunov2026fesom2}, are bringing the
same capability to state estimation, e.g., for data assimilation. More broadly, DJ4Earth\cite{Moses2026DJ4Earth} uses
compiler-level program transformations to make several Julia models (ShallowWaters.jl, SpeedyWeather.jl, Oceananigans.jl and
DJUICE.jl) differentiable and performance-portable, although these components are not yet coupled and differentiability has been
demonstrated for each component individually. For the land surface and hydrology,
differentiable models such as DifferLand\cite{Fang2026}, JAX-CanVeg\cite{Jiang2025CanVeg} and differentiable process-based
hydrology models\cite{Shen2023} have demonstrated that they can be used to train physical parameters or neural components end to end against
observations, building on the broader concept of differentiable programming in the
geosciences\cite{Shen2023,Gelbrecht2023,Kochkov2021}. Yet, at present, no unifying infrastructure exists that addresses all of those needs for the entire Earth system, rather than a targeted treatment of one piece in isolation.

Here, we present legoESM, a novel differentiable and composable modeling infrastructure for weather and climate science. 
Rather than a single ESM, it is an open, modular framework in
which differentiable bricks of individual processes or full Earth components can be assembled, coupled and
benchmarked to address diverse scientific questions or hypotheses. This modeling ecosystem is conceived as an evolving, agile, community-based infrastructure to test ideas and hypotheses, with clear ML-inspired metrics, across
weather and climate (Fig.~\ref{fig:arch}a). Distinctively, legoESM was developed almost entirely from the ground up using AI coding agents—autonomous software agents that implement, test, and revise code in response to a human-authored scientific specification and based on clear benchmarks. To ensure that agent-generated code remains physically consistent and scientifically faithful, development is constrained by explicit contracts, which specify the required interfaces, physical invariants, units, and expected behaviour of individual components. Implementations were further evaluated using faithfulness oracles (trusted reference solutions) for physical processes and systematic benchmarks against analytical solutions, established test cases and reference datasets. This approach allowed AI agents considerable autonomy in software implementation while keeping the scientific requirements and criteria for correctness under human control. legoESM is intended to change how we conduct research and teach weather and climate science through its ease of use, flexibility, and agility.

\section*{A model assembled from interchangeable bricks}
Conventional ESMs remain relatively rigid. Although components such as dynamical cores and grids can often be exchanged within established modelling frameworks\cite{Giorgetta2018,Hohenegger2023}, parameterizations and their interactions have evolved together over decades, making it difficult to modify or replace individual components without substantial effort elsewhere in the model. Moreover, obtaining a well-calibrated and physically credible simulation after such changes can be costly and time-consuming\cite{Hourdin2017,Couvreux2021,Bonnet2025}. legoESM aims to reduce these barriers through a modular, differentiable architecture in which alternative schemes and hypotheses can be exchanged and tested, and new model configurations could potentially be rapidly calibrated and evaluated.

\begin{figure*}[!htbp]
\centering
\includegraphics[width=\textwidth]{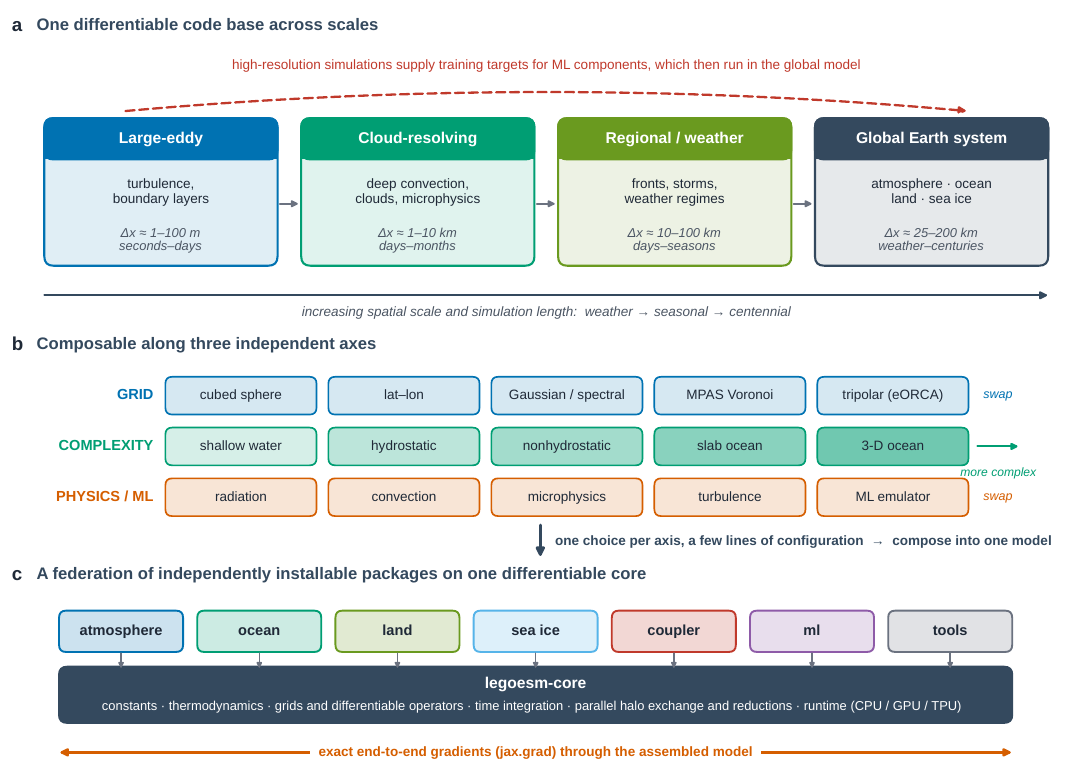}
\caption{\textbf{legoESM: one differentiable code base, composable across scales.}
\textbf{a}, The same code runs from metre-scale large-eddy simulation, through cloud-resolving and
regional configurations, to the global Earth system, spanning weather to centennial timescales.
High-resolution configurations supply training targets for machine-learned components, which are
then deployed in the global model (red dashed arrow). \textbf{b}, Any configuration is assembled by
choosing independently along three axes: the numerical grid, the level of complexity, and the set of
physics schemes or machine-learned emulators. Complexity runs along a ladder of
increasing realism, while physics tendencies are additive to machine precision, so processes can
be switched on and off freely. \textbf{c}, The selected components assemble into one model, a
federation of independently installable packages that share the \texttt{legoesm-core} substrate
and depend on it rather than on each other. Because the whole stack is a single differentiable
program, exact gradients propagate through the assembled model and identical code runs on CPU, GPU
and TPU. Package contents and the full scheme inventory are given in Methods and Extended Data
Tables~1--10.}
\label{fig:arch}
\end{figure*}

Written in Python with JAX\cite{Bradbury2018} (a differentiable machine-learning library) mostly by AI coding agents, legoESM is assembled from interchangeable
bricks (Fig.~\ref{fig:arch}c) at levels ranging from individual parameterizations to entire model components. For example, the atmosphere can range from a fully physical to a fully machine-learned (emulated) component, as in recent AI weather models\cite{Pathak2022,Bi2023,Lam2023,Kochkov2024}. Differentiability allows access to the model gradients (with respect to parameters or variables) at numerical precision and seamless interoperability and training between physics and ML, with end-to-end optimization\cite{Gelbrecht2023,Shen2023}. 
The atmosphere, ocean, land, and sea ice are independent and composable packages. Each
package can run on its own, so a single command yields a working ocean or atmosphere model, and the same code allows composing components as a fully coupled  system (e.g., ocean-atmosphere) without modification through a differentiable coupler. All of those packages build on one shared foundation (the core) composed of grids, differentiable operators, thermodynamics and time integration, and depend on that substrate but not on each other. 
A scheme, a grid or an entire component can therefore be swapped seamlessly (Methods), 
and the composed model can be optimized end-to-end in any configuration, unlike standard ESMs.

Writing the model in JAX enables two key properties: the same code runs natively on any current hardware (CPUs, GPUs and TPUs), and will be able to compile to accelerators that do not yet exist without any rewriting, sidestepping the multi-year multi-person effort of porting a legacy model to new hardware. Second, the entire model is differentiable: the gradient of any output with respect to any parameter or variable is
obtained automatically, with no hand-written adjoint. This enables efficient gradient-based optimization of any set of variables or parameters, as well as native data assimilation (shipped for the atmosphere in this first version, see below). 

The model configuration is defined through three axes that can vary independently: (i) the numerical grid; (ii) the model complexity; and (iii) the set of physics schemes or machine-learning emulation (Fig.~\ref{fig:arch}b). 
A grid is first chosen: legoESM supports cubed-sphere (FV3-style, with a duo-grid treatment of panel edges)\cite{Lin2004,PutmanLin2007,Mouallem2023}, latitude--longitude\cite{Sadourny1972}, Gaussian/spectral\cite{Bourke1972,HoskinsSimmons1975}, unstructured Voronoi (Model for Prediction Across Scales, MPAS)\cite{Thuburn2009,Ringler2010,Skamarock2012} and NEMO-style tripolar\cite{Madec2017,Murray1996} grids, as well as doubly periodic Cartesian planes for large-eddy and cloud-resolving simulations. 
The model complexity is flexible, spanning a ladder of physical complexity, from shallow-water to hydrostatic to
nonhydrostatic atmospheres, and from a prescribed sea surface temperature (SST) to a slab to a
full three-dimensional ocean, and a single framework ranges from metre-scale turbulence to the
entire Earth circulation (Supplementary Figs.~S1--S3 and~S5 for shallow water, S4 for a hydrostatic baroclinic wave, S9--S11 for nonhydrostatic large-eddy and cloud-resolving simulations, S13 for an atmosphere over prescribed SST and S14--S18 for the three-dimensional ocean). The physics are additive and composable: each process computes its tendency from the same model state and the tendencies are summed, so a parameterization can be added or removed cleanly (Supplementary Information~\S3). Mass is conserved by construction, and energy and
momentum are tracked with explicit budget closure. legoESM currently offers more than 160 selectable parameterizations, with tens of convection, turbulence and microphysics choices (Extended Data Tables~1--10), and any module or Earth component can be exchanged for a machine-learned alternative (e.g., 
a spherical Fourier neural operator (SFNO) emulation of the entire atmosphere), giving a continuous ladder from fully
physical to fully AI based instantiation of the model. Full equations and numerics are presented in Supplementary Information~\S1, and
the source is openly available at \url{https://github.com/climate-federation/legoESM}.

\paragraph{Development by AI agents under a scientific contract.}
legoESM was built through a human--agent workflow that we summarize here (details in Methods). We, human scientists, first
wrote, with an agentic coding environment, a global plan and a specification (a project ``constitution'') stating the
vision of the model (composable, swappable, differentiable) and the model requirements (e.g., conserved quantities and unit consistency). Throughout, we, humans, dictated the scientific logic and the agents implemented
and reviewed the code. We used coding agents built on independent large-language-model (LLM) families (Claude, Anthropic;
Codex, OpenAI; and GLM, Zhipu AI), with clear rules to limit duplication. Shared physical constants and core routines (for
example, thermodynamics) were centralized in the core package (the model substrate) and never re-derived elsewhere.
For the hardest problems, such as multi-node scaling,
grid-imprint artifacts at cubed-sphere panel edges, and faithful high-resolution ocean and atmosphere-only simulations, we used
a persistent-iteration (``Ralph'') loop\cite{Huntley2025Ralph}, which re-ran an agent on the same task and fed test and
diagnostic results back at each cycle until a target criterion was met, so that the agent did not stop prematurely. We note also that agents not only coded but also leveraged a vast array of benchmarks, from the
published literature to open-source datasets and codes, and were used to assess visual and physical realism (based on the
published literature) and to relate biases to physical processes. This use of the multi-modal capabilities of LLMs therefore
went well beyond code generation, to building, testing and validating a new model.

The AI agents rapidly produced code of good quality, yet also generated hallucinations, which were at times subtle. Because credible climate simulation requires accurate code, every component was validated along a ladder of increasing complexity. First, every substantial piece of code was reviewed by at least one adversarial code-review agent built on an LLM family (Codex, OpenAI) trained separately from that of the main coding agent (Claude). The review often found subtle errors or incorrect judgments, or challenged the choices made by the coding agent, which accelerated convergence and reduced defects. Second, the agents wrote unit tests covering most of the code, testing the accuracy and physical realism of small portions of code (several of them are available in the code repository). Third, physical equations were tested in asymptotic single-physics cases against well-known theoretical behaviour, such as convergence at the design order under Courant--Friedrichs--Lewy (CFL)-controlled refinement (Supplementary Figs.~S24 and~S25). Finally, components were tested with several interacting processes switched on, against increasingly stringent, larger-scale benchmarks, from dynamical-core tests of rising complexity (detailed below) to full atmosphere-only and ocean-only integrations
compared against established models such as the ICON ESM\cite{Giorgetta2018}, the Nucleus for European Modelling of the Ocean (NEMO)
model\cite{Madec2017} and the GFDL FV3 dynamical core\cite{Lin2004,PutmanLin2007}. When available, a code ``oracle'', i.e.,
a trusted reference implementation (typically the original Fortran or C/C++ code), was used as the target, with agreement required
within numerical tolerances rather than bit for bit, because differentiability and hardware portability imposed a different
code structure and, in places, smoother formulations.

For the atmosphere, large-eddy simulations were validated, and the code corrected, against oracles based on the MicroHH\cite{vanHeerwaarden2017} large-eddy simulation code in the canonical GABLS1 stable boundary layer\cite{Beare2006}, neutral Ekman layer\cite{Andren1994}, Wangara diurnal cycle\cite{Clarke1971}, and DYCOMS-II\cite{Stevens2005} and RICO\cite{vanZanten2011} cloud-topped boundary-layer cases, swapping all subgrid-scale turbulence options (the constant-coefficient Smagorinsky closure\cite{Smagorinsky1963} and its scale-dependent dynamic variant\cite{Germano1991,PorteAgel2000}) and all microphysics options\cite{Lamb2026} (Supplementary Figs.~S9--S12). The cloud-resolving configuration followed the design of the System for Atmospheric Modeling (SAM) cloud-resolving model (CRM)\cite{Khairoutdinov2003}, and tested against this model and validated on multiple cases, including the RCEMIP\cite{Wing2018} protocol, with our different microphysics schemes. The global atmospheric model and its various grids were validated through a suite of dycore benchmarks of increasing complexity: first shallow-water cases (Williamson cases~1, 2, 5 and~6\cite{Williamson1992}, Supplementary Figs.~S1--S3 and~S5; the Galewsky barotropic instability\cite{Galewsky2004}; and colliding modons\cite{Lin2017Modons}), then three-dimensional hydrostatic and non-hydrostatic cases (including the Jablonowski--Williamson baroclinic wave\cite{JablonowskiWilliamson2006}, Supplementary Fig.~S4, and the Held--Suarez test\cite{HeldSuarez1994}), and finally radiative--convective equilibrium (RCE; Supplementary Figs.~S7, S8 and~S11) and AMIP (Supplementary Fig.~S13) experiments. For these benchmarks, the model had to be faithful to the known reference solution and give consistent results across grid types. This was generally reached within a few persistent-iteration cycles, except for the cubed sphere (aimed to replicate the FV3 duo-grid\cite{Lin2004,PutmanLin2007,Mouallem2023}), on which subtle edge gradient defects produced artifacts along the cube-panel boundaries. Removing them was challenging even with the original FV3 code as an oracle, but a newer generation of coding agent, combined with adversarial review and persistent-iteration (Ralph) loops, ultimately removed these artifacts (Supplementary Fig.~S4c).

The ocean followed a different validation strategy, as fewer clear benchmarks existed than for the atmosphere. Beyond unit tests and adversarial review, idealized asymptotic tests verified, for both the ocean and the atmosphere, that the operators (many of which were shared between the two) reproduced their analytic targets: vector-invariant (Burgers) advection (Supplementary Fig.~S24a), Laplacian viscous decay with the expected $k^2$ scaling (Supplementary Fig.~S24b), the Coriolis tendency (Supplementary Fig.~S24c), and undamped and damped inertial oscillations (Supplementary Fig.~S24d,e), while finite-volume transport and gravity-wave terms converged at their design orders under
CFL-controlled refinement in both the atmosphere (Supplementary Fig.~S25a) and the ocean (Supplementary Fig.~S25b). Idealized and then realistic ocean cases were then run against ocean-model oracles (NEMO\cite{Madec2017}, MITgcm\cite{Marshall1997} and Oceananigans\cite{Ramadhan2020}) on latitude--longitude, cubed-sphere, unstructured
(MPAS/Voronoi\cite{Ringler2013} and Finite-volumE Sea ice--Ocean Model (FESOM)-type\cite{Danilov2017}) and NEMO-style tripolar grids (Supplementary Figs.~S14--S17). In an idealized basin twin experiment, legoESM stayed within $5\times10^{-3}$\,$^\circ$C (r.m.s.) of NEMO~5.0.2 in sea-surface temperature after one year (Supplementary Fig.~S15).
Even with an oracle's source in hand, validation did not reduce to line-by-line comparison: the codebases decomposed the equations differently, and tendencies matching the oracle to 3--4 digits over the interior could still hide order-one errors in a few critical cells (e.g., topography-adjacent faces controlling the form stress that sets circumpolar transport). These residuals were instead exposed by comparing term-by-term budgets, integrated over dynamically relevant regions at identical model states, against the oracle's native trend diagnostics, which amplified what pointwise metrics hid and localized the responsible discretization. This iteration revealed missing processes (freshwater runoff), stability-motivated coefficient substitutions, and boundary-adjacent stencil differences that were successively corrected to obtain faithful simulations.

Validation of the land component was less direct, as many of its processes are biological and represented by empirical relationships rather than first-principles equations, and hence lacked benchmark cases. Soil water flow (Richards' equation) and soil heat diffusion were checked for mass and energy conservation and physically consistent behavior, using agent-based diagnostics and visual inspection. Simulated fluxes were then evaluated against eddy-covariance measurements at several flux-tower sites and at the global scale, including the seasonality of the turbulent heat fluxes by biome, the global pattern of latent heat flux, and the zonal distribution of gross primary productivity (Supplementary Figs.~S19--S22). Evaluation criteria were met within only a few iterations, with the exception of soil evaporation where high biases desiccated soil. Therefore, a modernized soil-evaporation parameterization was implemented. The surface parameters were then calibrated by gradient descent at eddy-covariance sites (Fig.~\ref{fig:landcal}; described in the section on composability and differentiable calibration below). Finally, the initial pools of the dynamic carbon model, based on DifferLand\cite{Fang2026}, were estimated either through an analytical linearized solution or by differentiable optimization, with comparable results (Supplementary Fig.~S23). Comparing the pools with observational estimates revealed, for example, too little soil carbon in northern high-latitude soils\cite{Hugelius2014}, and the agents iterated on the parameters and structure of the carbon model until the initial pools were realistic.

The resulting modeling infrastructure, developed over 9 months, comprised about 620{,}000 lines of model code in the packages,
about 824{,}000 lines of tests, through which automated continuous integration enforced the physical contracts (Methods),
and about 555{,}000 lines of simulation and analysis scripts.

Two features of this workflow mattered as much as the end result. First, code generation was fast: with agents handling
implementation and testing, adding a new parameterization or dynamical core (historically a months-to-a-year undertaking in
an established model) shrank to days, compressing the development cycle between a scientific idea or hypothesis and a
tested, benchmarked implementation. Second, human experts in the loop were critical. They set the scope and strategy at all
stages of the project, and without supervision the agents tended to diverge, coding unneeded or even unintended features or spending compute on
irrelevant items, and often losing track of the ultimate objective. Human input was needed for physical choices and for
setting main project directions, for example deciding which reference solution or oracle a component should reproduce, which parameterizations to include, the overall architecture of the model, or whether to keep or replace a simple parameterization with a more complex one based on human-defined objective, or requesting denial experiments that remove a single
process to assess its effect on the result. This back-and-forth between the agents' questions and the scientists' responses
or hypotheses reduced the overall speed, but in our opinion largely improved the quality and clarified the scope of the end software. 
Our workflow was therefore closer to recent multi-agent ``co-scientist'' systems,
in which scientists specify goals and steer agents that generate and test hypotheses\cite{Gottweis2026}, than to a fully
autonomous research pipeline.

\section*{Many grids, one set of equations}
A central design choice in legoESM is that the governing equations can be solved on different grids tailored to best suit the problem. In the atmosphere, legoESM provides shallow-water, hydrostatic primitive-equation and nonhydrostatic compressible-Euler dynamics on four global grid families: the cubed sphere\cite{Lin2004,PutmanLin2007,Mouallem2023}, latitude--longitude finite-volume grids\cite{Sadourny1972}, Gaussian/spectral transforms\cite{Bourke1972,HoskinsSimmons1975} and unstructured Voronoi (MPAS) meshes\cite{Thuburn2009,Ringler2010,Skamarock2012}, together with a doubly periodic Cartesian plane on which the nonhydrostatic core runs large-eddy and cloud-resolving simulations. The same physics therefore runs through a controlled hierarchy of dynamical cores and grids. The ocean solves the Boussinesq hydrostatic primitive equations\cite{ShchepetkinMcWilliams2005}; nonhydrostatic effects are negligible at basin-to-global scales, so no nonhydrostatic ocean core has been included yet. The ocean equations run on the latitude--longitude, cubed-sphere, MPAS/Voronoi and tripolar grids. Production simulations use the latitude--longitude C-grid, in its curvilinear form on the eORCA1 tripole mesh\cite{Madec2017,Murray1996}. The MPAS, cubed-sphere and spectral ocean configurations serve idealized and benchmark experiments. 
The specific discretizations, transport operators and their community lineages are listed in Supplementary Table~S1.
As an illustration, the same shallow-water equations solved on the cubed-sphere, lat--lon,
icosahedral and spectral grids give near-identical (within numerical bounds) solutions across all four discretizations,
including flow over an isolated mountain (Williamson~5) and the propagating Rossby--Haurwitz wave (Williamson~6)\cite{Williamson1992} (Supplementary Figs.~S1, S3 and S5), and the
cubed-sphere C-D core reproduces the Jablonowski--Williamson baroclinic
wave\cite{JablonowskiWilliamson2006} (Supplementary Fig.~S4). Removing a residual grid imprint at cube-panel boundaries,
visible only on the meridional wind and not in error norms, was one of the problems for which
the persistent-iteration (Ralph) loop described above proved essential, leading to the implementation of the duo-grid treatment\cite{Mouallem2023}
that eliminates it (Supplementary Fig.~S4c). Our advection schemes across all grids (dycores) were validated against a substantial suite of dycore benchmarks, given in Supplementary Information~\S2, ensuring that the formulations were numerically sound across all cases.

While many of the grids, numerical schemes and parameterizations can be used as composable elements, legoESM's ocean component builds on the community's knowledge and provides recipes that closely shadow the combinations in existing production models, reproducing their numerical schemes and, where verified against the reference model's own output, their trajectories. A NEMO shadow recipe is certified statement by statement against NEMO 5.0.2 on its idealized test cases, and shadow recipes exist for Veros and Oceananigans, so that a legoESM configuration can be pinned to a trusted production model before its components are swapped.

\section*{Composability and differentiable multiscale calibration}
Swapping physical schemes or model components to assess their impact has been a long-standing scientific need. Although many contemporary ESMs have adopted increasingly modular software architectures, most process representations remain tightly coupled to other model components, reflecting in part technical debt accumulated over decades of development. Exchanging such components therefore still requires source-code modifications and careful bookkeeping. In legoESM, all model components follow a common tensor-in/tendency-out interface, taking the model state as input and returning a tendency. This standardized interface allows individual process representations, or larger parts of the system (e.g., the entire atmosphere), to be exchanged without modifying the rest of the model.

We illustrate this capability along with legoESM multiscale capability with atmospheric turbulence and deep-convection parameterizations (Supplementary Figs.~S6--S8), which govern turbulent exchange in the lower atmosphere and much of tropical rainfall, respectively, and remain important sources of uncertainty and error in climate models. Nine turbulence and ten convection schemes (Extended Data Tables~4 and~2) are first run unchanged in single-column configurations: the turbulence schemes are evaluated against reference turbulent cases in legoESM large-eddy simulation mode, while the convection schemes are evaluated in radiative--convective equilibrium (RCE) at three sea-surface temperatures (295, 300 and 305\,K) against legoESM cloud-resolving simulations of the same cases. All experiments therefore use the same model infrastructure, which can itself operate in large-eddy or cloud-resolving configurations, conserving other exactly similar components (e.g., radiation scheme or microphysics). The schemes exhibit clear differences in skill (Supplementary Information~\S3), enabling controlled comparisons of their off-the-shelf performance. All ten convection schemes reach a statistically steady equilibrium, but their errors against the cloud-resolving temperature and humidity profiles differ by up to a factor of about seven, from the eddy-diffusivity/mass-flux (EDMF) scheme, which is closest to the reference, to the Kuo and deep-convective-adjustment schemes (Supplementary Figs.~S7 and~S8). Once implemented, switching among schemes requires only a one-line configuration change. We then calibrate the parameters of each turbulence and convection scheme against the corresponding large-eddy and cloud-resolving reference high-resolution simulations; a single parameter set per scheme is optimized jointly across all cases for generality. Comparing default and calibrated performance (Supplementary Fig.~S6 for the turbulence closures against LES and Supplementary Figs.~S7 and~S8 for the convection schemes before and after calibration) demonstrates how legoESM’s differentiable infrastructure complements its modular design: process representations can not only be interchanged and compared under consistent conditions, but also systematically optimized against reference data. Because differentiability is built into legoESM, parameter or initial-condition calibration becomes a gradient-based optimization problem\cite{Gelbrecht2023,Shen2023}. Gradients propagated through the model via \jaxgrad{} allow parameters to be optimized directly against data using modern gradient-based methods\cite{KingmaBa2015,Jordan2024}, rather than derivative-free searches\cite{Hourdin2017,Cleary2021}, which are nonetheless still available. The benefit of calibration differs strongly between convection schemes (Supplementary Fig.~S7a). The Tiedtke scheme improves about threefold (joint score from about 8.2 to 2.8), rising from the lower half of the ranking to third, whereas most schemes change little; for the mass-flux and Zhang--McFarlane schemes the jointly tuned parameters even score slightly worse than the defaults and differ between optimization seeds, a reminder that fitting one parameter set to several climates is a non-convex compromise (see below). Even after calibration, skill depends on the climate: the Kuo, Zhang--McFarlane and deep-convective-adjustment schemes degrade markedly as the surface warms from 295 to 305\,K, whereas EDMF improves (Supplementary Fig.~S7b). A scheme that performs well in the present-day tropics therefore need not remain skillful in a warmer climate, which is precisely the regime that matters for projections, and legoESM makes such cross-climate tests a routine part of scheme evaluation.

\begin{figure*}[!htbp]
\centering
\includegraphics[width=0.99\textwidth]{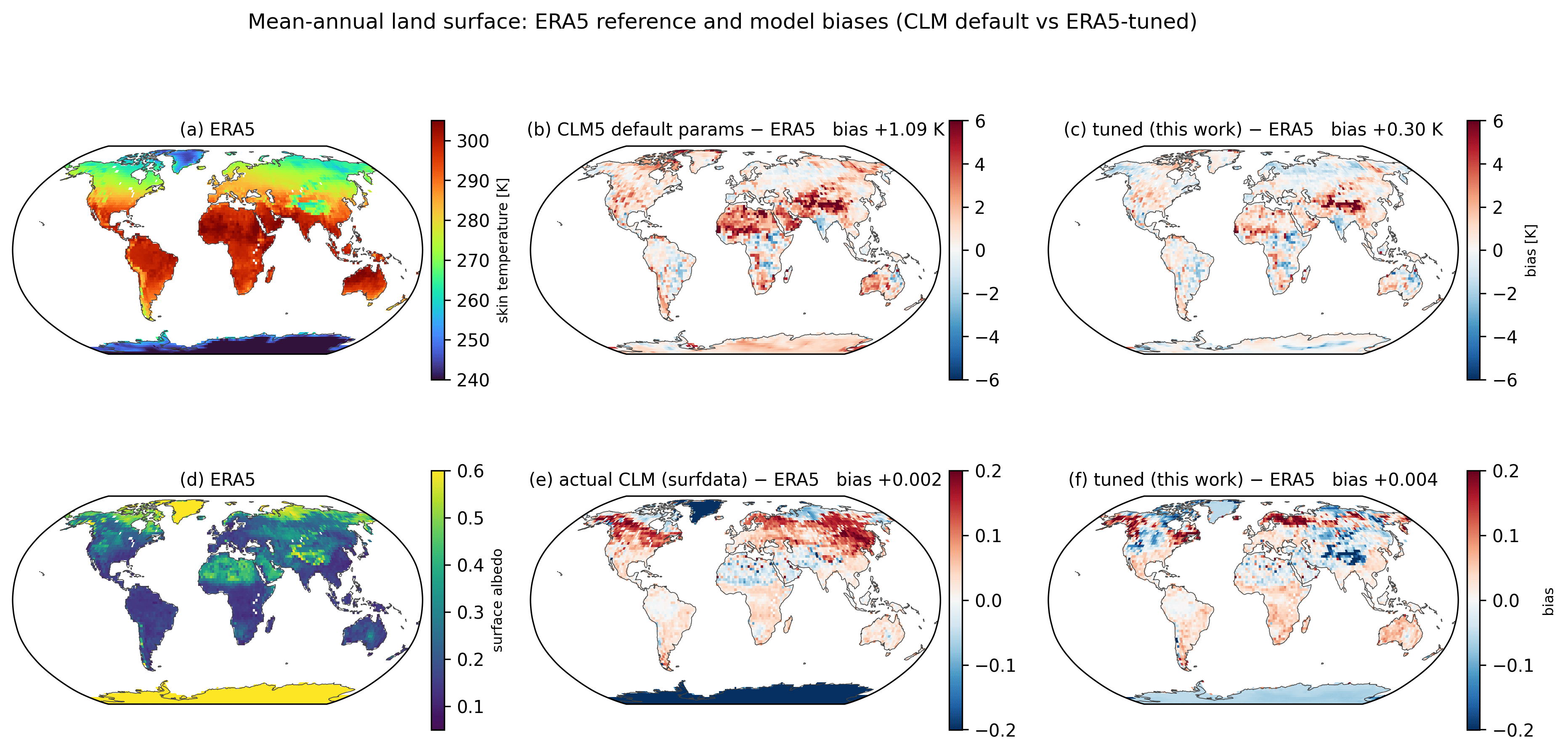}
\caption{\textbf{Differentiability: gradient-based calibration of the land surface.}
Per-plant-functional-type surface parameters are tuned by gradient descent \emph{through the
differentiable land model} against ERA5\cite{Hersbach2020} mean-annual fields, with physical bounds imposed by
differentiable (sigmoid) transforms. Top: skin temperature. Starting from CLM5\cite{Lawrence2019} default
parameters (b), the global mean warm bias against ERA5 (a) falls from $+1.09$ to $+0.30$\,K after
tuning (c), with the largest gains over arid and semi-arid land. Bottom: surface albedo (d--f)
stays essentially unbiased through the calibration ($+0.002$ to $+0.004$), confirming that the
temperature fit is not bought at the cost of the radiative surface state. Gradients are obtained
by reverse-mode differentiation of the land model at the global scale, per plant functional type; experimental set-up is in Methods and Supplementary
Information~\S6.}
\label{fig:landcal}
\end{figure*}

We further extend this approach to the global land surface by tuning per-plant-functional-type surface parameters globally by gradient descent against ERA5\cite{Hersbach2020} mean-annual
skin temperature and MODIS albedo (Fig.~\ref{fig:landcal}). We reduce the global mean warm temperature bias from $+1.09$ to $+0.30$\,K while keeping the surface albedo essentially unbiased ($+0.002$ to $+0.004$), with the largest improvements over the previously too-warm arid and semi-arid land. This is the same machinery that was used to calibrate turbulent- and convection-scheme parameters against the
high-resolution references, and that we also used to adjust a cloud-fraction critical-relative-humidity threshold
using gradients of the top-of-atmosphere shortwave and longwave fluxes so that the planetary
albedo falls in its observed $\sim$30\% range, and to estimate the initial land-carbon pools (Supplementary Fig.~S23; see the development section above).
The same gradient-based strategy can also replace an entire component: a classical physics suite, a
column neural network, or a full AI emulation, such as spherical neural operator\cite{Bonev2023} for the atmosphere (a learned
dynamical core) can be swapped,  which is demonstrated for the WeatherBench AI-weather model intercomparison (Supplementary
Information~\S3), and the ocean can run as a prescribed sea surface,
a simple single or multi-layer slab, or a full three-dimensional model, all from one driver.
The differentiable parallelization layer makes this
workable at scale. Exact reverse-mode gradients flow through distributed halo exchange (via a custom
adjoint that swaps send/receive ranks) and through the global reductions used by the mass-conservation fixers (Methods).

The same adjoint capability also underpins variational data assimilation, which is natively enabled in the model through differentiability. A three-dimensional
variational (3D-Var) analysis is implemented for the atmosphere: the analysis increment is
obtained by differentiating the observation operator and the background term through the model,
and it extends to 4D-Var through the differentiable time-stepping kernel, with no separately
coded adjoint model. Single-observation experiments for both 3D-Var and 4D-Var, including the
upstream propagation of the analysis increment that only 4D-Var recovers, are shown in
Supplementary Figs.~S26 and~S27 (Supplementary Information~\S10), demonstrating faithful 3D- and 4D-Var capability of the model for weather to climate data assimilation.

Accessibility to gradients is also useful for diagnostics and process understanding, legoESM makes it natural to pose
questions that are challenging in conventional models, for example one can ask which processes a given bias is
most sensitive to. The gradient
of any scalar diagnostic $J$ (a global-mean radiative imbalance, a regional precipitation bias, a
measure of tropical variability) with respect to every internal parameter and boundary
input is obtained in a single reverse pass, at a cost essentially independent of the number of
parameters; ranking those sensitivities attributes the diagnostic to the processes that control it,
where the brute-force equivalent would demand thousands of perturbed parameter integrations\cite{Murphy2004,Dagon2020,Eidhammer2024}. This shifts the
emphasis from solely using the model for predictive skill to also using the model
as a tool for understanding.

We note that these gradients are exact for the discretized model, but they are not always useful for calibration. For chaotic,
autoregressive dynamics, gradients through long integrations can vanish or explode\cite{Metz2021}, and implicit solvers,
iterative fixers and threshold-like physics can make them noisy or ill-conditioned, although probabilistic or ensemble formulations can mitigate many of these issues. 
There can also be a trade-off
between forward accuracy and the stability of the reverse pass; NeuralGCM, for instance, builds on a comparatively simple
spectral dynamical core\cite{Kochkov2024}. legoESM does not remove these limitations, but it lets users check gradients
(for example, against finite differences) and choose, for each application, the grid, dynamical core, rollout length and
loss (for example, time-averaged or ensemble statistics) that best balance forward fidelity against gradient quality.

\section*{Bridging scales and the model hierarchy}
Weather and climate processes span an enormous range of scales, from turbulent eddies at metre scale to circulations spanning the entire planet. Traditionally, different scales and regimes have often required distinct model configurations and representations of unresolved processes. legoESM instead provides a modular, swappable framework designed to facilitate experimentation across a range of spatial scales and resolutions. At sufficiently high resolution, the framework can support large-eddy and cloud-resolving configurations \cite{Smagorinsky1963}, while retaining the flexibility to reuse, adapt, or replace physical parameterizations as appropriate. This enables systematic investigation of which parameterizations and process representations remain applicable across scales, which require retuning, and which must be reformulated or replaced as processes become explicitly resolved\cite{Lamb2026}.

The model is validated across scales: first as large-eddy simulation against reference cases and against reference codes (MicroHH\cite{vanHeerwaarden2017}), reproducing the expected mean profiles and turbulence structure (Supplementary Figs.~S9 and~S10). Enlarging the domain and adding moist physics turns the model into a cloud-resolving model tested in radiative--convective equilibrium\cite{Wing2018} and against the SAM model\cite{Khairoutdinov2003}, and the model can
spontaneously organize convection into clusters (``self-aggregation''; Supplementary Fig.~S11). The warm-rain microphysics closely matches a particle-based (super-droplet)
reference\cite{Shima2009,Bartman2022} (Supplementary Fig.~S12). Microphysics schemes were additionally checked against their original Fortran or C implementations, used as oracles for the code translation, following a systematic LLM-based translation and verification
strategy\cite{Lahlou2026} and a systematic approach to microphysics-scheme development\cite{Lamb2026}.
At the global scale, the
atmosphere runs idealized cases ranging from shallow water (Supplementary Figs.~S1--S3 and~S5), 3D RCEs coupled to ocean or land, Held--Suarez\cite{HeldSuarez1994} or prescribed-ocean (Atmospheric Model Intercomparison Project, AMIP-style) simulations on any grid (Supplementary Fig.~S13), and is evaluated with ClimateEval, which builds on ESMValTool\cite{Eyring2020ESMValTool,Righi2020}, along with agents investigating the physical realism of the model (e.g., stability, mean annual precipitation, mean top of the atmosphere radiation fluxes...), and forced three-dimensional ocean integrations (Ocean Model Intercomparison Project, OMIP-style; Fig.~\ref{fig:ocean} and Supplementary Figs.~S14--S18) and offline land integrations (Supplementary Figs.~S19--S22)
were compared to reference models (next section). Some established frameworks already span large-eddy to global scales\cite{Hohenegger2023}; spanning a
metre-scale eddy to a global simulation within one differentiable code base is, to our
knowledge, not available in existing frameworks, and allows systematic hypothesis testing and facilitates model parameter tuning using more realistic fine-scale models as training data for the coarser-resolution parameterized or learned scheme\cite{Lamb2026}.

\section*{Global simulations against established models}
\begin{figure*}[!htbp]
\centering
\begin{subfigure}{\textwidth}
  \includegraphics[width=\textwidth]{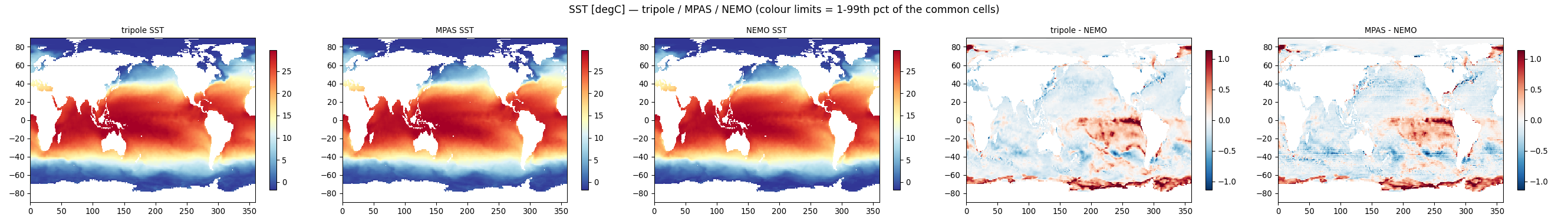}
  \caption{Sea-surface temperature: tripolar, MPAS/Voronoi and NEMO, and differences from NEMO}
\end{subfigure}\\[3pt]
\begin{subfigure}{0.9\textwidth}
  \centering
  \includegraphics[width=\textwidth]{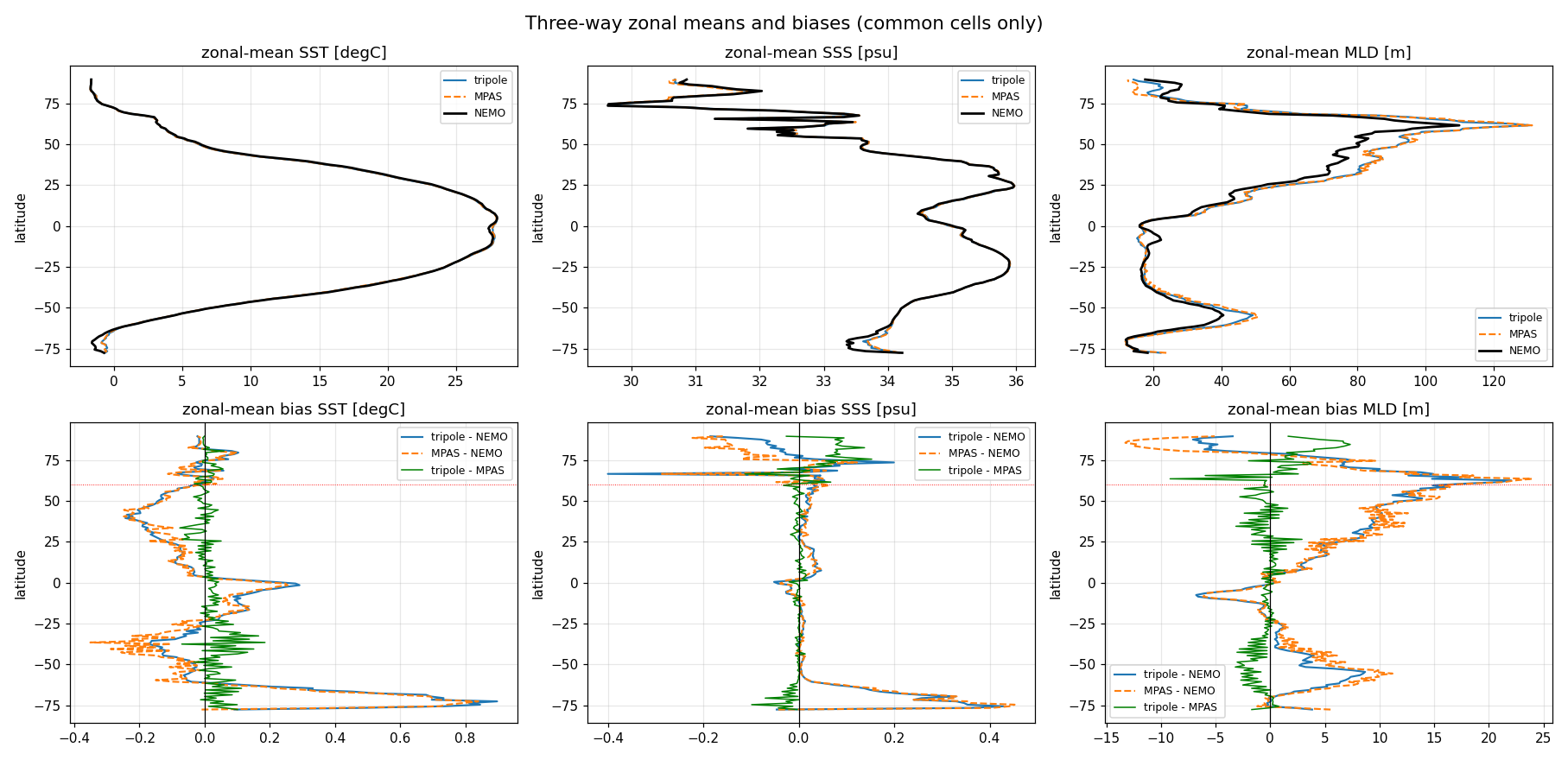}
  \caption{Zonal means (top) and zonal-mean differences (bottom) of SST, SSS and mixed-layer depth}
\end{subfigure}
\caption{\textbf{A global ocean validated against NEMO on complex grids.}
(a) Sea-surface temperature from forced legoESM ocean integrations on the NEMO-style tripolar
(eORCA) grid and on the unstructured MPAS/Voronoi icosahedral grid, from NEMO, and the differences
tripolar$-$NEMO and MPAS$-$NEMO (colour limits: 1st--99th percentiles of the cells common to all
three grids). (b) Zonal means of sea-surface temperature, sea-surface salinity and mixed-layer
depth (top) and their differences (bottom; tripolar$-$NEMO, MPAS$-$NEMO and tripolar$-$MPAS),
computed on the common cells. The two legoESM grids agree with each other much more closely than
either does with NEMO. Maps of salinity and mixed-layer depth, the idealized benchmark suite, a
NEMO twin experiment and eddy-permitting configurations are in Supplementary
Figs.~S14--S17.}
\label{fig:ocean}
\end{figure*}

A stringent test of the framework is whether its component, built from the same
composable elements, can reproduce a mature model on a complex grid, such as an
operational ocean model. For this case, the three-dimensional Boussinesq ocean uses a split-explicit baroclinic/barotropic
solver\cite{ShchepetkinMcWilliams2005} with K-profile parameterization (KPP) vertical mixing\cite{LMD1994} and Gent--McWilliams/Redi (GM--Redi) eddy
parameterization\cite{GM1990,Redi1982}, and runs on all supported grids.
Figure~\ref{fig:ocean} compares forced legoESM ocean integrations against
NEMO\cite{Madec2017} on two of the most demanding meshes: the NEMO-style tripolar (eORCA) grid,
whose folded northern boundary is a classic source of numerical artifacts, and the unstructured
MPAS/Voronoi icosahedral grid. On both grids the simulated sea-surface temperature closely tracks
NEMO (Fig.~\ref{fig:ocean}a): zonal-mean differences stay within about $\pm0.3$\,$^\circ$C between
60$^\circ$S and 60$^\circ$N, and the largest departures (up to $\sim$0.9\,$^\circ$C) are confined to
the Antarctic margin (Fig.~\ref{fig:ocean}b). Zonal-mean sea-surface salinity agrees to within a
few hundredths of a psu outside the polar regions, and the mixed layer is somewhat deeper than in
NEMO at northern mid-latitudes. The two legoESM configurations differ from each other far less
than either differs from NEMO, so the residual differences reflect model formulation, relative to NEMO, rather than
the choice of grid (Supplementary Fig.~S16). A centennial spin-up
library, an Atlantic meridional overturning circulation (AMOC) tracker, conservation diagnostics, and a peer-comparison fidelity harness
support quantitative model intercomparison (Supplementary Information~\S5), and the same code runs
eddy-permitting global configurations on 15-km MPAS and FESOM2\cite{Danilov2017} meshes (Supplementary Fig.~S17). That an
agent-built ocean reproduces a mature community model on geometrically awkward
grids is a strong test of the framework. 
Similar comparative tests have also been done on simpler configurations that are representative of different aspects of the global circulation at different resolution, e.g. the DINO \cite{kamm2025dino}.

The same drivers run the standard uncoupled
CMIP6 protocols \cite{Eyring2016} end to end. In atmosphere-only (AMIP) mode, the atmosphere with an interactive land surface are integrated over the CMIP6 historical period (1979--2014) with sea-surface temperatures and sea ice concentrations prescribed from observations and with transient CMIP6 solar, greenhouse-gas, ozone, tropospheric
aerosol and volcanic forcing. The AMIP simulation is initialized from ERA5\cite{Hersbach2020} and
written as CMOR-compliant output that ESMValTool\cite{Righi2020, Eyring2020ESMValTool} ingests with
the recipes used for CMIP6 models. The simulation is evaluated with ClimateEval\cite{WatsonParris2026ClimateBench2}: it reproduces the observed large-scale patterns of near-surface temperature, precipitation, water vapour and cloud cover, with a global-mean cold bias and a too-weak shortwave cloud radiative effect as the main biases (Supplementary Fig.~S13) without calibration. We leave the full calibration of AMIP runs for future work as it is bound to computing availability. In ocean-only (OMIP) mode, the three-dimensional ocean is driven by prescribed surface fluxes and tracks NEMO, as shown above. In land-only mode (LMIP-style, following the land-only experiments of the Land Surface, Snow and Soil Moisture Model Intercomparison Project), the land model is forced offline by CRU-JRA v2.5 reanalysis meteorology on a 2$^\circ$ grid at an hourly timestep over 1985--2014, in a two-leaf canopy biophysics-only configuration. It demonstrates comparable fidelity to a CLM5 satellite-phenology simulation \cite{Lawrence2019} driven by identical forcing, surface data and leaf-area index (Supplementary Figs.~S20--S22). All three are described, with their forcing and evaluation
metrics, in Supplementary Information~\S\S4--6. They all demonstrate physically realistic behaviour, with the remaining biases documented there.

\section*{Performance and scalability}
Writing legoESM in JAX enabled flexible portability; identical code runs unmodified on a laptop and on distributed multi-node high-performance computing systems, at single or double precision. A configuration can therefore be developed and tested on a laptop and run at production scale, and prototyping is fast enough to be scientifically useful: a ~2.8° dry Held–Suarez integration completes four simulated years in under an hour on an eight-core laptop, and in under six minutes on a single A100 GPU. At production scale, a single A100 GPU runs the $\sim$20-km lat--lon atmosphere $7.2\times$ faster than a full 128-core CPU node. Because the two architectures differ, such comparisons are best normalized by energy or allocation cost; using nominal thermal design powers (400\,W for one A100 and $2\times280$\,W for the two CPUs of a node), this corresponds to roughly a tenfold advantage per unit of nominal processor power. Laptop-GPU speed-ups are given in Supplementary Information~\S7.

\begin{figure*}[!htbp]
\centering
\includegraphics[width=\textwidth]{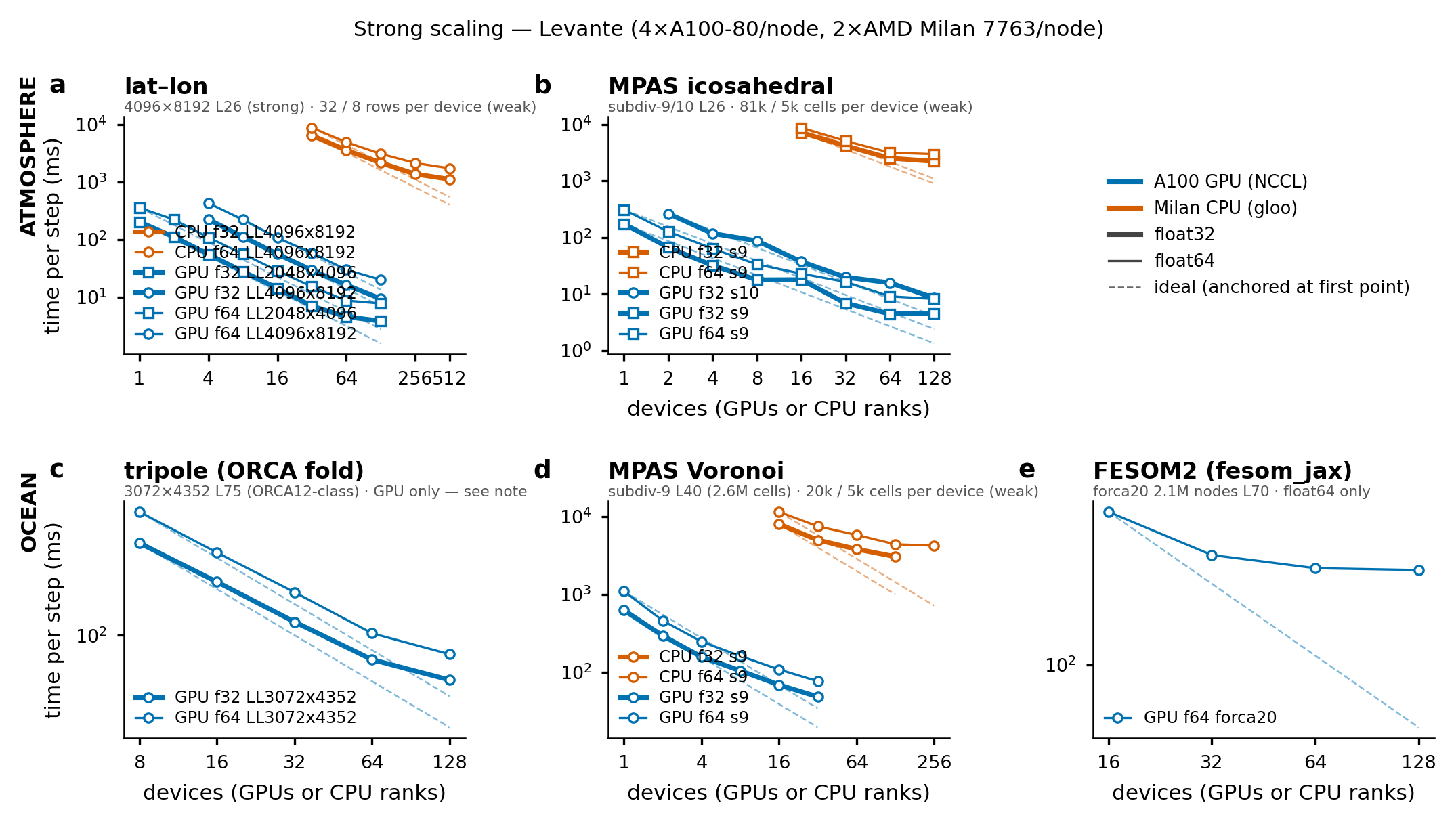}
\caption{\textbf{Strong scaling of the atmosphere and ocean at production resolution.} Wall-clock
time per model step as a function of the number of devices on the DKRZ Levante system (A100-80GB
GPUs, four per node, NCCL communication, blue; AMD EPYC 7763 Milan CPU ranks, 128 per node, gloo
communication, orange) at single (thick lines) and double (thin lines) precision; dashed lines
show ideal scaling anchored at the first point. Atmosphere: \textbf{a}, lat--lon grid at
$4096\times8192$ ($\sim$5\,km) and $2048\times4096$, 26 levels; \textbf{b}, MPAS icosahedral mesh
(subdivision levels 9 and 10, 26 levels). Ocean: \textbf{c}, ORCA12-class tripolar grid
($3072\times4352$, 75 levels; GPU only); \textbf{d}, MPAS Voronoi mesh (2.6 million cells, 40
levels); \textbf{e}, FESOM2 FORCA20 mesh (2.1 million nodes, 70 levels; double precision).
Structured grids follow the ideal line to large device counts, whereas the unstructured meshes
roll off once per-device work is too small to amortize the halo exchange. Single-device
throughput and laptop-GPU speed-ups are in Supplementary Information~\S7.}
\label{fig:scaling}
\end{figure*}

To further improve the scaling on CPUs and GPUs, we used agents to tune distributed scaling on MPI and multi GPUs for all grids, for the atmosphere and ocean. As much as possible, operators were shared between the ocean and atmosphere.
Two algorithmic changes found with agentic optimization dominated the
gains: parallelizing the correlated-$k$ radiation over its $g$-points led to a $\sim$6$\times$ speed up on the
radiation kernel and a split-explicit barotropic ocean solver was iterated on to avoid the global-reduction
latency wall of the implicit solver, being up to $1.65\times$ faster at 64 ranks. At production resolutions on the DKRZ Levante system the model uses added hardware efficiently (Fig.~\ref{fig:scaling}): the $\sim$5-km lat--lon atmosphere retains roughly 75\% parallel efficiency from 4 to 128 A100 GPUs at single precision, and the ORCA12-class tripolar ocean scales close to ideal up to 64 GPUs. Node for node (four A100 GPUs versus 128 CPU cores), the GPU partition is roughly an order of magnitude faster for the atmosphere, although, as noted above, such node-to-node comparisons do not account for differences in cost or energy use. Strong scaling is limited by the work available per device, not by total problem size, and a GPU needs a much larger subdomain than a CPU core to stay saturated: on the unstructured MPAS and FESOM2 meshes, speed-up flattens once the per-GPU subdomain becomes small and halo exchange can no longer be hidden behind computation (Fig.~\ref{fig:scaling}b,d,e). Documented limits remain (Supplementary Information~\S7): spectral transforms are communication-bound and admit no domain-decomposed path, and the structured grids outperform the unstructured MPAS core in per-cell throughput, as expected from their memory-access patterns. For context, a GPU-optimized hydrostatic ocean dynamical core (Oceananigans) simulates about a decade of near-global ocean per wall-clock day at 8-km resolution on 64 A100 GPUs\cite{Silvestri2025}; a like-for-like comparison with the legoESM ocean, at matched resolution, time step and hardware, is left for future work.

The throughput reported in Fig.~\ref{fig:scaling} is for forward model steps only; critically, reverse-mode differentiability is preserved: gradients flow correctly through MPI halo exchange and global sum reductions, verified against single-rank references to machine precision (Methods). A full gradient step for a dry dynamical core costs only 5.4x a forward step on a single GPU, and gradient checkpointing cuts peak memory effectively (from 8.0 to 0.73 GB) at no measurable wall-clock cost, making end-to-end gradients practical for short rollouts on a single device.

\section*{Accessibility, reproducibility and physical consistency}
legoESM is written in Python so that it can reach the broadest group of users, as Python is the language most widely used by
students, early-career researchers and scientists in adjacent disciplines. The model is modular and composable by construction,
so that specific parameterizations or hypotheses can be tested in isolation or coupled to specific components, and an entire
experiment is typically defined in a few lines of readable code (selecting a grid, a complexity rung, a set of physics schemes
and a forcing; Supplementary Information~\S8). The model runs without modification or manual compilation on hardware ranging
from laptops to supercomputers, which is essential for a genuine community model with global reach, for both research and
education. It is also designed for fast prototyping of new ideas or hypotheses, whether by coding directly or by delegating
the implementation to agents, which greatly shortens the model development cycle. As coding agents mature, we expect writing
model code by hand to become increasingly the exception, shifting the scientist's role towards specifying, testing and
interpreting. Finally, to make results reproducible, every run automatically records a manifest of everything needed to
repeat it (the fully resolved configuration, a checksum of the model state, the git commit and the library versions), so that
a simulation can be re-run with a single command and reproduced bit for bit on the same hardware and software stack.

\section*{Conclusion}
We present legoESM, a novel modelling framework for the Earth system made of composable, differentiable building blocks. 
legoESM shows that a single, high-level, differentiable code base can span the ESM
hierarchy, from large-eddy and cloud-resolving simulation to global atmosphere and ocean
simulations, while remaining composable enough to swap dynamical cores, grids, physics schemes and
complexity levels as interchangeable bricks, and differentiable end-to-end to calibrate variables, physical parameters, and
train embedded neural networks by gradient descent through the model. Concretely, the architecture makes practical
several things that are challenging or impossible in a conventional model: differentiating a
whole-model back to the parameter or input that controls it (end-to-end sensitivity
and attribution); training an embedded neural component online, through the model's own
evolution rather than on offline single-step tendencies; and, although not new in itself, running models across scales from a
large-eddy column to a global simulation within the same differentiable code, which facilitates clean cross-scale comparison. These capabilities,
rather than any single simulation, are the novelty we wish to demonstrate.

Two points deserve attention. First, treating differentiability and composability as
architectural building blocks, rather than features added onto an existing model, makes the
hierarchy continuous and the model flexible. Any of the parameterizations or component can be coupled to other parts of the system based on either physics or AI and, importantly, trained in tandem. Second, the model was built largely by AI
agents, but its trustworthiness rested on a human-authored scientific definition, choices and validation. 
We found adversarial agent review and persistent iteration loops essential precisely for the
problems where physical intuition is hard to encode, such as resolving grid-imprint artifacts at
cubed-sphere edges (Supplementary Fig.~S4c). A key aspect of the model concerns the role of benchmarks. Agentic development of legoESM was possible only
because the community has, over decades, distilled its understanding into open, well-posed
benchmark cases and datasets, from the standard dynamical-core tests to intercomparison projects. 
These were used to give the agents unambiguous targets against which the model could be checked, and iteratively corrected. 
Such benchmarks remain sparse for some components, notably the land surface, sea ice, biogeochemistry and the coupled
system, where fewer well-posed, openly shared reference cases exist. Community efforts to create and share such baselines
would directly benefit agent-assisted model development. 

We believe that legoESM represents a shift in computational science: while a model's source code was once
the primary repository of physical understanding, in an agentic workflow the code becomes
comparatively inexpensive to regenerate, and it is the tests, benchmarks and reference datasets,
the accumulated work of the whole community, as well as the methodology to reach the final model, 
that increasingly encode what we know and that must be curated and preserved most carefully. This is an important point for the weather and climate community in the future and likely requires a community effort to fully exploit the potential. Finally, legoESM builds on the immense body of physical understanding encoded in community models
developed over decades (from the parameterizations of the Community Earth System
Model\cite{Danabasoglu2020}, the ocean physics of NEMO\cite{Madec2017}, the dynamical cores
pioneered at GFDL, and many others), repackaged in a modern,
differentiable composable and swappable infrastructure. This paper is a demonstration of the potential of this new type of 
architecture rather than a finished ESM. The model as of now is not as complex as established
operational models. The framework is not meant to be finished but rather to be an agile, evolving infrastructure for and by the community.
This raises governance questions for a federated framework: a core team can curate, test and integrate contributions into a
common reference version, or users can fork the code and adapt it for their own purposes, which requires less long-term
maintenance but makes it harder to integrate and evaluate improvements across the framework. A combination, in which
contributions that pass the shared benchmark and contract tests are merged into a reference version, seems most promising. 

Two caveats apply. Gradient-based calibration allows data assimilation, initial state estimation and optimal parameter tuning to reduce model bias and errors, 
but it does not remove structural error in physics-based parameterizations. A well-tuned simulation can therefore still carry
substantial biases due to its inherent model structure. The flexibility of the model nonetheless allows replacing defaulting structures with pure machine learning models with many degrees of freedom and see whether those can address the biases. Moreover, gradients through long rollouts of chaotic dynamics can become noisy or explode\cite{Metz2021}, so calibration windows, time-averaged losses or ensemble statistics must be chosen with care to address those issues. Further, the model capability to swap parameterization and to use gradients for sensitivity can be used to pinpoint sources of biases. We expect that this will become an important strategy in the future to refine parameterizations. While the
differentiable coupler makes assembling components straightforward, we anticipate the same hard
coupled-system problems that challenge established models, such as spurious deep-water formation
in the Labrador Sea, drift in the Atlantic meridional overturning circulation\cite{Danabasoglu2014}, or biases in
tropical interannual variability\cite{Planton2021,Brown2026} that are likely to remain difficult to resolve. However, the capacity to calibrate parts of the model much faster should enable more rapid testing of solutions for coupled model problems. Such
differentiable, AI-ready infrastructure is a concrete step toward the hybrid ESMs
that are increasingly seen as a route to sharper, more actionable climate
projections\cite{Eyring2024}. By making the full coupled system transparent, with common coding and inline-documentation standards set from the outset, easy to
differentiate and easy to run on modern hardware, we hope to lower the barrier to
experimentation and to invite the community to build on and contribute to it.

To make the modelling framework usable and verifiable by the community, we include a set of curated reference
configurations, idealized dynamical-core tests, radiative--convective equilibrium, and test cases (from large-eddy simulations to single column models) as well as AMIP, OMIP and LMIP and cross-reference, so that a student or a new
group can reproduce a well known baseline before modifying it. As such, legoESM is
intended as an open, evolving research and educational tool, for theoretical insight, hypothesis
testing and clean counterfactual experiments, and, in time and with
sustained community contributions, potentially evolving towards a full-fledged weather and climate model. 

Finally, our modelling strategy aims to be a template that can extend beyond weather and climate. The design choices underlying legoESM
(composability, differentiability, hardware portability, and agent-assisted development under a
human-defined scientific contract) are responses to structural challenges shared by many
multiscale physical systems: astrophysics, materials science, plasma physics, fluid dynamics, and neurosciences. 
Those systems all couple processes across vastly different scales, all benefit from gradient-based inverse methods, and all face the
same tension between legacy code and modern accelerators. A composable, differentiable,
agent-built framework in which domain scientists specify the governing logic, conserved
quantities and criteria while AI agents write and mechanically verify the
implementation offers a template for simulation-based science more broadly, wherever
physical models are complex and data are abundant.

\section*{Methods}
\small
The complete governing equations and numerical methods of all components are given in
Supplementary Information~\S1; this section summarises the computational implementation.
End-to-end differentiability, conservation and stable-shape (non-retracing) JAX execution are
treated throughout as hard constraints.

\subsection*{Agentic development methodology}
legoESM was implemented by large-language-model coding agents in an agentic coding environment
under a human-authored specification. Most of the code was written by Claude (Anthropic; the Fable, Opus
and Sonnet model families) driving edits, test execution and debugging, and a second agent based
on Codex (OpenAI), a separately trained LLM family, performed adversarial review of every substantial change; GLM (Zhipu AI) was used for part of the implementation. A human-maintained
specification fixed the conserved quantities, sign and unit conventions, valid scheme sets,
reference literature and per-component acceptance criteria, and domain scientists reviewed and
adjudicated the scientific choices at each step (expert-in-the-loop). For hard,
slowly-converging problems (multi-node scaling; cube-edge grid artifacts) the persistent-iteration
loop described in the main text\cite{Huntley2025Ralph} was used; community agent-skill libraries (\texttt{agent-skills}\cite{Osmani2025AgentSkills},
\texttt{superpowers}\cite{Vincent2025Superpowers}) were used in the final hardening stages to reach production-level code quality and to surface subtle bugs. Agents also iteratively improved weak and strong CPU and GPU scaling, including multi-node MPI and multi-GPU performance, on the NSF NCAR Derecho and DKRZ Levante supercomputers. Physical contracts were enforced
mechanically throughout development by continuous-integration tripwires (see Verification harness below).

\paragraph{Use-case configurations.}
The convection sweep (Supplementary Information~\S3) ran each scheme, unchanged, in the 40-level
single-column model to a statistically steady radiative--convective equilibrium at sea-surface
temperatures of 295, 300 and 305\,K and scored the equilibrated temperature and water-vapour
profiles against cloud-resolving simulations of the same cases, aggregated into a joint score over
the three cases. One parameter set per scheme was then calibrated jointly across the three cases by
gradient descent, repeated with two random seeds. The land calibration
(Fig.~\ref{fig:landcal}) optimized per-plant-functional-type soil-water and surface parameters by
gradient descent (Adam\cite{KingmaBa2015}) against ERA5\cite{Hersbach2020} annual-mean skin
temperature, with physical bounds imposed by sigmoid transforms and gradients taken through the
differentiable land integration. Ocean comparisons (Fig.~\ref{fig:ocean}) integrated the
three-dimensional ocean on MPAS/Voronoi and tripolar grids against NEMO\cite{Madec2017}.

\paragraph{JAX implementation and differentiability.}
All components are pure-functional JAX, integrated in time with \texttt{lax.scan} and using
\texttt{jnp.where}/\texttt{lax.cond} for data-dependent branching so that the program is
differentiable and JIT-compilable. State objects and fields are registered as pytrees, so
\texttt{jit}, \texttt{grad}, \texttt{vmap} and \texttt{scan} apply through the whole coupled
model. Implementation details that preserve differentiability and avoid recompilation (for example, static
feature gates, a non-donating time-stepping kernel for use inside gradient computations, and time-varying
forcings passed as explicit traced arguments) are documented with the code.

\paragraph{Differentiable distributed execution.}
Distributed runs use \texttt{initialize\_distributed} to build the MPI topology, then scatter to
rank-local tiles, step, and gather only for I/O. Halo exchange uses a four-dimensional path that
sends all vertical levels in one message per neighbour. Reverse-mode gradients flow through
communication: every \texttt{sendrecv} is wrapped in a custom vector--Jacobian product that
swaps source and destination ranks in the backward pass, and the mass-conservation fixers
differentiate through \texttt{allreduce(SUM)}. Only summation reductions are AD-safe;
\texttt{max}/\texttt{min}/\texttt{allgather} are restricted to diagnostics. Multi-rank results
are verified against single-rank references to $\sim$$10^{-9}$, and gradients are checked the
same way.

\paragraph{Verification harness.}
The human-authored specification was enforced by mechanical tripwires in continuous integration:
physical constants could come only from a central registry; saturation curves could come only from
the shared thermodynamics module; every scheme factory had to raise on an unknown selection rather
than silently default; and additive physics composition was asserted to machine precision. Each
physics module declared a contract (units, signs, conserved quantities, differentiability,
reference, idealized test), and a self-test proved each tripwire to be non-vacuous. Dynamical cores
were gated on the standard sphere test set\cite{Williamson1992,JablonowskiWilliamson2006,HeldSuarez1994};
spatial/grid artifacts additionally required visual regression because error norms are necessary
but not sufficient. Substantial changes additionally passed an adversarial code-review agent.

\paragraph{Configurations for the figures.}
Shallow-water tests used cubed-sphere C36, lat--lon $72\times144$, icosahedral level-5 and
spectral T21 grids (Supplementary Information~\S2). RCE convection sweeps
(Supplementary Information~\S3) used the single-column model with shared physics at sea-surface
temperatures of 295, 300 and 305\,K. LES and CRM cases
used the nonhydrostatic plane core, the latter following the RCEMIP RCE300 protocol\cite{Wing2018}
(Supplementary Information~\S4). Ocean comparisons (Fig.~\ref{fig:ocean}) used MPAS/Voronoi
(icosahedral) and tripolar (eORCA) grids against NEMO\cite{Madec2017}. Single-device throughput and speed-ups were measured on a consumer-class RTX GPU, and strong
scaling (Fig.~\ref{fig:scaling}) on the DKRZ Levante system with four NVIDIA A100-80GB GPUs or two
64-core AMD EPYC 7763 CPUs per node (Supplementary Information~\S7). Full settings are in
Supplementary Information~\S1--\S10.

\normalsize

\section*{Data availability}
The ERA5 reanalysis is available from the Copernicus Climate Data Store. NEMO reference output and
the idealized-test, LES, CRM and OMIP datasets used for the figures are available from the authors
and via the model repository.

\section*{Code availability}
legoESM is open source (MIT licence) and under active development at \url{\repourl}. A separate,
version-frozen companion repository (legoESM v1.0), archived on Zenodo with a citable DOI on acceptance,
contains the notebooks, configurations and reference data needed to reproduce every figure and
experiment in this paper, so that the results remain reproducible independently of the evolving
model code.

\section*{Acknowledgements}
We thank Owen Hughes and Christiane Jablonowski for valuable discussions and Owen Hughes for pointing us to several atmospheric dycore benchmarks we did not have. legoESM builds on
parameterizations and numerical methods developed over decades by the broader climate modelling
community, including contributions from the CESM, GFDL, ECMWF, NCAR, CICE and NEMO development
teams, and we gratefully acknowledge that heritage (Extended Data Tables~1--10). We also thank
the CMIP community for the forcing datasets used in the atmosphere-, ocean- and land-only
simulations. This work was supported in part by the U.S.\ National Science Foundation Science and
Technology Center ``Learning the Earth with Artificial Intelligence and Physics'' (LEAP) award \#2019625-STC,
and by the European Research Council (ERC) Synergy Grant ``Understanding and Modelling the Earth System with Machine Learning''
(USMILE) under the European Union's Horizon 2020 research and innovation programme (grant agreement No.\ 855187).
V.E.\ and A.P.\ were supported by the Deutsche
Forschungsgemeinschaft (DFG, German Research Foundation) through the Gottfried Wilhelm Leibniz Prize awarded to Veronika Eyring (reference no.\ EY 22/2-1). legoESM was developed with the assistance of AI coding agents
(Claude, Codex and GLM) operating under a human-authored scientific specification, and made use of
the open-source \texttt{agent-skills} (\url{https://github.com/addyosmani/agent-skills}) and
\texttt{superpowers} (\url{https://github.com/obra/superpowers}) libraries. This work used resources of the Deutsches Klimarechenzentrum (DKRZ), granted by its Scientific Steering Committee (WLA) under project ID no.\ 1596. D.M.L. is supported by the NSF National Center for Atmospheric Research, which is a major facility sponsored by the U.S. National Science Foundation under Cooperative Agreement No. 1852977. 

\section*{Author contributions}
P.G.\ conceived the scientific vision, led the project, specified the scientific contract and
wrote the paper. V.E.\ co-shaped the scientific vision and helped guide the modelling and
evaluation approach. D.B., L.H.\ and A.P.\ led the OMIP, LMIP and AMIP developments and
simulations, respectively. P.G., D.B., L.H., A.P., A.A., J.F., A.L., K.D.L., D.M.L., J.L.S., J.M., D.W.-P.\ and
V.E.\ contributed to model design, component development, validation and analysis across the
atmosphere, ocean, land, sea ice, coupler, machine-learning and high-performance-computing
subsystems. All authors discussed the results and edited the manuscript.

\section*{Competing interests}
The authors declare no competing interests.

\section*{Extended Data}
\noindent legoESM re-implements, within a single differentiable framework, numerical methods and
parameterizations developed by the weather, climate, ocean and machine-learning communities over
many decades. Extended Data Tables~1--10 credit the principal schemes available in each component
together with their community lineage; the lists are representative rather than exhaustive.

\footnotesize
\renewcommand{\arraystretch}{1.2}
\newcommand{\edthead}{\toprule \textbf{Scheme} & \textbf{Description} & \textbf{Community lineage} \\ \midrule}

\smallskip
\noindent\textbf{Extended Data Table 1 $\vert$ Radiation schemes and community lineage.}\\[3pt]
\begin{tabular}{@{}p{3.0cm} p{5.7cm} p{6.3cm}@{}}
\edthead
Gray (Frierson) & Two-stream gray radiation with tunable optical depth & Frierson et al. (2006, JAS) \\
RRTMGP correlated-$k$ & 128 LW $+$ 112 SW $g$-points; full gas optics & Pincus et al. (2019, JAMES); Mlawer et al. (1997, JGR); jax-rrtmgp (Parker, Watson-Parris, Nathaniel) \\
\bottomrule
\end{tabular}

\smallskip
\noindent\textbf{Extended Data Table 2 $\vert$ Convection schemes and community lineage.}\\[3pt]
\begin{tabular}{@{}p{3.0cm} p{5.7cm} p{6.3cm}@{}}
\edthead
Simplified Betts--Miller & Relaxation to reference profile with CAPE trigger & Betts \& Miller (1986, QJRMS); Frierson (2007, JAS) \\
Bechtold & Mass-flux deep/shallow with CAPE closure & Bechtold et al. (2001, QJRMS); ECMWF IFS lineage \\
Tiedtke & Mass-flux with moisture-convergence closure & Tiedtke (1989, MWR); ECHAM lineage \\
Kain--Fritsch & Mass-flux with dilute CAPE closure & Kain \& Fritsch (1990/1993, JAS); WRF/NCAR \\
Zhang--McFarlane & CAPE-based mass-flux with downdrafts & Zhang \& McFarlane (1995, Atmos.--Ocean); CESM/CAM \\
Emanuel & Spectral mass-flux with sub-cloud quasi-equilibrium & Emanuel (1991, JAS); MIT lineage \\
Kuo & Moisture-convergence closure & Kuo (1965/1974, JAS) \\
Deep convective adjustment & Column instability removal by $T$/$q$ mixing & Manabe et al. (1965, MWR) \\
Unified mass-flux & Entraining--detraining bulk plume sharing the updraft core of the mass-flux family & Tiedtke (1989, MWR); Bechtold et al. (2001, QJRMS) lineage \\
EDMF & Eddy-diffusivity/mass-flux unified BL--shallow & Siebesma et al. (2007, JAS); ECMWF/Siebesma lineage \\
\bottomrule
\end{tabular}

\smallskip
\noindent\textbf{Extended Data Table 3 $\vert$ Microphysics schemes and community lineage.}\\[3pt]
\begin{tabular}{@{}p{3.0cm} p{5.7cm} p{6.3cm}@{}}
\edthead
Kessler & Warm-rain autoconversion, accretion, evaporation & Kessler (1969) \\
Sundqvist large-scale condensation & Prognostic cloud water with diagnostic precipitation & Sundqvist et al. (1989, MWR); ECMWF/ECHAM lineage \\
Morrison double-moment & Two-moment ice--liquid with five hydrometeor categories & Morrison et al. (2005, JAS); NCAR/WRF \\
Seifert--Beheng & Two-moment warm $+$ ice microphysics & Seifert \& Beheng (2006, Meteorol. Atmos. Phys.) \\
Thompson & Mixed-phase with aerosol-aware option & Thompson et al. (2008, MWR); WRF/NOAA \\
Predicted Particle Properties (P3) & Two-moment ice--liquid with predicted particle properties for ice phase & Morrison \& Milbrandt (2015); NCAR/WRF \\
Fast Spectral Bin Microphysics (SBM) & Fast SBM scheme & Hebrew University Cloud Model (HUCM); WRF fast-SBM (Khain et al. 2010); Shpund et al. (2019) \\
Lagrangian super-droplet microphysics & Monte-Carlo super-droplet microphysics & Shima et al. (2009, QJRMS); PySDM (Bartman et al. 2022); ERF \\
\bottomrule
\end{tabular}

\smallskip
\noindent\textbf{Extended Data Table 4 $\vert$ Boundary-layer turbulence schemes.}\\[3pt]
\begin{tabular}{@{}p{3.0cm} p{5.7cm} p{6.3cm}@{}}
\edthead
Louis & First-order stability-dependent $K(\mathrm{Ri})$ & Louis (1979, Boundary-Layer Meteorol.) \\
Holtslag--Boville & Nonlocal countergradient transport & Holtslag \& Boville (1993, J. Climate); NCAR/CESM \\
YSU & Nonlocal $K$-profile with explicit entrainment & Hong et al. (2006, MWR); WRF/Yonsei \\
TKE 1.5-order & Prognostic TKE with diagnostic length scale & Mellor \& Yamada (1982, Rev. Geophys.) \\
EDMF & Eddy-diffusivity/mass-flux unified boundary layer--shallow convection & Siebesma et al. (2007, JAS); ECMWF/Siebesma lineage \\
MYNN 2.5 & Level-2.5 TKE closure with improved mixing length & Nakanishi \& Niino (2009, J. Meteorol. Soc. Jpn); WRF/NOAA \\
CLUBB & Assumed-PDF higher-order closure & Golaz et al. (2002, JAS); CESM/E3SM \\
CLUBB-lite & Higher-order closure (simplified) & Golaz et al. (2002, JAS); CESM/E3SM \\
Smagorinsky & Diagnostic eddy viscosity for LES/CRM (constant-coefficient and scale-dependent dynamic) & Smagorinsky (1963, MWR); Germano et al. (1991); Port\'e-Agel et al. (2000) \\
\bottomrule
\end{tabular}

\smallskip
\noindent\textbf{Extended Data Table 5 $\vert$ Gravity-wave drag schemes.}\\[3pt]
\begin{tabular}{@{}p{3.0cm} p{5.7cm} p{6.3cm}@{}}
\edthead
Lindzen & Orographic and convective gravity-wave breaking & Lindzen (1981, JGR) \\
Hines & Doppler-spread non-orographic GWD & Hines (1997, JAS); ECHAM/DLR \\
McFarlane & Orographic wave breaking with saturation & McFarlane (1987, JAS); CESM/CCCma \\
Rayleigh & Sponge-layer damping & Standard implementation \\
\bottomrule
\end{tabular}

\smallskip
\noindent\textbf{Extended Data Table 6 $\vert$ Ocean physics parameterizations.}\\[3pt]
\begin{tabular}{@{}p{3.0cm} p{5.7cm} p{6.3cm}@{}}
\edthead
KPP & $K$-profile parameterization for the surface boundary layer & Large et al. (1994, Rev. Geophys.); NCAR/CESM/MOM \\
Gent--McWilliams & Isopycnal eddy transport with skew flux & Gent \& McWilliams (1990, JPO) \\
Redi diffusion & Along-isopycnal tracer diffusion & Redi (1982, JPO) \\
Visbeck adaptive GM & Flow-dependent GM coefficient & Visbeck et al. (1997, JPO) \\
Jayne--St Laurent tidal mixing & Abyssal tidal dissipation parameterization & Jayne \& St Laurent (2001, GRL) \\
Split-explicit barotropic & Free-surface sub-cycling & Shchepetkin \& McWilliams (2005, Ocean Model.) \\
\bottomrule
\end{tabular}

\smallskip
\noindent\textbf{Extended Data Table 7 $\vert$ Land-surface and carbon-cycle schemes.}\\[3pt]
\begin{tabular}{@{}p{3.0cm} p{5.7cm} p{6.3cm}@{}}
\edthead
Slab energy balance & Single-layer energy balance with bucket hydrology & Manabe (1969, MWR) \\
Multi-layer Richards & Mixed-form Picard solver with six retention curves & van Genuchten (1980, SSSAJ); CLM/CESM lineage \\
DALEC carbon & Six-pool carbon cycle (labile/foliage/root/wood/litter/SOM) & Williams et al. (2005, Global Change Biol.) \\
Farquhar photosynthesis & C3 biochemical model with stomatal conductance & Farquhar et al. (1980, Planta); Medlyn et al. (2011, Global Change Biol.) \\
\bottomrule
\end{tabular}

\smallskip
\noindent\textbf{Extended Data Table 8 $\vert$ Sea-ice schemes.}\\[3pt]
\begin{tabular}{@{}p{3.0cm} p{5.7cm} p{6.3cm}@{}}
\edthead
Zero-layer thermodynamics & Semtner zero-layer ice--ocean heat balance & Semtner (1976, JPO) \\
EVP/mEVP rheology & Elastic--viscous--plastic ice dynamics & Hunke \& Dukowicz (1997, JPO); Bouillon et al. (2013); CICE lineage \\
Multi-category ITD & Ice-thickness distribution with linear remapping & Lipscomb (2001, JGR); CICE/CESM \\
\bottomrule
\end{tabular}

\smallskip
\noindent\textbf{Extended Data Table 9 $\vert$ Coupler and surface-flux schemes.}\\[3pt]
\begin{tabular}{@{}p{3.0cm} p{5.7cm} p{6.3cm}@{}}
\edthead
COARE 3.0 bulk fluxes & State-of-the-art air--sea bulk-flux algorithm & Fairall et al. (2003, J. Climate) \\
Large \& Yeager bulk fluxes & Alternative bulk flux for climate-model coupling & Large \& Yeager (2004, NCAR Tech. Note); CESM \\
Briegleb ocean albedo & Zenith-angle-dependent ocean surface albedo & Briegleb et al. (1986); Briegleb (1992) \\
\bottomrule
\end{tabular}

\smallskip
\noindent\textbf{Extended Data Table 10 $\vert$ Machine learning and data assimilation.}\\[3pt]
\begin{tabular}{@{}p{3.0cm} p{5.7cm} p{6.3cm}@{}}
\edthead
SFNO (spherical FNO) & Spherical Fourier neural operator for learned dynamics on the sphere & Bonev et al. (2023, ICML); Pathak et al. (2022), FourCastNet \\
NeuralGCM coupling & Learned atmospheric emulator coupled to legoESM physics and ocean & Kochkov et al. (2024, Nature) \\
4D-Var DA & Incremental variational assimilation with exact adjoint from reverse-mode AD & Courtier et al. (1994, QJRMS) \\
\bottomrule
\end{tabular}
\normalsize

\bibliographystyle{unsrtnat}
\bibliography{references}

@article{Eyring2016,
  author = {Eyring, V. and Bony, S. and Meehl, G. A. and Senior, C. A. and Stevens, B. and Stouffer, R. J. and Taylor, K. E.},
  title = {Overview of the {Coupled Model Intercomparison Project Phase 6 (CMIP6)} experimental design and organization},
  journal = {Geoscientific Model Development}, volume = {9}, pages = {1937--1958}, year = {2016},
  doi = {10.5194/gmd-9-1937-2016}}

@incollection{eyringHumanInfluenceClimate2021,
  type = {Book {{Section}}},
  title = {Human {{Influence}} on the {{Climate System}}},
  booktitle = {Climate {{Change}} 2021: {{The Physical Science Basis}}. {{Contribution}} of {{Working Group I}} to the {{Sixth Assessment Report}} of the {{Intergovernmental Panel}} on {{Climate Change}}},
  author = {Eyring, V. and Gillett, N.P. and Achuta Rao, K.M. and Barimalala, R. and Barreiro Parrillo, M. and Bellouin, N. and Cassou, C. and Durack, P.J. and Kosaka, Y. and McGregor, S. and Min, S. and Morgenstern, O. and Sun, Y.},
  editor = {{Masson-Delmotte}, V. and Zhai, P. and Pirani, A. and Connors, S.L. and P{\'e}an, C. and Berger, S. and Caud, N. and Chen, Y. and Goldfarb, L. and Gomis, M.I. and Huang, M. and Leitzell, K. and Lonnoy, E. and Matthews, J.B.R. and Maycock, T.K. and Waterfield, T. and Yelek{\c c}i, O. and Yu, R. and Zhou, B.},
  year = 2021,
  pages = {423--552},
  publisher = {Cambridge University Press},
  address = {Cambridge, United Kingdom and New York, NY, USA},
  doi = {10.1017/9781009157896.005}
}

@article{WatsonParris2026ClimateBench2,
  author  = {Watson-Parris, Duncan and Tobin, Willa and Pa{\c{c}}al, Ayta{\c{c}} and Schlund, Manuel and
             Balaji, V. and Bowman, Kevin and Bretherton, Chris and Caldwell, Peter M. and
             Chapman, Will and Collins, William D. and Elsaesser, Gregory S. and Gentine, Pierre and
             Hewitt, Helene and Hoyer, Stephan and Keeling, Ralph and Koldunov, Nikolay and
             Lawrence, David M. and Lessig, Christian and Lunt, Daniel J. and Neelin, J. David and
             Pritchard, Mike and Purkey, Sarah and Schmidt, Gavin and Schneider, Tapio and
             Schulz, Michael and Shaw, Tiffany and Simpson, Isla R. and Stephens, Graeme and
             Subramanian, Aneesh C. and Teixeira, Joao and Tierney, Jessica and
             Williams, Andrew I. L. and Zanna, Laure and Eyring, Veronika and Yu, Rose},
  title   = {{ClimateBench v2.0: Probabilistic Climate Model Benchmarking}},
  journal = {Journal of Advances in Modeling Earth Systems},
  note    = {Submitted},
  year    = {2026}
}

@article{balwada2026design,
  title={Design and implementation of a data-driven parameterization for mesoscale eddy-induced transport: Idealized two-layer configurations},
  author={Balwada, Dhruv and Perezhogin, Pavel and Adcroft, Alistair and Zanna, Laure},
  journal={Journal of Advances in Modeling Earth Systems},
  volume={18},
  number={9},
  pages={e2025MS005245},
  year={2026},
  publisher={Wiley Online Library}
}

@article{perezhogin2025generalizable,
  title={Generalizable neural-network parameterization of mesoscale eddies in idealized and global ocean models},
  author={Perezhogin, Pavel and Adcroft, Alistair and Zanna, Laure},
  journal={Geophysical Research Letters},
  volume={52},
  number={19},
  pages={e2025GL117046},
  year={2025},
  publisher={Wiley Online Library}
}

@article{kamm2025dino,
  title={DINO: a diabatic model of pole-to-pole ocean dynamics to assess subgrid parameterizations across horizontal scales},
  author={Kamm, David and Deshayes, Julie and Madec, Gurvan},
  journal={Geoscientific Model Development},
  volume={18},
  number={21},
  pages={8091--8107},
  year={2025},
  publisher={Copernicus Publications G{\"o}ttingen, Germany}
}

@article{donoho2024data,
  title={Data science at the singularity},
  author={Donoho, David},
  journal={Harvard Data Science Review},
  volume={6},
  number={1},
  year={2024},
  publisher={The MIT Press}
}

@article{koldunov2026fesom2,
  title={FESOM2-JAX v1. 0: a differentiable shadow of the ocean-sea-ice model FESOM2, cast onto GPUs},
  author={Koldunov, Nikolay V and Danilov, Sergey and Cheedela, Suvarchal and Sidorenko, Dmitry and Beyer, Sebastian and Scholz, Patrick and Kuznetsov, Ivan and Streffing, Jan and Koldunov, Aleksei and Pantiukhin, Dmitrii and others},
  journal={arXiv preprint arXiv:2608.01546},
  year={2026}
}

@article{Eyring2024,
  author = {Eyring, V. and Gentine, P. and Camps-Valls, G. and Lawrence, D. M. and Reichstein, M.},
  title = {{AI}-empowered next-generation multiscale climate modelling for mitigation and adaptation},
  journal = {Nature Geoscience}, volume = {17}, pages = {963--971}, year = {2024},
  doi = {10.1038/s41561-024-01527-w}}

@article{Sherwood2020,
  author = {Sherwood, S. C. and Webb, M. J. and Annan, J. D. and Armour, K. C. and Forster, P. M. and Hargreaves, J. C. and Hegerl, G. and Klein, S. A. and Marvel, K. D. and Rohling, E. J. and others},
  title = {An assessment of {Earth's} climate sensitivity using multiple lines of evidence},
  journal = {Reviews of Geophysics}, volume = {58}, number = {4}, pages = {e2019RG000678}, year = {2020},
  doi = {10.1029/2019RG000678}}

@book{IPCC2021,
  author = {{IPCC}},
  title = {Climate Change 2021: The Physical Science Basis. Contribution of Working Group I to the Sixth Assessment Report of the Intergovernmental Panel on Climate Change},
  publisher = {Cambridge University Press}, year = {2021}, doi = {10.1017/9781009157896}}

@article{OGorman2015,
  author = {O'Gorman, P. A.},
  title = {Precipitation extremes under climate change},
  journal = {Current Climate Change Reports}, volume = {1}, pages = {49--59}, year = {2015},
  doi = {10.1007/s40641-015-0009-3}}

@article{Arora2020,
  author = {Arora, V. K. and Katavouta, A. and Williams, R. G. and Jones, C. D. and Brovkin, V. and Friedlingstein, P. and others},
  title = {Carbon--concentration and carbon--climate feedbacks in {CMIP6} models and their comparison to {CMIP5} models},
  journal = {Biogeosciences}, volume = {17}, pages = {4173--4222}, year = {2020}, doi = {10.5194/bg-17-4173-2020}}

@article{Friedlingstein2022,
  author = {Friedlingstein, P. and O'Sullivan, M. and Jones, M. W. and Andrew, R. M. and others},
  title = {Global Carbon Budget 2022},
  journal = {Earth System Science Data}, volume = {14}, pages = {4811--4900}, year = {2022}, doi = {10.5194/essd-14-4811-2022}}

@article{Held2005,
  author = {Held, I. M.},
  title = {The gap between simulation and understanding in climate modeling},
  journal = {Bulletin of the American Meteorological Society}, volume = {86}, pages = {1609--1614}, year = {2005},
  doi = {10.1175/BAMS-86-11-1609}}

@article{Jeevanjee2017,
  author = {Jeevanjee, N. and Hassanzadeh, P. and Hill, S. and Sheshadri, A.},
  title = {A perspective on climate model hierarchies},
  journal = {Journal of Advances in Modeling Earth Systems}, volume = {9}, pages = {1760--1771}, year = {2017},
  doi = {10.1002/2017MS001038}}

@article{Maher2019,
  author = {Maher, P. and Gerber, E. P. and Medeiros, B. and Merlis, T. M. and Sherwood, S. and others},
  title = {Model hierarchies for understanding atmospheric circulation},
  journal = {Reviews of Geophysics}, volume = {57}, pages = {250--280}, year = {2019}, doi = {10.1029/2018RG000607}}

@article{Gentine2018,
  author = {Gentine, P. and Pritchard, M. and Rasp, S. and Reinaudi, G. and Yacalis, G.},
  title = {Could machine learning break the convection parameterization deadlock?},
  journal = {Geophysical Research Letters}, volume = {45}, pages = {5742--5751}, year = {2018},
  doi = {10.1029/2018GL078202}}

@article{Rasp2018,
  author = {Rasp, S. and Pritchard, M. S. and Gentine, P.},
  title = {Deep learning to represent subgrid processes in climate models},
  journal = {Proceedings of the National Academy of Sciences}, volume = {115}, pages = {9684--9689}, year = {2018},
  doi = {10.1073/pnas.1810286115}}

@article{BrenowitzBretherton2018,
  author = {Brenowitz, N. D. and Bretherton, C. S.},
  title = {Prognostic validation of a neural network unified physics parameterization},
  journal = {Geophysical Research Letters}, volume = {45}, pages = {6289--6298}, year = {2018},
  doi = {10.1029/2018GL078510}}

@article{YuvalOGorman2020,
  author = {Yuval, J. and O'Gorman, P. A.},
  title = {Stable machine-learning parameterization of subgrid processes for climate modeling at a range of resolutions},
  journal = {Nature Communications}, volume = {11}, pages = {3295}, year = {2020}, doi = {10.1038/s41467-020-17142-3}}

@article{BoltonZanna2019,
  author = {Bolton, T. and Zanna, L.},
  title = {Applications of deep learning to ocean data inference and subgrid parameterization},
  journal = {Journal of Advances in Modeling Earth Systems}, volume = {11}, pages = {376--399}, year = {2019},
  doi = {10.1029/2018MS001472}}

@article{ZannaBolton2020,
  author = {Zanna, L. and Bolton, T.},
  title = {Data-driven equation discovery of ocean mesoscale closures},
  journal = {Geophysical Research Letters}, volume = {47}, pages = {e2020GL088376}, year = {2020},
  doi = {10.1029/2020GL088376}}

@article{Ott2020,
  author = {Ott, J. and Pritchard, M. and Best, N. and Linstead, E. and Curcic, M. and Baldi, P.},
  title = {A {Fortran--Keras} deep learning bridge for scientific computing},
  journal = {Scientific Programming}, volume = {2020}, pages = {8888811}, year = {2020}, doi = {10.1155/2020/8888811}}

@article{Pathak2022,
  author = {Pathak, J. and Subramanian, S. and Harrington, P. and Raja, S. and Chattopadhyay, A. and Mardani, M. and Kurth, T. and Hall, D. and Li, Z. and Azizzadenesheli, K. and Hassanzadeh, P. and Kashinath, K. and Anandkumar, A.},
  title = {{FourCastNet}: A global data-driven high-resolution weather model using adaptive {Fourier} neural operators},
  journal = {arXiv preprint arXiv:2202.11214}, year = {2022}, doi = {10.48550/arXiv.2202.11214}}

@article{Bi2023,
  author = {Bi, K. and Xie, L. and Zhang, H. and Chen, X. and Gu, X. and Tian, Q.},
  title = {Accurate medium-range global weather forecasting with {3D} neural networks},
  journal = {Nature}, volume = {619}, pages = {533--538}, year = {2023}, doi = {10.1038/s41586-023-06185-3}}

@article{Lam2023,
  author = {Lam, R. and Sanchez-Gonzalez, A. and Willson, M. and Wirnsberger, P. and Fortunato, M. and Alet, F. and Ravuri, S. and Ewalds, T. and Eaton-Rosen, Z. and Hu, W. and others},
  title = {Learning skillful medium-range global weather forecasting},
  journal = {Science}, volume = {382}, pages = {1416--1421}, year = {2023}, doi = {10.1126/science.adi2336}}

@article{Lehmann2026,
  author = {Lehmann, F. and Ozdemir, F. and Cheng, Y. and Hoefler, T. and Schemm, S. and Soja, B. and Mishra, S.},
  title = {Can {AI} weather models predict beyond two weeks? {A} quantitative benchmark and analysis of long rollouts},
  journal = {arXiv preprint arXiv:2605.30184}, year = {2026}, doi = {10.48550/arXiv.2605.30184}}

@article{Kochkov2024,
  author = {Kochkov, D. and Yuval, J. and Langmore, I. and Norgaard, P. and Smith, J. and Mooers, G. and Kl\"ower, M. and Lottes, J. and Rasp, S. and D\"uben, P. and others},
  title = {Neural general circulation models for weather and climate},
  journal = {Nature}, volume = {632}, pages = {1060--1066}, year = {2024}, doi = {10.1038/s41586-024-07744-y}}

@inproceedings{Li2021,
  author = {Li, Z. and Kovachki, N. and Azizzadenesheli, K. and Liu, B. and Bhattacharya, K. and Stuart, A. and Anandkumar, A.},
  title = {Fourier neural operator for parametric partial differential equations},
  booktitle = {International Conference on Learning Representations (ICLR)}, year = {2021}}

@inproceedings{Bonev2023,
  author = {Bonev, B. and Kurth, T. and Hundt, C. and Pathak, J. and Baust, M. and Kashinath, K. and Anandkumar, A.},
  title = {Spherical {Fourier} neural operators: Learning stable dynamics on the sphere},
  booktitle = {Proceedings of the 40th International Conference on Machine Learning (ICML)},
  series = {PMLR}, volume = {202}, pages = {2806--2823}, year = {2023}}

@misc{Bradbury2018,
  author = {Bradbury, J. and Frostig, R. and Hawkins, P. and Johnson, M. J. and Leary, C. and Maclaurin, D. and Necula, G. and Paszke, A. and Vander{P}las, J. and Wanderman-Milne, S. and Zhang, Q.},
  title = {{JAX}: composable transformations of {Python+NumPy} programs}, year = {2018}, howpublished = {\url{http://github.com/google/jax}}}

@article{Kochkov2021,
  author = {Kochkov, D. and Smith, J. A. and Alieva, A. and Wang, Q. and Brenner, M. P. and Hoyer, S.},
  title = {Machine learning--accelerated computational fluid dynamics},
  journal = {Proceedings of the National Academy of Sciences}, volume = {118}, pages = {e2101784118}, year = {2021},
  doi = {10.1073/pnas.2101784118}}

@article{Hafner2018,
  author = {H\"afner, D. and Jacobsen, R. L. and Eden, C. and Kristensen, M. R. B. and Jochum, M. and Nuterman, R. and Vinter, B.},
  title = {{Veros} v0.1 -- a fast and versatile ocean simulator in pure {Python}},
  journal = {Geoscientific Model Development}, volume = {11}, pages = {3299--3312}, year = {2018}, doi = {10.5194/gmd-11-3299-2018}}

@article{Gelbrecht2023,
  author = {Gelbrecht, M. and White, A. and Bathiany, S. and Boers, N.},
  title = {Differentiable programming for {Earth} system modeling},
  journal = {Geoscientific Model Development}, volume = {16}, pages = {3123--3135}, year = {2023}, doi = {10.5194/gmd-16-3123-2023}}

@article{Shen2023,
  author = {Shen, C. and Appling, A. P. and Gentine, P. and Bandai, T. and Gupta, H. and Tartakovsky, A. and Baity-Jesi, M. and Fenicia, F. and Kifer, D. and Li, L. and others},
  title = {Differentiable modelling to unify machine learning and physical models for geosciences},
  journal = {Nature Reviews Earth \& Environment}, volume = {4}, pages = {552--567}, year = {2023}, doi = {10.1038/s43017-023-00450-9}}

@article{Klower2024,
  author = {Kl\"ower, M. and Gelbrecht, M. and Hotta, D. and Willmert, J. and Silvestri, S. and Wagner, G. L. and White, A. and Hatfield, S. and Kimpson, T. and Constantinou, N. C. and Hill, C.},
  title = {{SpeedyWeather.jl}: Reinventing atmospheric general circulation models towards interactivity and differentiability},
  journal = {Journal of Open Source Software}, volume = {9}, number = {98}, pages = {6323}, year = {2024}, doi = {10.21105/joss.06323}}

@article{Hourdin2017,
  author = {Hourdin, F. and Mauritsen, T. and Gettelman, A. and Golaz, J.-C. and Balaji, V. and Duan, Q. and Folini, D. and Ji, D. and Klocke, D. and Qian, Y. and others},
  title = {The art and science of climate model tuning},
  journal = {Bulletin of the American Meteorological Society}, volume = {98}, pages = {589--602}, year = {2017},
  doi = {10.1175/BAMS-D-15-00135.1}}

@article{Couvreux2021,
  author = {Couvreux, F. and Hourdin, F. and Williamson, D. and Roehrig, R. and Volod{\'i}na, V. and Villefranque, N. and Rio, C. and Audouin, O. and Salter, J. and Bazile, E. and others},
  title = {Process-based climate model development harnessing machine learning: {I.} A calibration tool for parameterization improvement},
  journal = {Journal of Advances in Modeling Earth Systems}, volume = {13}, pages = {e2020MS002217}, year = {2021}, doi = {10.1029/2020MS002217}}

@article{Cleary2021,
  author = {Cleary, E. and Garbuno-Inigo, A. and Lan, S. and Schneider, T. and Stuart, A. M.},
  title = {Calibrate, emulate, sample},
  journal = {Journal of Computational Physics}, volume = {424}, pages = {109716}, year = {2021}, doi = {10.1016/j.jcp.2020.109716}}

@article{Schneider2017,
  author = {Schneider, T. and Lan, S. and Stuart, A. and Teixeira, J.},
  title = {Earth system modeling 2.0: A blueprint for models that learn from observations and targeted high-resolution simulations},
  journal = {Geophysical Research Letters}, volume = {44}, pages = {12396--12417}, year = {2017}, doi = {10.1002/2017GL076101}}

@article{Fuhrer2018,
  author = {Fuhrer, O. and Chadha, T. and Hoefler, T. and Kwasniewski, G. and Lapillonne, X. and Leutwyler, D. and L\"uthi, D. and Osuna, C. and Sch\"ar, C. and Schulthess, T. C. and Vogt, H.},
  title = {Near-global climate simulation at 1\,km resolution: establishing a performance baseline on 4888 {GPUs} with {COSMO} 5.0},
  journal = {Geoscientific Model Development}, volume = {11}, pages = {1665--1681}, year = {2018}, doi = {10.5194/gmd-11-1665-2018}}

@article{Bauer2021,
  author = {Bauer, P. and Stevens, B. and Hazeleger, W.},
  title = {A digital twin of {Earth} for the green transition},
  journal = {Nature Climate Change}, volume = {11}, pages = {80--83}, year = {2021}, doi = {10.1038/s41558-021-00986-y}}

@article{Hersbach2020,
  author = {Hersbach, H. and Bell, B. and Berrisford, P. and Hirahara, S. and Hor{\'a}nyi, A. and Mu{\~n}oz-Sabater, J. and Nicolas, J. and Peubey, C. and Radu, R. and Schepers, D. and others},
  title = {The {ERA5} global reanalysis},
  journal = {Quarterly Journal of the Royal Meteorological Society}, volume = {146}, pages = {1999--2049}, year = {2020}, doi = {10.1002/qj.3803}}

@article{Williamson1992,
  author = {Williamson, D. L. and Drake, J. B. and Hack, J. J. and Jakob, R. and Swarztrauber, P. N.},
  title = {A standard test set for numerical approximations to the shallow water equations in spherical geometry},
  journal = {Journal of Computational Physics}, volume = {102}, pages = {211--224}, year = {1992}, doi = {10.1016/S0021-9991(05)80016-6}}

@article{Galewsky2004,
  author = {Galewsky, J. and Scott, R. K. and Polvani, L. M.},
  title = {An initial-value problem for testing numerical models of the global shallow-water equations},
  journal = {Tellus A}, volume = {56}, pages = {429--440}, year = {2004}, doi = {10.3402/tellusa.v56i5.14436}}

@article{JablonowskiWilliamson2006,
  author = {Jablonowski, C. and Williamson, D. L.},
  title = {A baroclinic instability test case for atmospheric model dynamical cores},
  journal = {Quarterly Journal of the Royal Meteorological Society}, volume = {132}, pages = {2943--2975}, year = {2006}, doi = {10.1256/qj.06.12}}

@article{HeldSuarez1994,
  author = {Held, I. M. and Suarez, M. J.},
  title = {A proposal for the intercomparison of the dynamical cores of atmospheric general circulation models},
  journal = {Bulletin of the American Meteorological Society}, volume = {75}, pages = {1825--1830}, year = {1994}, doi = {10.1175/1520-0477(1994)075<1825:APFTIO>2.0.CO;2}}

@article{Lin2004,
  author = {Lin, S.-J.},
  title = {A ``vertically {Lagrangian}'' finite-volume dynamical core for global models},
  journal = {Monthly Weather Review}, volume = {132}, pages = {2293--2307}, year = {2004}, doi = {10.1175/1520-0493(2004)132<2293:AVLFDC>2.0.CO;2}}

@article{PutmanLin2007,
  author = {Putman, W. M. and Lin, S.-J.},
  title = {Finite-volume transport on various cubed-sphere grids},
  journal = {Journal of Computational Physics}, volume = {227}, pages = {55--78}, year = {2007}, doi = {10.1016/j.jcp.2007.07.022}}

@article{Ronchi1996,
  author = {Ronchi, C. and Iacono, R. and Paolucci, P. S.},
  title = {The ``cubed sphere'': A new method for the solution of partial differential equations in spherical geometry},
  journal = {Journal of Computational Physics}, volume = {124}, pages = {93--114}, year = {1996}, doi = {10.1006/jcph.1996.0047}}

@article{Mouallem2023,
  author = {Mouallem, J. and Harris, L. M. and Benson, R.},
  title = {Implementation of the novel duo-grid in {GFDL}'s {FV3} dynamical core},
  journal = {Journal of Advances in Modeling Earth Systems}, volume = {15}, pages = {e2023MS003712}, year = {2023}, doi = {10.1029/2023MS003712}}

@article{Sadourny1972,
  author = {Sadourny, R.},
  title = {Conservative finite-difference approximations of the primitive equations on quasi-uniform spherical grids},
  journal = {Monthly Weather Review}, volume = {100}, pages = {136--144}, year = {1972}, doi = {10.1175/1520-0493(1972)100<0136:CFAOTP>2.3.CO;2}}

@article{Thuburn2009,
  author = {Thuburn, J. and Ringler, T. D. and Skamarock, W. C. and Klemp, J. B.},
  title = {Numerical representation of geostrophic modes on arbitrarily structured {C}-grids},
  journal = {Journal of Computational Physics}, volume = {228}, pages = {8321--8335}, year = {2009}, doi = {10.1016/j.jcp.2009.08.006}}

@article{Ringler2010,
  author = {Ringler, T. D. and Thuburn, J. and Klemp, J. B. and Skamarock, W. C.},
  title = {A unified approach to energy conservation and potential vorticity dynamics for arbitrarily-structured {C}-grids},
  journal = {Journal of Computational Physics}, volume = {229}, pages = {3065--3090}, year = {2010}, doi = {10.1016/j.jcp.2009.12.007}}

@article{Skamarock2012,
  author = {Skamarock, W. C. and Klemp, J. B. and Duda, M. G. and Fowler, L. D. and Park, S.-H. and Ringler, T. D.},
  title = {A multiscale nonhydrostatic atmospheric model using centroidal {Voronoi} tessellations and {C}-grid staggering},
  journal = {Monthly Weather Review}, volume = {140}, pages = {3090--3105}, year = {2012}, doi = {10.1175/MWR-D-11-00215.1}}

@article{SimmonsBurridge1981,
  author = {Simmons, A. J. and Burridge, D. M.},
  title = {An energy and angular-momentum conserving vertical finite-difference scheme and hybrid vertical coordinates},
  journal = {Monthly Weather Review}, volume = {109}, pages = {758--766}, year = {1981}, doi = {10.1175/1520-0493(1981)109<0758:AEAAMC>2.0.CO;2}}

@article{HoskinsSimmons1975,
  author = {Hoskins, B. J. and Simmons, A. J.},
  title = {A multi-layer spectral model and the semi-implicit method},
  journal = {Quarterly Journal of the Royal Meteorological Society}, volume = {101}, pages = {637--655}, year = {1975}, doi = {10.1002/qj.49710142918}}

@article{Bourke1972,
  author = {Bourke, W.},
  title = {An efficient, one-level, primitive-equation spectral model},
  journal = {Monthly Weather Review}, volume = {100}, pages = {683--689}, year = {1972}, doi = {10.1175/1520-0493(1972)100<0683:AEOPSM>2.3.CO;2}}

@article{ColellaWoodward1984,
  author = {Colella, P. and Woodward, P. R.},
  title = {The Piecewise Parabolic Method ({PPM}) for gas-dynamical simulations},
  journal = {Journal of Computational Physics}, volume = {54}, pages = {174--201}, year = {1984}, doi = {10.1016/0021-9991(84)90143-8}}

@techreport{Madec2017,
  author = {Madec, G. and {the NEMO team}},
  title = {{NEMO} ocean engine, version 3.6},
  institution = {Institut Pierre-Simon Laplace (IPSL), Note du P\^ole de mod\'elisation 27}, year = {2017}}

@article{LMD1994,
  author = {Large, W. G. and McWilliams, J. C. and Doney, S. C.},
  title = {Oceanic vertical mixing: A review and a model with a nonlocal boundary layer parameterization},
  journal = {Reviews of Geophysics}, volume = {32}, pages = {363--403}, year = {1994}, doi = {10.1029/94RG01872}}

@article{GM1990,
  author = {Gent, P. R. and McWilliams, J. C.},
  title = {Isopycnal mixing in ocean circulation models},
  journal = {Journal of Physical Oceanography}, volume = {20}, pages = {150--155}, year = {1990}, doi = {10.1175/1520-0485(1990)020<0150:IMIOCM>2.0.CO;2}}

@article{ShchepetkinMcWilliams2005,
  author = {Shchepetkin, A. F. and McWilliams, J. C.},
  title = {The Regional Oceanic Modeling System ({ROMS}): a split-explicit, free-surface, topography-following-coordinate oceanic model},
  journal = {Ocean Modelling}, volume = {9}, pages = {347--404}, year = {2005}, doi = {10.1016/j.ocemod.2004.08.002}}

@article{Marshall1997,
  author = {Marshall, J. and Adcroft, A. and Hill, C. and Perelman, L. and Heisey, C.},
  title = {A finite-volume, incompressible {Navier--Stokes} model for studies of the ocean on parallel computers},
  journal = {Journal of Geophysical Research}, volume = {102}, number = {C3}, pages = {5753--5766}, year = {1997}, doi = {10.1029/96JC02775}}

@article{HunkeDukowicz1997,
  author = {Hunke, E. C. and Dukowicz, J. K.},
  title = {An elastic--viscous--plastic model for sea ice dynamics},
  journal = {Journal of Physical Oceanography}, volume = {27}, pages = {1849--1867}, year = {1997}, doi = {10.1175/1520-0485(1997)027<1849:AEVPMF>2.0.CO;2}}

@article{Farquhar1980,
  author = {Farquhar, G. D. and von Caemmerer, S. and Berry, J. A.},
  title = {A biochemical model of photosynthetic {CO$_2$} assimilation in leaves of {C$_3$} species},
  journal = {Planta}, volume = {149}, pages = {78--90}, year = {1980}, doi = {10.1007/BF00386231}}

@article{Frierson2006,
  author = {Frierson, D. M. W. and Held, I. M. and Zurita-Gotor, P.},
  title = {A gray-radiation aquaplanet moist {GCM}. {Part I}: Static stability and eddy scale},
  journal = {Journal of the Atmospheric Sciences}, volume = {63}, pages = {2548--2566}, year = {2006}, doi = {10.1175/JAS3753.1}}

@article{Pincus2019,
  author = {Pincus, R. and Mlawer, E. J. and Delamere, J. S.},
  title = {Balancing accuracy, efficiency, and flexibility in radiation calculations for dynamical models},
  journal = {Journal of Advances in Modeling Earth Systems}, volume = {11}, pages = {3074--3089}, year = {2019}, doi = {10.1029/2019MS001621}}

@article{Tiedtke1989,
  author = {Tiedtke, M.},
  title = {A comprehensive mass flux scheme for cumulus parameterization in large-scale models},
  journal = {Monthly Weather Review}, volume = {117}, pages = {1779--1800}, year = {1989}, doi = {10.1175/1520-0493(1989)117<1779:ACMFSF>2.0.CO;2}}

@article{ZhangMcFarlane1995,
  author = {Zhang, G. J. and McFarlane, N. A.},
  title = {Sensitivity of climate simulations to the parameterization of cumulus convection in the {Canadian Climate Centre} general circulation model},
  journal = {Atmosphere--Ocean}, volume = {33}, pages = {407--446}, year = {1995}, doi = {10.1080/07055900.1995.9649539}}

@article{Smagorinsky1963,
  author = {Smagorinsky, J.},
  title = {General circulation experiments with the primitive equations. {I. The} basic experiment},
  journal = {Monthly Weather Review}, volume = {91}, pages = {99--164}, year = {1963}, doi = {10.1175/1520-0493(1963)091<0099:GCEWTP>2.3.CO;2}}

@article{Wing2018,
  author = {Wing, A. A. and Reed, K. A. and Satoh, M. and Stevens, B. and Bony, S. and Ohno, T.},
  title = {Radiative--convective equilibrium model intercomparison project ({RCEMIP})},
  journal = {Geoscientific Model Development}, volume = {11}, pages = {793--813}, year = {2018}, doi = {10.5194/gmd-11-793-2018}}

@article{Beare2006,
  author = {Beare, R. J. and Macvean, M. K. and Holtslag, A. A. M. and Cuxart, J. and Esau, I. and Golaz, J.-C. and Jimenez, M. A. and Khairoutdinov, M. and Kosovic, B. and Lewellen, D. and others},
  title = {An intercomparison of large-eddy simulations of the stable boundary layer},
  journal = {Boundary-Layer Meteorology}, volume = {118}, pages = {247--272}, year = {2006}, doi = {10.1007/s10546-004-2820-6}}

@article{Stevens2005,
  author = {Stevens, B. and Moeng, C.-H. and Ackerman, A. S. and Bretherton, C. S. and Chlond, A. and de Roode, S. and Edwards, J. and Golaz, J.-C. and Jiang, H. and Khairoutdinov, M. and others},
  title = {Evaluation of large-eddy simulations via observations of nocturnal marine stratocumulus},
  journal = {Monthly Weather Review}, volume = {133}, pages = {1443--1462}, year = {2005}, doi = {10.1175/MWR2930.1}}

@article{vanZanten2011,
  author = {van Zanten, M. C. and Stevens, B. and Nuijens, L. and Siebesma, A. P. and Ackerman, A. S. and Burnet, F. and Cheng, A. and Couvreux, F. and Jiang, H. and Khairoutdinov, M. and others},
  title = {Controls on precipitation and cloudiness in simulations of trade-wind cumulus as observed during {RICO}},
  journal = {Journal of Advances in Modeling Earth Systems}, volume = {3}, pages = {M06001}, year = {2011}, doi = {10.1029/2011MS000056}}

@article{Shima2009,
  author = {Shima, S. and Kusano, K. and Kawano, A. and Sugiyama, T. and Kawahara, S.},
  title = {The super-droplet method for the numerical simulation of clouds and precipitation: a particle-based and probabilistic microphysics model coupled with a non-hydrostatic model},
  journal = {Quarterly Journal of the Royal Meteorological Society}, volume = {135}, pages = {1307--1320}, year = {2009}, doi = {10.1002/qj.441}}

@article{Bartman2022,
  author = {Bartman, P. and Bulenok, O. and G{\'o}rski, K. and Jaruga, A. and {\L}azarski, G. and Olesik, M. and Piasecki, B. and Singer, C. and Talar, A. and Arabas, S.},
  title = {{PySDM} v1: particle-based cloud modelling package for warm-rain microphysics and aqueous chemistry},
  journal = {Journal of Open Source Software}, volume = {7}, number = {72}, pages = {3219}, year = {2022}, doi = {10.21105/joss.03219}}

@inproceedings{KingmaBa2015,
  author = {Kingma, D. P. and Ba, J.},
  title = {{Adam}: A method for stochastic optimization},
  booktitle = {International Conference on Learning Representations (ICLR)}, year = {2015}, doi = {10.48550/arXiv.1412.6980}}

@article{Redi1982,
  author = {Redi, M. H.}, title = {Oceanic isopycnal mixing by coordinate rotation},
  journal = {Journal of Physical Oceanography}, volume = {12}, pages = {1154--1158}, year = {1982}, doi = {10.1175/1520-0485(1982)012<1154:OIMBCR>2.0.CO;2}}

@phdthesis{Johansen1975,
  author = {Johansen, O.}, title = {Thermal Conductivity of Soils}, school = {University of Trondheim}, year = {1975}}

@article{Ball1987,
  author = {Ball, J. T. and Woodrow, I. E. and Berry, J. A.},
  title = {A model predicting stomatal conductance and its contribution to the control of photosynthesis under different environmental conditions},
  journal = {Progress in Photosynthesis Research}, volume = {4}, pages = {221--224}, year = {1987}, doi = {10.1007/978-94-017-0519-6_48}}

@article{Grundner2025,
  author = {Grundner, A. and Beucler, T. and Savre, J. and Lauer, A. and Schlund, M. and Eyring, V.},
  title = {Reduced cloud cover errors in a hybrid {AI}-climate model through equation discovery and automatic tuning},
  journal = {Scientific Reports}, volume = {15}, pages = {43836}, year = {2025}, doi = {10.1038/s41598-025-29155-3}}

@article{Heuer2026,
  author = {Heuer, H. and Beucler, T. and Schwabe, M. and Savre, J. and Schlund, M. and Eyring, V.},
  title = {Beyond the training data: confidence-guided mixing of parameterizations in a hybrid {AI}-climate model},
  journal = {Journal of Advances in Modeling Earth Systems}, volume = {18}, pages = {e2025MS005544}, year = {2026}, doi = {10.1029/2025MS005544}}

@article{IglesiasSuarez2024,
  author = {Iglesias-Suarez, F. and Gentine, P. and Solino-Fernandez, B. and Beucler, T. and Pritchard, M. and Runge, J. and Eyring, V.},
  title = {Causally-informed deep learning to improve climate models and projections},
  journal = {Journal of Geophysical Research: Atmospheres}, volume = {129}, pages = {e2023JD039202}, year = {2024}, doi = {10.1029/2023JD039202}}

@article{Bonnet2025,
  author = {Bonnet, P. and Pastori, L. and Schwabe, M. and Giorgetta, M. and Iglesias-Suarez, F. and Eyring, V.},
  title = {Tuning the {ICON-A} 2.6.4 climate model with machine-learning-based emulators and history matching},
  journal = {Geoscientific Model Development}, volume = {18}, pages = {3681--3706}, year = {2025}, doi = {10.5194/gmd-18-3681-2025}}

@article{Lawrence2019,
  author = {Lawrence, D. M. and Fisher, R. A. and Koven, C. D. and Oleson, K. W. and Swenson, S. C. and Bonan, G. and others},
  title = {The {Community Land Model} version 5: description of new features, benchmarking, and impact of forcing uncertainty},
  journal = {Journal of Advances in Modeling Earth Systems}, volume = {11}, pages = {4245--4287}, year = {2019}, doi = {10.1029/2018MS001583}}

@article{Davenport2026,
  title={{JCM} v1.0: A Differentiable, Intermediate-Complexity Atmospheric Model},
  author={Davenport, Ellen H. and Madan, J. Varan and Watson-Parris, Duncan},
  journal={EGUsphere [preprint]},
  year={2026},
  doi={10.5194/egusphere-2025-6266}
}

@article{Jiang2025CanVeg,
  title={{JAX-CanVeg}: A Differentiable Land Surface Model},
  author={Jiang, Peishi and others},
  journal={Water Resources Research},
  volume={61},
  number={1},
  year={2025},
  doi={10.1029/2024WR038116}
}

@article{Moses2026DJ4Earth,
  title={{DJ4Earth}: Differentiable, and performance-portable {Earth} system modeling via program transformations},
  author={Moses, William S. and Cheng, Gong and Churavy, Valentin and Gelbrecht, Maximilian and Kl{\"o}wer, Milan and Kump, Joseph and Morlighem, Mathieu and Williamson, Sarah and others and Heimbach, Patrick},
  journal={Journal of Advances in Modeling Earth Systems},
  pages={e2025MS005615},
  year={2026},
  doi={10.1029/2025MS005615}
}

@article{Bodnar2024,
  title={A foundation model for the {Earth} system ({Aurora})},
  author={Bodnar, Cristian and Bruinsma, Wessel P. and Lucic, Ana and Stanley, Megan and others},
  journal={Nature},
  year={2025},
  note={arXiv:2405.13063}
}

@article{Lang2024,
  title={{AIFS}: {ECMWF}'s data-driven forecasting system},
  author={Lang, Simon and Alexe, Mihai and Chantry, Matthew and others},
  journal={arXiv preprint arXiv:2406.01465},
  year={2024}
}

@article{Giorgetta2018,
  title={{ICON-A}, the atmosphere component of the {ICON} {Earth} system model: {I}. {Model} description},
  author={Giorgetta, Marco A. and others and Stevens, Bjorn},
  journal={Journal of Advances in Modeling Earth Systems},
  volume={10}, number={7}, pages={1613--1637}, year={2018},
  doi={10.1029/2017MS001242}
}

@article{vanHeerwaarden2017,
  title={{MicroHH} 1.0: a computational fluid dynamics code for direct numerical simulation and large-eddy simulation of atmospheric boundary layer flows},
  author={van Heerwaarden, Chiel C. and van Stratum, Bart J. H. and Heus, Thijs and Gibbs, Jeremy A. and Fedorovich, Evgeni and Mellado, Juan Pedro},
  journal={Geoscientific Model Development},
  volume={10}, number={8}, pages={3145--3165}, year={2017},
  doi={10.5194/gmd-10-3145-2017}
}

@article{Khairoutdinov2003,
  title={Cloud resolving modeling of the {ARM} summer 1997 {IOP}: model formulation, results, uncertainties, and sensitivities},
  author={Khairoutdinov, Marat F. and Randall, David A.},
  journal={Journal of the Atmospheric Sciences},
  volume={60}, number={4}, pages={607--625}, year={2003}
}

@article{Fang2026,
  title={A differentiable land model ({DifferLand}) reveals global environmental controls on the terrestrial carbon cycle},
  author={Fang, J. and others},
  journal={Nature Communications},
  year={2026},
  doi={10.1038/s41467-026-73395-4}
}

@misc{Jordan2024,
  title={{Muon}: an optimizer for hidden layers in neural networks},
  author={Jordan, K. and Jin, Y. and Boza, V. and You, J. and Cesista, F. and Newhouse, L. and Bernstein, J.},
  year={2024},
  howpublished={\url{https://kellerjordan.github.io/posts/muon/}}
}

@article{Stevens2019,
  title={{DYAMOND}: the {DYnamics} of the {Atmospheric} general circulation {Modeled} {On} {Non-hydrostatic} {Domains}},
  author={Stevens, Bjorn and Satoh, Masaki and Auger, Ludovic and others},
  journal={Progress in Earth and Planetary Science},
  volume={6}, number={1}, pages={61}, year={2019},
  doi={10.1186/s40645-019-0304-z}
}

@article{Morrison2009,
  title={Impact of cloud microphysics on the development of trailing stratiform precipitation in a simulated squall line: comparison of one- and two-moment schemes},
  author={Morrison, H. and Thompson, G. and Tatarskii, V.},
  journal={Monthly Weather Review},
  volume={137}, number={3}, pages={991--1007}, year={2009},
  doi={10.1175/2008MWR2556.1}
}

@article{Huang2009WRF4DVar,
  author  = {Huang, X.-Y. and Xiao, Q. and Barker, D. M. and Zhang, X. and Michalakes, J. and Huang, W. and Henderson, T. and Bray, J. and Chen, Y. and Ma, Z. and Dudhia, J. and Guo, Y. and Zhang, X. and Won, D.-J. and Lin, H.-C. and Kuo, Y.-H.},
  title   = {Four-Dimensional Variational Data Assimilation for {WRF}: Formulation and Preliminary Results},
  journal = {Monthly Weather Review},
  year    = {2009},
  volume  = {137},
  number  = {1},
  pages   = {299--314},
  doi     = {10.1175/2008MWR2577.1}
}

@article{ParrishDerber1992NMC,
  author  = {Parrish, D. F. and Derber, J. C.},
  title   = {The {National Meteorological Center}'s spectral statistical-interpolation analysis system},
  journal = {Monthly Weather Review},
  year    = {1992},
  volume  = {120},
  number  = {8},
  pages   = {1747--1763},
  doi     = {10.1175/1520-0493(1992)120<1747:TNMCSS>2.0.CO;2}
}

@article{Courtier1994Incremental4DVar,
  author  = {Courtier, P. and Th{\'e}paut, J.-N. and Hollingsworth, A.},
  title   = {A Strategy for Operational Implementation of {4D-Var}, Using an Incremental Approach},
  journal = {Quarterly Journal of the Royal Meteorological Society},
  year    = {1994},
  volume  = {120},
  number  = {519},
  pages   = {1367--1387},
  doi     = {10.1002/qj.49712051912}
}

@article{Solvik2025Backprop4DVar,
  author  = {Solvik, K. and Penny, S. G. and Hoyer, S.},
  title   = {{4D-Var} Using {Hessian} Approximation and Backpropagation Applied to Automatically Differentiable Numerical and Machine Learning Models},
  journal = {Journal of Advances in Modeling Earth Systems},
  year    = {2025},
  volume  = {17},
  number  = {4},
  pages   = {e2024MS004608},
  doi     = {10.1029/2024MS004608}
}

@article{Massart2010ThreeDFGAT,
  author  = {Massart, S. and Pajot, B. and Piacentini, A. and Pannekoucke, O.},
  title   = {On the Merits of Using a {3D-FGAT} Assimilation Scheme with an Outer Loop for Atmospheric Situations Governed by Transport},
  journal = {Monthly Weather Review},
  year    = {2010},
  volume  = {138},
  number  = {12},
  pages   = {4509--4522},
  doi     = {10.1175/2010MWR3237.1}
}

@misc{USMMS,
  author       = {Novick, K. and Phillips, R.},
  title        = {{AmeriFlux FLUXNET-1F US-MMS Morgan Monroe State Forest, Ver. 3-5}},
  howpublished = {AmeriFlux AMP (Dataset)},
  year         = {2022},
  doi          = {10.17190/AMF/1854369}
}

@misc{USTon,
  author       = {Ma, S. and Xu, L. and Verfaillie, J. and Baldocchi, D.},
  title        = {{AmeriFlux FLUXNET-1F US-Ton Tonzi Ranch, Ver. 4-6}},
  howpublished = {AmeriFlux AMP (Dataset)},
  year         = {2025},
  doi          = {10.17190/AMF/2204880}
}

@misc{DEObe,
  author       = {Bernhofer, C. and Gr{\"u}nwald, T.},
  title        = {{Fluxnet Archive Product from Oberb{\"a}renburg, 2008--2023}},
  year         = {2024},
  url          = {https://hdl.handle.net/11676/HCnUq6q2u2Dl4-7GlKgFVmkn}
}

@misc{DEHai,
  author       = {Knohl, A. and Tiedemann, F. and Klosterhalfen, A. and Markwitz, C. and Fellert, D. and Herbst, M. and Kolle, O. and Kreilein, H. and Kutsch, W. and Rebmann, C. and Siebicke, L.},
  title        = {{Fluxnet Archive Product from Hainich, 2000--2023}},
  year         = {2024},
  url          = {https://hdl.handle.net/11676/11Ur73qeVFIknyMheaAXZX0_}
}

@article{Pastorello2020,
  author  = {Pastorello, G. and Trotta, C. and Canfora, E. and others},
  title   = {The {FLUXNET2015} dataset and the {ONEFlux} processing pipeline for
             eddy covariance data},
  journal = {Scientific Data},
  volume  = {7},
  pages   = {225},
  year    = {2020},
  doi     = {10.1038/s41597-020-0534-3}
}

@article{Myneni2002,
  author  = {Myneni, R. B. and Hoffman, S. and Knyazikhin, Y. and others},
  title   = {Global products of vegetation leaf area and fraction absorbed {PAR}
             from year one of {MODIS} data},
  journal = {Remote Sensing of Environment},
  volume  = {83},
  pages   = {214--231},
  year    = {2002},
  doi     = {10.1016/S0034-4257(02)00074-3}
}

@article{Schaaf2002,
  author  = {Schaaf, C. B. and Gao, F. and Strahler, A. H. and others},
  title   = {First operational {BRDF}, albedo nadir reflectance products from
             {MODIS}},
  journal = {Remote Sensing of Environment},
  volume  = {83},
  pages   = {135--148},
  year    = {2002},
  doi     = {10.1016/S0034-4257(02)00091-3}
}

@article{Ryu2018,
  author  = {Ryu, Y. and Jiang, C. and Kobayashi, H. and Detto, M.},
  title   = {{MODIS}-derived global land products of shortwave radiation and
             diffuse and total photosynthetically active radiation at 5 km
             resolution from 2000},
  journal = {Remote Sensing of Environment},
  volume  = {204},
  pages   = {812--825},
  year    = {2018},
  doi     = {10.1016/j.rse.2017.09.021}
}

@article{LiFang2022,
  author  = {Li, Y. and Fang, H.},
  title   = {Real-time software for the efficient generation of the clumping
             index and its application based on the {Google Earth Engine}},
  journal = {Remote Sensing},
  volume  = {14},
  number  = {15},
  pages   = {3837},
  year    = {2022},
  doi     = {10.3390/rs14153837}
}

@article{Wei2019,
  author  = {Wei, S. and Fang, H. and Schaaf, C. B. and He, L. and Chen, J. M.},
  title   = {Global 500 m clumping index product derived from {MODIS BRDF} data
             (2001--2017)},
  journal = {Remote Sensing of Environment},
  volume  = {232},
  pages   = {111296},
  year    = {2019},
  doi     = {10.1016/j.rse.2019.111296}
}

@article{Ryu2011,
  author  = {Ryu, Y. and Baldocchi, D. D. and Kobayashi, H. and others},
  title   = {Integration of {MODIS} land and atmosphere products with a
             coupled-process model to estimate gross primary productivity and
             evapotranspiration from 1 km to global scales},
  journal = {Global Biogeochemical Cycles},
  volume  = {25},
  pages   = {GB4017},
  year    = {2011},
  doi     = {10.1029/2011GB004053}
}

@article{Celia1990,
  author  = {Celia, M. A. and Bouloutas, E. T. and Zarba, R. L.},
  title   = {A general mass-conservative numerical solution for the unsaturated
             flow equation},
  journal = {Water Resources Research},
  volume  = {26},
  pages   = {1483--1496},
  year    = {1990},
  doi     = {10.1029/WR026i007p01483}
}

@article{Sellers1985,
  author  = {Sellers, P. J.},
  title   = {Canopy reflectance, photosynthesis and transpiration},
  journal = {International Journal of Remote Sensing},
  volume  = {6},
  pages   = {1335--1372},
  year    = {1985},
  doi     = {10.1080/01431168508948283}
}

@article{KattgeKnorr2007,
  author  = {Kattge, J. and Knorr, W.},
  title   = {Temperature acclimation in a biochemical model of photosynthesis: a
             reanalysis of data from 36 species},
  journal = {Plant, Cell \& Environment},
  volume  = {30},
  pages   = {1176--1190},
  year    = {2007},
  doi     = {10.1111/j.1365-3040.2007.01690.x}
}

@article{Collatz1992,
  author  = {Collatz, G. J. and Ribas-Carbo, M. and Berry, J. A.},
  title   = {Coupled photosynthesis--stomatal conductance model for leaves of
             {C$_4$} plants},
  journal = {Australian Journal of Plant Physiology},
  volume  = {19},
  pages   = {519--538},
  year    = {1992},
  doi     = {10.1071/PP9920519}
}

@article{Reichstein2005,
  author  = {Reichstein, M. and Falge, E. and Baldocchi, D. and others},
  title   = {On the separation of net ecosystem exchange into assimilation and
             ecosystem respiration: review and improved algorithm},
  journal = {Global Change Biology},
  volume  = {11},
  number  = {9},
  pages   = {1424--1439},
  year    = {2005},
  doi     = {10.1111/j.1365-2486.2005.001002.x}
}

@article{jung2023three,
  title={Three-dimensional variational assimilation with a multivariate background error covariance for the {Model for Prediction Across Scales--Atmosphere} with the {Joint Effort for Data assimilation Integration (JEDI-MPAS 2.0.0-beta)}},
  author={Jung, B.-J. and M{\'e}n{\'e}trier, B. and Snyder, C. and Liu, Z. and Guerrette, J. J. and Ban, J. and Hern{\'a}ndez Ba{\~n}os, I. and Yu, Y. G. and Skamarock, W. C.},
  journal={Geoscientific Model Development},
  volume={17},
  number={9},
  pages={3879--3895},
  year={2024},
  doi={10.5194/gmd-17-3879-2024}
}

@article{Courtier1998ECMWF3DVar,
  author  = {Courtier, P. and Andersson, E. and Heckley, W. A. and Pailleux, J. and Vasiljevic, D. and Hamrud, M. and Hollingsworth, A. and Rabier, F. and Fisher, M.},
  title   = {The {ECMWF} implementation of three-dimensional variational
             assimilation ({3D-Var}). {I}: Formulation},
  journal = {Quarterly Journal of the Royal Meteorological Society},
  year    = {1998},
  volume  = {124},
  number  = {550},
  pages   = {1783--1807},
  doi     = {10.1002/qj.49712455002}
}

@Article{Matthes2017,
AUTHOR = {Matthes, K. and Funke, B. and Andersson, M. E. and Barnard, L. and Beer, J. and Charbonneau, P. and Clilverd, M. A. and Dudok de Wit, T. and Haberreiter, M. and Hendry, A. and Jackman, C. H. and Kretzschmar, M. and Kruschke, T. and Kunze, M. and Langematz, U. and Marsh, D. R. and Maycock, A. C. and Misios, S. and Rodger, C. J. and Scaife, A. A. and Sepp\"al\"a, A. and Shangguan, M. and Sinnhuber, M. and Tourpali, K. and Usoskin, I. and van de Kamp, M. and Verronen, P. T. and Versick, S.},
TITLE = {Solar forcing for {CMIP6} (v3.2)},
JOURNAL = {Geoscientific Model Development},
VOLUME = {10},
YEAR = {2017},
NUMBER = {6},
PAGES = {2247--2302},
URL = {https://gmd.copernicus.org/articles/10/2247/2017/},
DOI = {10.5194/gmd-10-2247-2017}
}

@Article{Eyring2020ESMValTool,
AUTHOR = {Eyring, V. and Bock, L. and Lauer, A. and Righi, M. and Schlund, M. and Andela, B. and Arnone, E. and Bellprat, O. and Br\"otz, B. and Caron, L.-P. and Carvalhais, N. and Cionni, I. and Cortesi, N. and Crezee, B. and Davin, E. L. and Davini, P. and Debeire, K. and de Mora, L. and Deser, C. and Docquier, D. and Earnshaw, P. and Ehbrecht, C. and Gier, B. K. and Gonzalez-Reviriego, N. and Goodman, P. and Hagemann, S. and Hardiman, S. and Hassler, B. and Hunter, A. and Kadow, C. and Kindermann, S. and Koirala, S. and Koldunov, N. and Lejeune, Q. and Lembo, V. and Lovato, T. and Lucarini, V. and Massonnet, F. and M\"uller, B. and Pandde, A. and P\'erez-Zan\'on, N. and Phillips, A. and Predoi, V. and Russell, J. and Sellar, A. and Serva, F. and Stacke, T. and Swaminathan, R. and Torralba, V. and Vegas-Regidor, J. and von Hardenberg, J. and Weigel, K. and Zimmermann, K.},
TITLE = {Earth System Model Evaluation Tool (ESMValTool) v2.0 -- an extended set of
large-scale diagnostics for quasi-operational and comprehensive evaluation
of Earth system models in CMIP},
JOURNAL = {Geoscientific Model Development},
VOLUME = {13},
YEAR = {2020},
NUMBER = {7},
PAGES = {3383--3438},
URL = {https://gmd.copernicus.org/articles/13/3383/2020/},
DOI = {10.5194/gmd-13-3383-2020}
}

@Article{Righi2020,
AUTHOR = {Righi, M. and Andela, B. and Eyring, V. and Lauer, A. and Predoi, V. and Schlund, M. and Vegas-Regidor, J. and Bock, L. and Br\"otz, B. and de Mora, L. and Diblen, F. and Dreyer, L. and Drost, N. and Earnshaw, P. and Hassler, B. and Koldunov, N. and Little, B. and Loosveldt Tomas, S. and Zimmermann, K.},
TITLE = {{Earth System Model Evaluation Tool (ESMValTool)} v2.0 -- technical overview},
JOURNAL = {Geoscientific Model Development},
VOLUME = {13},
YEAR = {2020},
NUMBER = {3},
PAGES = {1179--1199},
URL = {https://gmd.copernicus.org/articles/13/1179/2020/},
DOI = {10.5194/gmd-13-1179-2020}
}

@article{Berger1978,
  author = {Berger, A. L.},
  title = {Long-term variations of daily insolation and {Quaternary} climatic changes},
  journal = {Journal of the Atmospheric Sciences},
  volume = {35},
  number = {12},
  pages = {2362--2367},
  year = {1978},
  doi = {10.1175/1520-0469(1978)035<2362:LTVODI>2.0.CO;2}
}

@article{Kinne2013,
  author = {Kinne, S. and O'Donnell, D. and Stier, P. and Kloster, S. and Zhang, K. and Schmidt, H. and Rast, S. and Giorgetta, M. and Eck, T. F. and Stevens, B.},
  title = {{MAC-v1}: A new global aerosol climatology for climate studies},
  journal = {Journal of Advances in Modeling Earth Systems},
  volume = {5},
  number = {4},
  pages = {704--740},
  year = {2013},
  doi = {10.1002/jame.20035}
}

@techreport{TaylorWilliamsonZwiers2000,
  title={The sea surface temperature and sea-ice concentration boundary conditions for {AMIP II} simulations},
  author={Taylor, K. E. and Williamson, D. and Zwiers, F.},
  year={2000},
  type={PCMDI Report},
  number={60},
  institution={Program for Climate Model Diagnosis and Intercomparison, Lawrence Livermore National Laboratory},
  address={Livermore, CA},
  url={https://pcmdi.llnl.gov/report/ab60.html}
}

@Article{Meinshausen2017,
AUTHOR = {Meinshausen, M. and Vogel, E. and Nauels, A. and Lorbacher, K. and Meinshausen, N. and Etheridge, D. M. and Fraser, P. J. and Montzka, S. A. and Rayner, P. J. and Trudinger, C. M. and Krummel, P. B. and Beyerle, U. and Canadell, J. G. and Daniel, J. S. and Enting, I. G. and Law, R. M. and Lunder, C. R. and O'Doherty, S. and Prinn, R. G. and Reimann, S. and Rubino, M. and Velders, G. J. M. and Vollmer, M. K. and Wang, R. H. J. and Weiss, R.},
TITLE = {Historical greenhouse gas concentrations for climate modelling ({CMIP6})},
JOURNAL = {Geoscientific Model Development},
VOLUME = {10},
YEAR = {2017},
NUMBER = {5},
PAGES = {2057--2116},
URL = {https://gmd.copernicus.org/articles/10/2057/2017/},
DOI = {10.5194/gmd-10-2057-2017}
}

@misc{ChecaGarcia2018,
  author = {Checa-Garcia, R.},
  title = {{CMIP6} ozone forcing dataset: supporting information},
  howpublished = {Zenodo},
  year = {2018},
  doi = {10.5281/zenodo.1135127}
}

@article{Richards1931,
  title = {Capillary conduction of liquids through porous mediums},
  author = {Richards, L. A.},
  year = {1931},
  journal = {Physics},
  volume = {1},
  number = {5},
  pages = {318--333},
  doi = {10.1063/1.1745010}
}

@article{Danabasoglu2020,
  author = {Danabasoglu, G. and Lamarque, J.-F. and Bacmeister, J. and others},
  title = {The {Community Earth System Model Version 2 (CESM2)}},
  journal = {Journal of Advances in Modeling Earth Systems},
  volume = {12},
  pages = {e2019MS001916},
  year = {2020},
  doi = {10.1029/2019MS001916}
}

@article{Boucher2020,
  author = {Boucher, O. and Servonnat, J. and Albright, A. L. and others},
  title = {Presentation and evaluation of the {IPSL-CM6A-LR} climate model},
  journal = {Journal of Advances in Modeling Earth Systems},
  volume = {12},
  pages = {e2019MS002010},
  year = {2020},
  doi = {10.1029/2019MS002010}
}

@article{Sellar2019,
  author = {Sellar, A. A. and Jones, C. G. and Mulcahy, J. P. and others},
  title = {{UKESM1}: Description and evaluation of the {U.K.} {Earth} System Model},
  journal = {Journal of Advances in Modeling Earth Systems},
  volume = {11},
  pages = {4513--4558},
  year = {2019},
  doi = {10.1029/2019MS001739}
}

@misc{ECMWF2023IFS,
  author = {{ECMWF}},
  title = {{IFS} Documentation {CY48R1}},
  howpublished = {European Centre for Medium-Range Weather Forecasts, Reading, UK},
  year = {2023},
}

@misc{Meunier2025,
  author = {Meunier, E. and others},
  title = {Towards fully differentiable neural ocean model with {Veros}},
  howpublished = {arXiv:2511.17427},
  year = {2025}
}

@article{Lamb2026,
  author = {Lamb, K. D. and others},
  title = {Perspectives on systematic cloud microphysics scheme development with machine learning},
  journal = {Journal of Advances in Modeling Earth Systems},
  year = {2026},
  doi = {10.1029/2025MS005341}
}

@misc{Metz2021,
  author = {Metz, L. and Freeman, C. D. and Schoenholz, S. S. and Kachman, T.},
  title = {Gradients are not all you need},
  howpublished = {arXiv:2111.05803},
  year = {2021}
}

@misc{Brown2026,
  author = {Brown, F. and Bischof, S. and Pearson, B. and Sengupta, A. and Bonnet, P. and others},
  title = {Fresh eyes on {CMIP} model biases: diagnosing biases across {CMIP} generations and implications for {CMIP7}},
  howpublished = {EGUsphere [preprint]},
  year = {2026},
  doi = {10.5194/egusphere-2026-3510}
}

@article{Gottweis2026,
  author = {Gottweis, J. and Weng, W.-H. and Daryin, A. and others},
  title = {Accelerating scientific discovery with {Co-Scientist}},
  journal = {Nature},
  volume = {655},
  pages = {487--496},
  year = {2026},
  doi = {10.1038/s41586-026-10644-y}
}

@misc{Lahlou2026,
  author = {Lahlou, A. and Hawkins, L. and Gentine, P.},
  title = {Systematic {LLM} translation of legacy scientific code to differentiable frameworks: application to a land surface model},
  howpublished = {arXiv:2606.07681},
  year = {2026}
}

@article{Hohenegger2023,
  author = {Hohenegger, C. and Korn, P. and Linardakis, L. and others and Stevens, B.},
  title = {{ICON-Sapphire}: simulating the components of the {Earth} system and their interactions at kilometer and subkilometer scales},
  journal = {Geoscientific Model Development},
  volume = {16},
  pages = {779--811},
  year = {2023},
  doi = {10.5194/gmd-16-779-2023}
}

@article{Murray1996,
  author = {Murray, R. J.},
  title = {Explicit generation of orthogonal grids for ocean models},
  journal = {Journal of Computational Physics},
  volume = {126},
  pages = {251--273},
  year = {1996},
  doi = {10.1006/jcph.1996.0136}
}

@article{Andren1994,
  author = {Andr{\'e}n, A. and Brown, A. R. and Graf, J. and Mason, P. J. and Moeng, C.-H. and Nieuwstadt, F. T. M. and Schumann, U.},
  title = {Large-eddy simulation of a neutrally stratified boundary layer: a comparison of four computer codes},
  journal = {Quarterly Journal of the Royal Meteorological Society},
  volume = {120},
  pages = {1457--1484},
  year = {1994},
  doi = {10.1002/qj.49712052003}
}

@techreport{Clarke1971,
  author = {Clarke, R. H. and Dyer, A. J. and Brook, R. R. and Reid, D. G. and Troup, A. J.},
  title = {The {Wangara} experiment: boundary layer data},
  institution = {CSIRO Division of Meteorological Physics},
  type = {Technical Paper},
  number = {19},
  year = {1971}
}

@misc{Huntley2025Ralph,
  author = {Huntley, G.},
  title = {Ralph: a persistent-iteration loop for coding agents},
  howpublished = {\url{https://ghuntley.com/ralph/}},
  year = {2025}
}

@misc{Osmani2025AgentSkills,
  author = {Osmani, A.},
  title = {agent-skills},
  howpublished = {\url{https://github.com/addyosmani/agent-skills}},
  year = {2025}
}

@misc{Vincent2025Superpowers,
  author = {Vincent, J.},
  title = {superpowers},
  howpublished = {\url{https://github.com/obra/superpowers}},
  year = {2025}
}

@article{Hugelius2014,
  author = {Hugelius, G. and Strauss, J. and Zubrzycki, S. and others},
  title = {Estimated stocks of circumpolar permafrost carbon with quantified uncertainty ranges and identified data gaps},
  journal = {Biogeosciences},
  volume = {11},
  pages = {6573--6593},
  year = {2014},
  doi = {10.5194/bg-11-6573-2014}
}

@article{Murphy2004,
  author = {Murphy, J. M. and Sexton, D. M. H. and Barnett, D. N. and Jones, G. S. and Webb, M. J. and Collins, M. and Stainforth, D. A.},
  title = {Quantification of modelling uncertainties in a large ensemble of climate change simulations},
  journal = {Nature},
  volume = {430},
  pages = {768--772},
  year = {2004},
  doi = {10.1038/nature02771}
}

@article{Dagon2020,
  author = {Dagon, K. and Sanderson, B. M. and Fisher, R. A. and Lawrence, D. M.},
  title = {A machine learning approach to emulation and biophysical parameter estimation with the {Community Land Model}, version 5},
  journal = {Advances in Statistical Climatology, Meteorology and Oceanography},
  volume = {6},
  pages = {223--244},
  year = {2020},
  doi = {10.5194/ascmo-6-223-2020}
}

@article{Eidhammer2024,
  author = {Eidhammer, T. and Gettelman, A. and Thayer-Calder, K. and others},
  title = {An extensible perturbed parameter ensemble for the {Community Atmosphere Model} version 6},
  journal = {Geoscientific Model Development},
  volume = {17},
  pages = {7835--7853},
  year = {2024},
  doi = {10.5194/gmd-17-7835-2024}
}

@article{Germano1991,
  author = {Germano, M. and Piomelli, U. and Moin, P. and Cabot, W. H.},
  title = {A dynamic subgrid-scale eddy viscosity model},
  journal = {Physics of Fluids A},
  volume = {3},
  pages = {1760--1765},
  year = {1991},
  doi = {10.1063/1.857955}
}

@article{PorteAgel2000,
  author = {Port{\'e}-Agel, F. and Meneveau, C. and Parlange, M. B.},
  title = {A scale-dependent dynamic model for large-eddy simulation: application to a neutral atmospheric boundary layer},
  journal = {Journal of Fluid Mechanics},
  volume = {415},
  pages = {261--284},
  year = {2000},
  doi = {10.1017/S0022112000008776}
}

@article{Ramadhan2020,
  author = {Ramadhan, A. and Wagner, G. L. and Hill, C. and Campin, J.-M. and Churavy, V. and Besard, T. and Souza, A. and Edelman, A. and Ferrari, R. and Marshall, J.},
  title = {{Oceananigans.jl}: fast and friendly geophysical fluid dynamics on {GPUs}},
  journal = {Journal of Open Source Software},
  volume = {5},
  number = {53},
  pages = {2018},
  year = {2020},
  doi = {10.21105/joss.02018}
}

@article{Ringler2013,
  author = {Ringler, T. and Petersen, M. and Higdon, R. L. and Jacobsen, D. and Jones, P. W. and Maltrud, M.},
  title = {A multi-resolution approach to global ocean modeling},
  journal = {Ocean Modelling},
  volume = {69},
  pages = {211--232},
  year = {2013},
  doi = {10.1016/j.ocemod.2013.04.010}
}

@article{Danilov2017,
  author = {Danilov, S. and Sidorenko, D. and Wang, Q. and Jung, T.},
  title = {The {Finite-volumE Sea ice--Ocean Model} ({FESOM2})},
  journal = {Geoscientific Model Development},
  volume = {10},
  pages = {765--789},
  year = {2017},
  doi = {10.5194/gmd-10-765-2017}
}

@article{Danabasoglu2014,
  author = {Danabasoglu, G. and Yeager, S. G. and Bailey, D. and others},
  title = {North {Atlantic} simulations in {Coordinated Ocean-ice Reference Experiments} phase {II} ({CORE-II}). {Part I}: mean states},
  journal = {Ocean Modelling},
  volume = {73},
  pages = {76--107},
  year = {2014},
  doi = {10.1016/j.ocemod.2013.10.005}
}

@article{Planton2021,
  author = {Planton, Y. Y. and Guilyardi, E. and Wittenberg, A. T. and others},
  title = {Evaluating climate models with the {CLIVAR} 2020 {ENSO} metrics package},
  journal = {Bulletin of the American Meteorological Society},
  volume = {102},
  pages = {E193--E217},
  year = {2021},
  doi = {10.1175/BAMS-D-19-0337.1}
}

@article{Dheeshjith2025Samudra,
  author = {Dheeshjith, S. and Subel, A. and Adcroft, A. and Busecke, J. and Fernandez-Granda, C. and Gupta, S. and Zanna, L.},
  title = {Samudra: An {AI} global ocean emulator for climate},
  journal = {Geophysical Research Letters},
  volume = {52},
  number = {10},
  pages = {e2024GL114318},
  year = {2025},
  doi = {10.1029/2024GL114318}
}

@misc{Donno2026Neptune,
  author = {Donno, D. and Epicoco, I. and Cafaro, M. and Accarino, G. and Amirian, M. M. and Acquaviva, V. and Nassisi, P. and Iovino, D. and Bracco, A. and Masina, S. and Gentine, P.},
  title = {Neptune: An {AI} model for global ocean subseasonal prediction},
  howpublished = {arXiv:2609.08606},
  year = {2026}
}

@article{Lin2017Modons,
  author = {Lin, S.-J. and Harris, L. and Chen, X. and Yao, W. and Chai, J.},
  title = {Colliding modons: A nonlinear test for the evaluation of global dynamical cores},
  journal = {Journal of Advances in Modeling Earth Systems},
  volume = {9},
  number = {7},
  pages = {2483--2492},
  year = {2017},
  doi = {10.1002/2017MS000965}
}

@article{Silvestri2025,
  author  = {Silvestri, S. and Wagner, G. L. and Constantinou, N. C. and Hill, C. N. and Campin, J.-M. and Souza, A. N. and Bishnu, S. and Churavy, V. and Marshall, J. and Ferrari, R. and others},
  title   = {A {GPU}-based ocean dynamical core for routine mesoscale-resolving climate simulations},
  journal = {Journal of Advances in Modeling Earth Systems},
  volume  = {17},
  pages   = {e2024MS004465},
  year    = {2025},
  doi     = {10.1029/2024MS004465}
}

\end{document}


\maketitle
\vspace{-2.2em}
\tableofcontents
\bigskip
\hrule
\bigskip

\noindent This Supplement provides (\S\ref{model_desc}) a self-contained description of the main governing
equations and numerical methods; (\S\ref{s2_dycore}) the dynamical-core verification; (\S\ref{s3_composability}) the composability
details; (\S\ref{s4_multiscale}) the large-eddy, cloud-resolving and atmosphere-only (AMIP) simulations; (\S\ref{s5:ocean_val}) ocean
validation; (\S\ref{si:ecsite}) land-surface validation; (\S\ref{s7}) performance and scaling; (\S\ref{s8}) accessibility and a
code example; (\S\ref{s9}) operator-level physical-consistency tests; and (\S\ref{s10}) variational data
assimilation. A complete description is in the legoESM Scientific Guide that ships
with the source code on GitHub.

\section{Model description: governing equations and numerics}\label{model_desc}
Our strategy is to have a modular, brick-like structure for our Earth System Model (ESM). As such, legoESM is developed as a federation of packages sharing one namespace under a one-way dependency graph
(\texttt{core} $\prec$ \{\texttt{atmosphere}, \texttt{ocean}, \texttt{land}, \texttt{ice}\}
$\prec$ \{\texttt{coupler}, \texttt{ml}, \texttt{tools}\}; main-text Fig.~1). All code is
written in pure-functional JAX, with 32- or 64-bit floating-point precision.

\subsection{Horizontal grids}
Because the equation set and the spatial discretization are independent choices in legoESM, each
component may be integrated on any of several horizontal meshes. The atmosphere and the ocean draw
on a common operator library, so that a mesh implemented for one component is in principle
available to the other.

\paragraph{Atmospheric horizontal grids} Five families are currently supported.

\emph{(i) Cubed sphere} Six faces are mapped to the sphere by an equiangular gnomonic
projection\cite{Ronchi1996,PutmanLin2007} and discretized in the FV3-style C-D arrangement with
piecewise-parabolic transport. With equiangular coordinates
$(\alpha_x,\alpha_y)\in[-\pi/4,\pi/4]^2$, the cell area on a C$n$ face is
\begin{equation}
\Delta A=\frac{R^2\,\Delta\alpha^2}{\cos^2\!\alpha_x\,\cos^2\!\alpha_y\,D^3},\qquad
D=\sqrt{1+\tan^2\!\alpha_x+\tan^2\!\alpha_y},\qquad \Delta\alpha=\frac{\pi}{2n},
\end{equation}
so that integration over the six faces recovers $4\pi R^2$ to round-off and area is conserved
discretely. Vector halo exchange rotates to geographic components across panel seams, and a
duo-grid treatment of the edges suppresses the residual differential-operator
imprint\cite{Mouallem2023}. The duo-grid treatment was implemented following the published
formulation, with the original FV3 code as the reference implementation, and was verified
iteratively: after each code revision the shallow-water and baroclinic-wave tests (\S2) were
rerun and assessed with $L_2$ error norms, global mass and momentum budgets, and visual
inspection of the meridional wind on the native mesh for panel-edge artifacts, and the cycle was
repeated until no panel-edge imprint was visible (Fig.~\ref{si:baroclinic}c).

\emph{(ii) Latitude--longitude} A regular spherical mesh is discretized on the Arakawa C grid
with finite-volume operators\cite{Sadourny1972}; a polar filter relaxes the severe time-step
restriction that would otherwise be imposed by the convergence of the meridians.

\emph{(iii) Gaussian (spectral).} Fields are represented by spherical-harmonic coefficients under
triangular truncation at total wavenumber $n_{\max}$\cite{Bourke1972}. The transform consists of a
fast Fourier transform in longitude followed by Gaussian quadrature in latitude using the fully
normalized associated Legendre functions $P_n^m$. Because the Laplacian is diagonal in this basis,
$\nabla^2\hat f_{nm}=-n(n+1)/a^2\,\hat f_{nm}$, the linear operators are exact within the truncated spectral representation and free of grid
imprint, and hyperdiffusion takes the form $-\nu[n(n+1)/a^2]^p$.

\emph{(iv) Unstructured Voronoi (MPAS).} A centroidal Voronoi tessellation is combined with
mimetic TRiSK operators\cite{Thuburn2009,Ringler2010,Skamarock2012}, with scalars at cell centres,
normal velocities at edges and vorticity at vertices,
\begin{equation}
(\divg\vect{u})_c=\frac{1}{A_c}\sum_e n_{e,c}u_e l_e,\quad
(\grad\phi)_e=\frac{\phi_{c_2}-\phi_{c_1}}{d_e},\quad
(\curl\vect{u})_v=\frac{1}{A_v}\sum_e t_{e,v}u_e d_e .
\end{equation}

\emph{(v) Doubly periodic Cartesian plane} A planar mesh with periodic lateral boundaries is used
for the large-eddy and cloud-resolving configurations, for which the spherical metric terms reduce
to their Cartesian limits.

\paragraph{Oceanic horizontal grids} We distinguish grid families (a discretization and its
operator set) from the specific meshes and configurations built on them. Four families are
supported. (i) The structured Arakawa C grid provides the reference configuration; it is used on
regular latitude--longitude and Mercator meshes and, in curvilinear form, on the NEMO-style
tripolar mesh (eORCA), which displaces two northern poles over land in order to remove the polar
singularity from the ocean domain; its folded northern boundary is treated by a dedicated halo
exchange that enforces the correct parity of scalar and vector fields across the fold. (ii) The
cubed sphere is available in the same C-D arrangement used by the atmosphere. (iii) The
unstructured Voronoi (MPAS) mesh supports variable-resolution configurations. (iv) Unstructured
triangular meshes of the FESOM type\cite{Danilov2017} are used for the idealized benchmarks
(Fig.~\ref{si:oceanbench}) and the eddy-permitting FESOM2 FORCA20 configuration
(Fig.~\ref{si:oceanhr}b).

\subsection{Vertical coordinates}
\paragraph{Atmosphere} Three vertical coordinates are implemented. The first is a
terrain-following sigma coordinate, $\sigma=p/p_s$. The second is a hybrid sigma--pressure
coordinate, $p(k)=A(k)p_{\mathrm{ref}}+B(k)p_s$, which is terrain-following near the surface and
becomes pressure-like aloft. The third is a terrain-following height coordinate, used by the
nonhydrostatic core in cloud-resolving and large-eddy configurations. Geopotential is evaluated
with the energy-conserving Simmons--Burridge discretization\cite{SimmonsBurridge1981},
\begin{equation}
\Phi_k=\Phi_s+\Rd\!\!\sum_{k'=k+1}^{N-1}\!T_{k'}\ln\frac{\sigma_{k'+1/2}}{\sigma_{k'-1/2}}
+\alpha_k\Rd T_k,\qquad
\alpha_k=1-\frac{\sigma_{k-1/2}}{\Delta\sigma_k}\ln\frac{\sigma_{k+1/2}}{\sigma_{k-1/2}} .
\end{equation}

\paragraph{Ocean} The ocean employs a dynamic $z^\ast$ coordinate in which the free surface $\eta$
is distributed over the water column through the layer Jacobian $J(\eta)=(\eta+H)/H_{\max}$, so
that layer thicknesses vary smoothly with the free surface rather than being absorbed entirely by
the uppermost cell. Level spacing may be uniform, prescribed from a table of thicknesses, or
generated by an analytic stretching function that refines the near-surface resolution. The bottom
boundary is represented either by full steps or by partial cells, the latter allowing the deepest
cell of each column to be thinned so as to follow the bathymetry more faithfully.

\subsection{Atmospheric dynamics}
\paragraph{Shallow water (vector-invariant)}
\begin{equation}
\frac{\pp h}{\pp t}=-\divg(h\vect{v}),\quad
\frac{\pp u}{\pp t}=(\zeta+f)v-\frac{\pp B}{\pp x}+D_u,\quad
\frac{\pp v}{\pp t}=-(\zeta+f)u-\frac{\pp B}{\pp y}+D_v,
\end{equation}
with Bernoulli function $B=\half(u^2+v^2)+g(h+h_s)$ and relative vorticity $\zeta$. On the
cubed sphere the FV3-style C-D grid carries momentum on the D-grid and diagnoses C-grid mass
fluxes, with PPM transport\cite{ColellaWoodward1984,Lin2004,PutmanLin2007}; the
circulation-based corner vorticity removes the Hollingsworth--K\"allberg instability.

\paragraph{Hydrostatic primitive equations ($\sigma$)}
\begin{align}
\frac{\pp p_s}{\pp t}&=-\!\int_0^1\!\divg(p_s\vect{v})\,\dd\sigma,\\
\frac{\pp u}{\pp t}&=(\zeta+f)v-\frac{\pp B}{\pp x}-\Rd T\frac{\pp\ln p_s}{\pp x}+D_u,\\
\frac{\pp T}{\pp t}&=-\divg(T\vect{v})+T\divg\vect{v}-\dot\sigma\frac{\pp T}{\pp\sigma}
+\kappa T\frac{\omega}{p}+Q,
\end{align}
with $\kappa=\Rd/c_p$ and diabatic heating $Q$ from physics. The spectral core advances the
vorticity--divergence form with the fast gravity-wave terms treated semi-implicitly
\cite{HoskinsSimmons1975}, yielding a per-wavenumber Helmholtz solve pre-factored at
initialization. The nonhydrostatic core integrates the fully compressible Euler equations on a
terrain-following height coordinate for CRM/LES use.

\paragraph{Coupling} legoESM uses time-split coupling between dynamics and physics: the dynamical core updates the state,
then the physics pipeline evaluates the parameterizations in the order radiation $\rightarrow$ deep
convection $\rightarrow$ shallow convection/large-scale condensation $\rightarrow$ microphysics
$\rightarrow$ boundary-layer turbulence $\rightarrow$ gravity-wave drag $\rightarrow$ surface
fluxes, followed by conservation fixers and diagnostics. By default, each parameterization is evaluated from the same
post-dynamics state and the resulting tendencies are summed (process splitting), so that the evaluation order does not
affect the result (see \S\ref{s3_composability}). A grid-agnostic column adapter reshapes
any grid's 3D state to $(n_\mathrm{col},n_\mathrm{lev})$ for physics.

\subsection{Atmospheric physics (selected closures)}
Radiation is either gray (Frierson-style\cite{Frierson2006}) or correlated-$k$
RRTMGP\cite{Pincus2019}. Convection schemes include mass-flux closures (Tiedtke\cite{Tiedtke1989},
Zhang--McFarlane\cite{ZhangMcFarlane1995}, Bechtold, Kain--Fritsch, Emanuel, EDMF), which share a unified updraft core,
as well as Kuo, simplified Betts--Miller (SBM) and deep convective adjustment (DCA). Surface exchange uses Monin--Obukhov similarity; boundary-layer
mixing offers Louis, Holtslag--Boville, YSU, EDMF, TKE and Smagorinsky\cite{Smagorinsky1963}
options. Warm-rain microphysics ranges from Kessler to two-moment schemes and a super-droplet
method\cite{Shima2009}. Each scheme is a factory returning
$\texttt{physics\_fn}(\text{state},\text{grid},\text{coord})\!\to\!\text{tendencies}$, and the
combined tendency equals the sum of the enabled process tendencies to floating-point round-off
(\S3).

\subsection{Ocean}
The ocean solves the hydrostatic Boussinesq primitive equations with a split-explicit
free-surface barotropic solver\cite{ShchepetkinMcWilliams2005}, KPP vertical
mixing\cite{LMD1994}, GM--Redi isopycnal/eddy mixing\cite{GM1990}, and a TEOS-10-style
polynomial equation of state. Tracers obey
$\pp_t(h\theta)+\divg(h\vect{u}\theta)=\divg(h\,\mathbf{K}\grad\theta)+\mathcal{S}_\theta$,
with $h$ the layer thickness and $\mathbf{K}$ the (rotated) diffusivity tensor. The barotropic
subsystem is advanced either implicitly (Crank--Nicolson) or by split-explicit barotropic
sub-cycling\cite{ShchepetkinMcWilliams2005}, with the barotropic sub-step limited by a local CFL
criterion; both conserve volume and tracers to within the tolerances specified in the ocean
component's physical contract and enforced by the continuous-integration tests (main-text Methods,
Verification harness). The ocean offers more options (vertical coordinates, mixing
schemes, equation-of-state variants and boundary treatments) than are described here; the complete, up-to-date list is
maintained in the model documentation (Scientific Guide) distributed with the source code.

\subsection{Land, sea ice, coupler, ML}
The multi-layer land model solves the mixed-form Richards
equation\cite{van1980closed} for soil moisture, Johansen thermal diffusion for soil temperature\cite{Johansen1975}, Farquhar photosynthesis\cite{Farquhar1980} with Ball--Berry stomatal conductance,
and a DALEC-style carbon cycle. Sea ice uses EVP/mEVP
rheology\cite{HunkeDukowicz1997} with a multi-category thickness distribution and
energy-conserving thermodynamics. The coupler exchanges area-weighted surface fluxes over
sub-grid tiles with a Monin--Obukhov surface layer. The \texttt{ml} package provides spherical
Fourier neural operators\cite{Bonev2023} (built on Fourier neural operators\cite{Li2021}),
conservation-aware losses, differentiable training drivers, and variational (4DVar) data
assimilation that differentiates through the compiled time-stepping kernel.

\section{Dynamical-core verification}\label{s2_dycore}
\begin{figure}[!htbp]\centering
\begin{subfigure}{0.495\textwidth}\includegraphics[width=\textwidth]{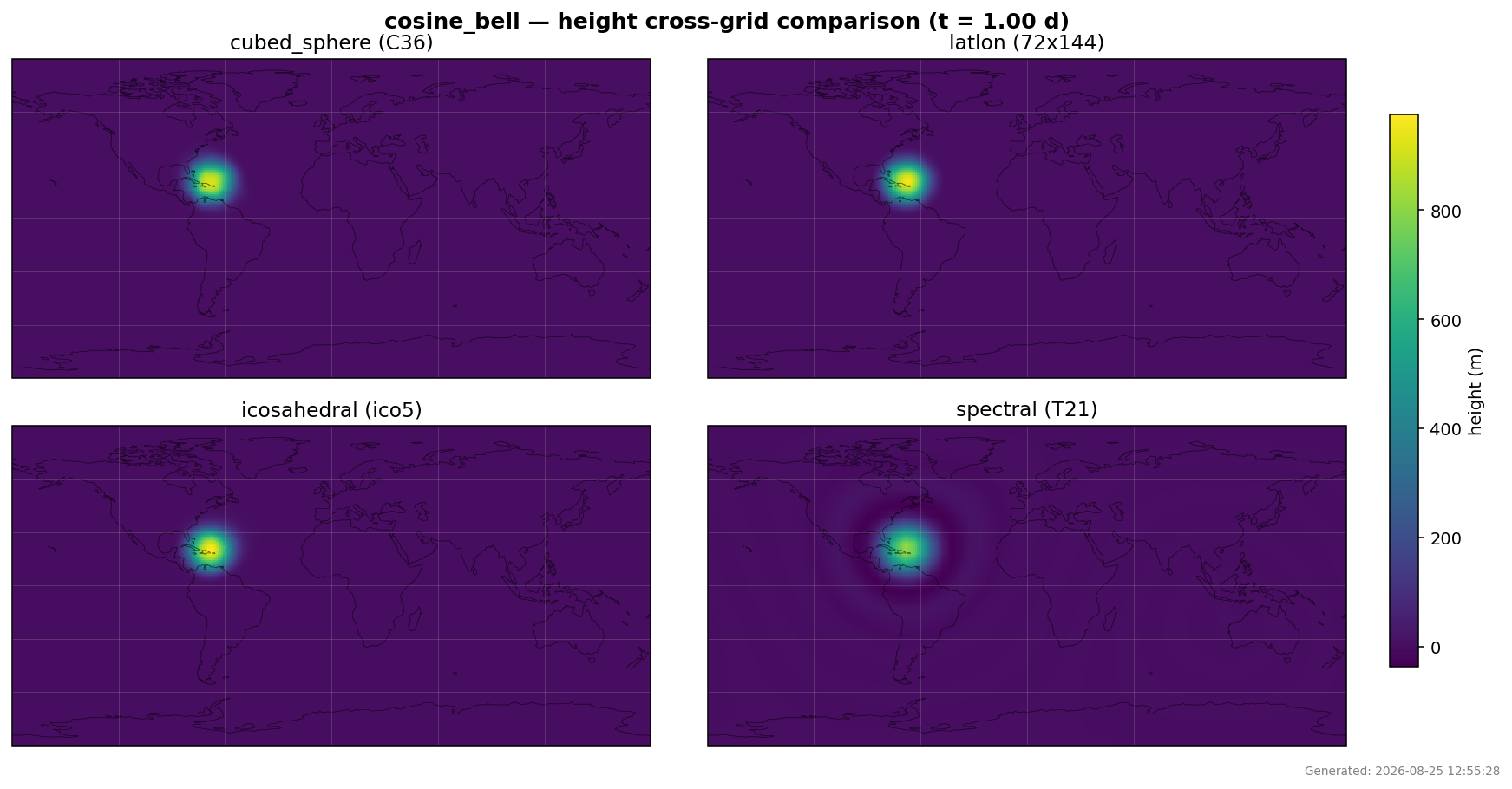}
\caption{Cosine-bell advection (Williamson 1)}\end{subfigure}\hfill
\begin{subfigure}{0.495\textwidth}\includegraphics[width=\textwidth]{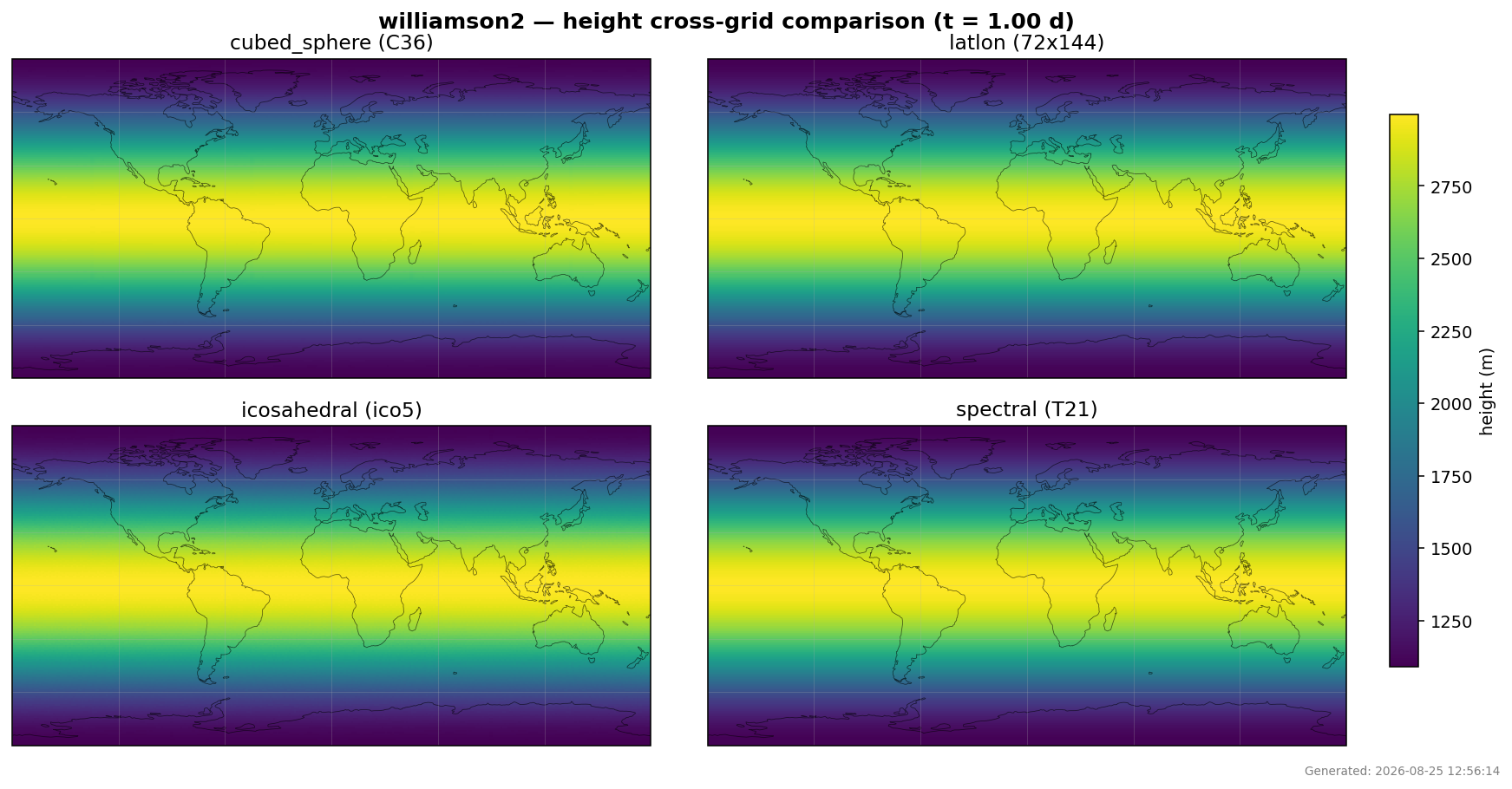}
\caption{Steady zonal flow (Williamson 2)}\end{subfigure}\\[3pt]
\begin{subfigure}{0.495\textwidth}\includegraphics[width=\textwidth]{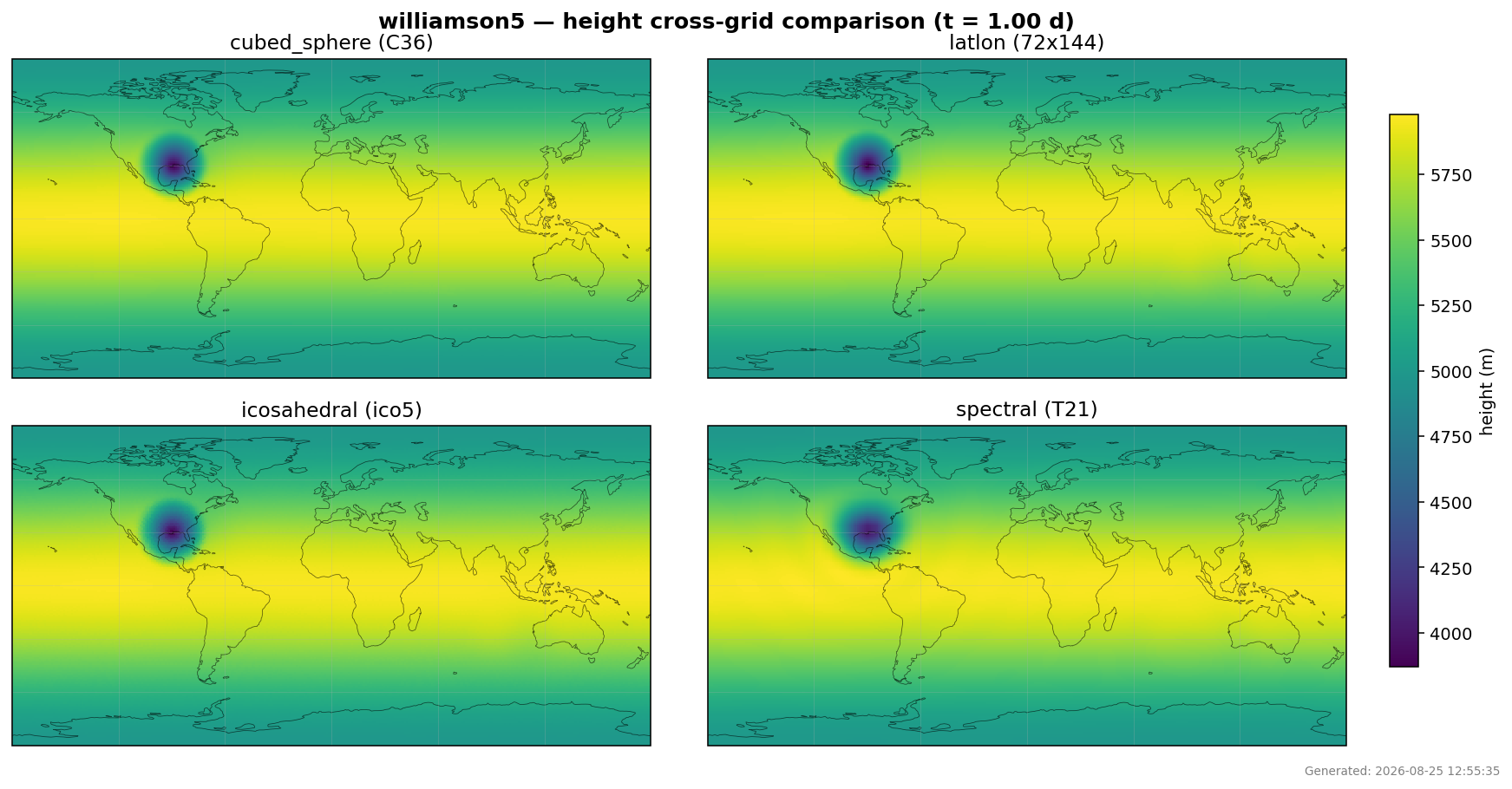}
\caption{Flow over a mountain (Williamson 5)}\end{subfigure}\hfill
\begin{subfigure}{0.495\textwidth}\includegraphics[width=\textwidth]{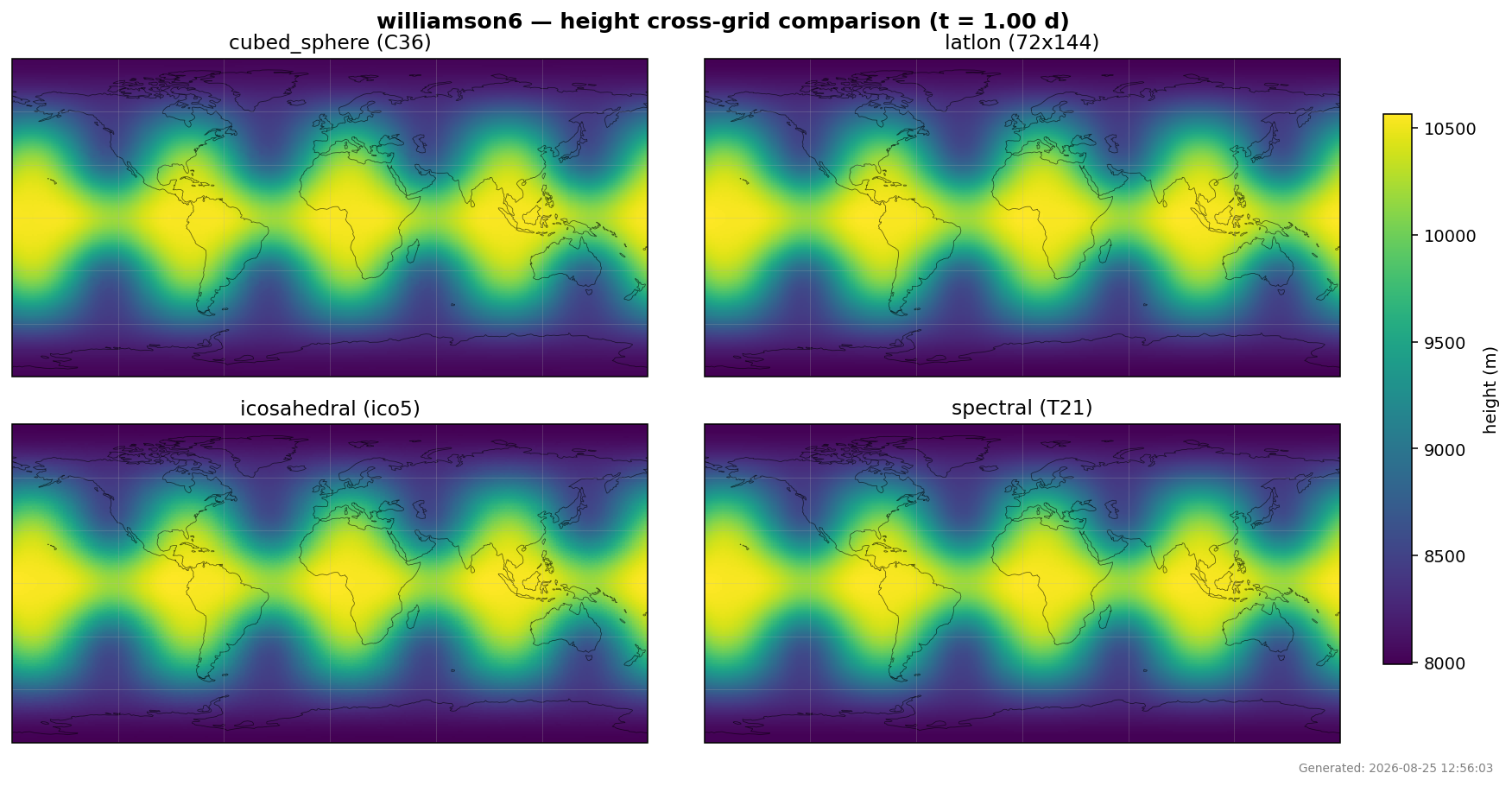}
\caption{Rossby--Haurwitz wave (Williamson 6)}\end{subfigure}\\[3pt]
\begin{subfigure}{\textwidth}\includegraphics[width=\textwidth]{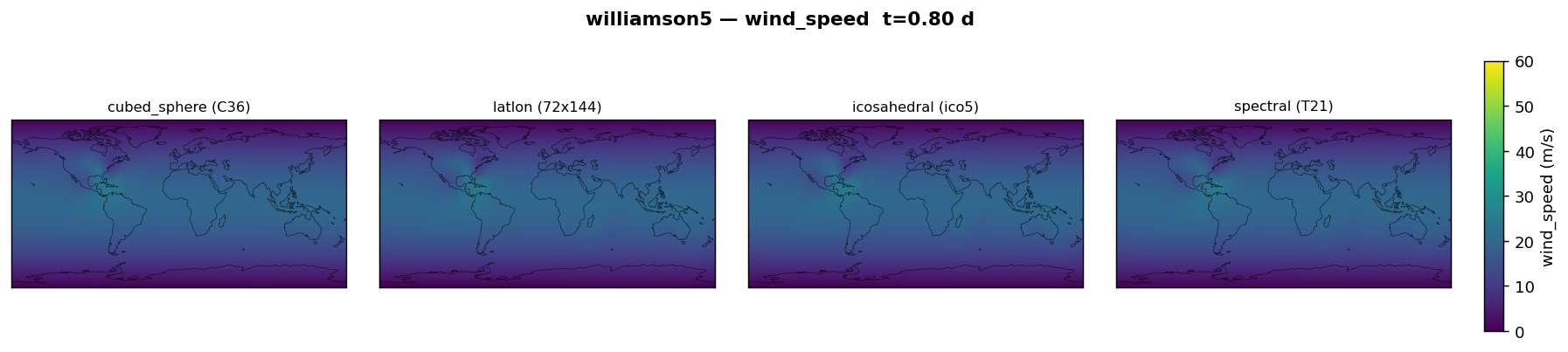}
\caption{Williamson 5 wind speed at day 0.8}\end{subfigure}
\caption{\textbf{The same shallow-water equations solved on four grids.} Height field after one
day for (a) cosine-bell advection, (b) steady zonal geostrophic flow, (c) flow over an isolated
mountain and (d) the Rossby--Haurwitz wave, and (e) wind speed for the mountain case, each
integrated on the cubed-sphere (C36), lat--lon ($72\times144$), icosahedral (level~5) and
spectral (T21) discretizations. The solutions show close visual agreement across grids, apart
from the Gibbs ringing of the T21 spectral transform around the cosine bell; the physical
equation set and the numerical method are selected independently\cite{Williamson1992}.}\label{si:gridcompare}
\end{figure}

\begin{figure}[h]\centering
\includegraphics[width=0.92\textwidth]{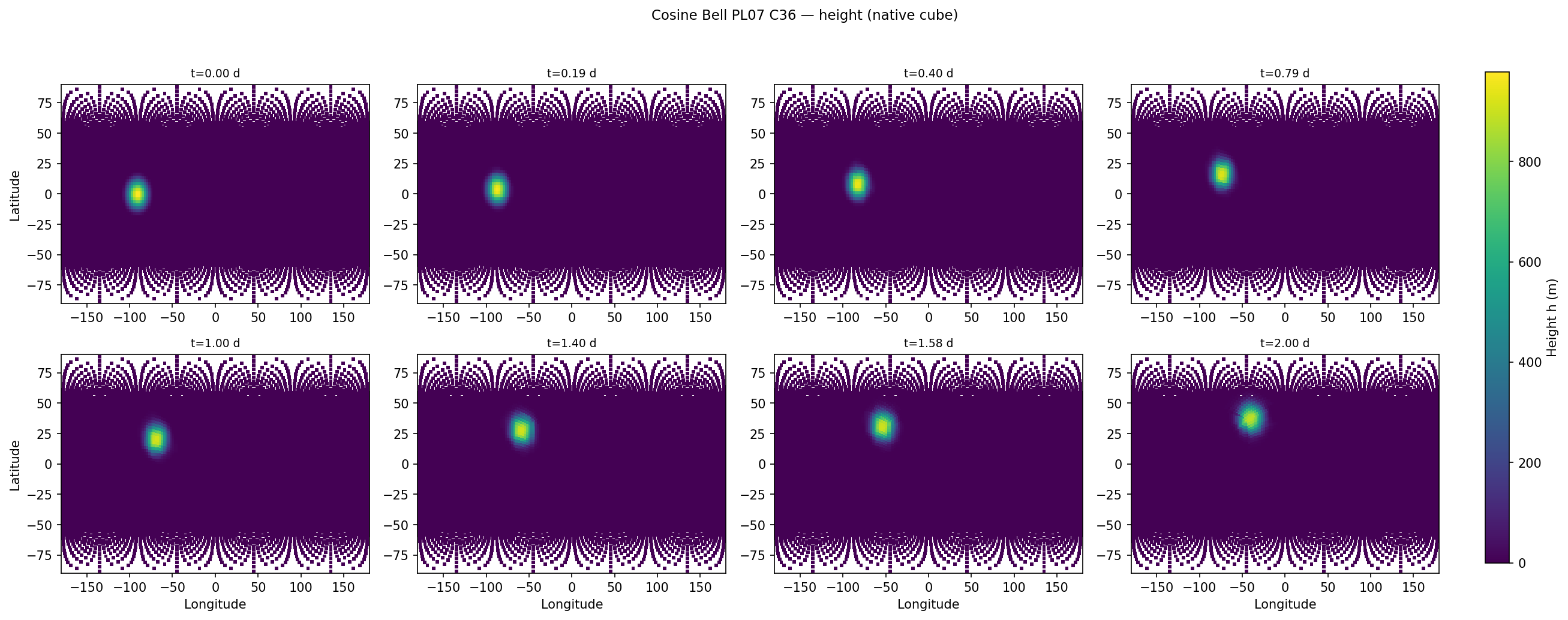}
\caption{Cosine-bell advection (Williamson case 1) on the cubed sphere (C36), shown in the
lat--lon projection at successive times. The tracer shows no visible change in shape or
amplitude as it crosses cube-panel edges.}\label{si:cosinebell}
\end{figure}

\begin{figure}[h]\centering
\begin{subfigure}{0.495\textwidth}\includegraphics[width=\textwidth]{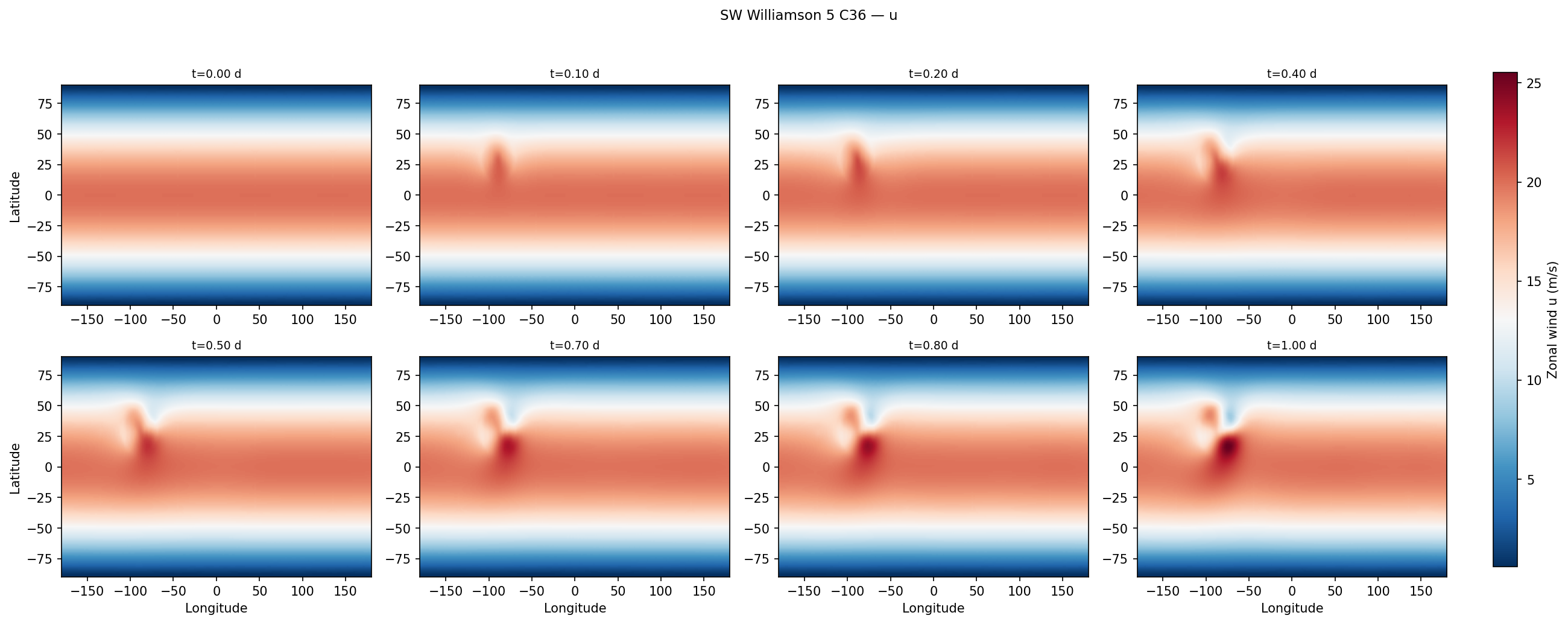}
\caption{Williamson 5 (flow over a mountain)}\end{subfigure}\hfill
\begin{subfigure}{0.495\textwidth}\includegraphics[width=\textwidth]{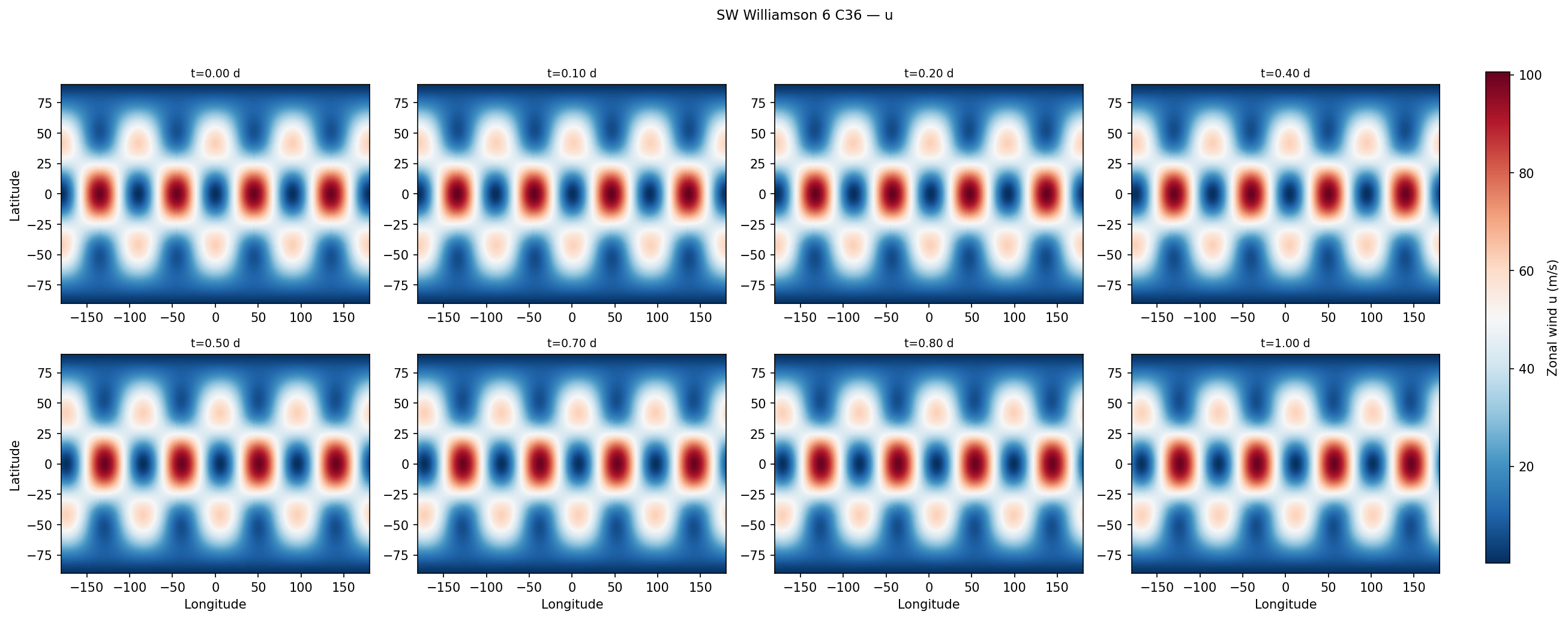}
\caption{Williamson 6 (Rossby--Haurwitz wave)}\end{subfigure}
\caption{\textbf{Shallow-water test cases on the cubed sphere (C36).} Height field for (a) flow
over an isolated mountain (Williamson case~5) and (b) the Rossby--Haurwitz wave (Williamson
case~6)\cite{Williamson1992}.}\label{si:w5w6}
\end{figure}

\begin{figure}[h]\centering
\begin{subfigure}{0.495\textwidth}\includegraphics[width=\textwidth]{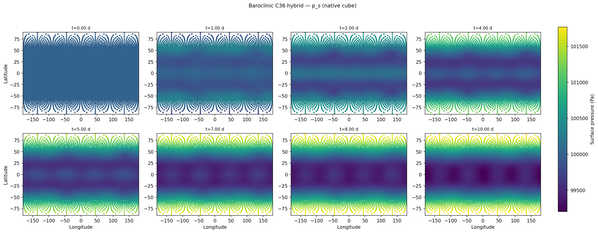}
\caption{Surface pressure}\end{subfigure}\hfill
\begin{subfigure}{0.495\textwidth}\includegraphics[width=\textwidth]{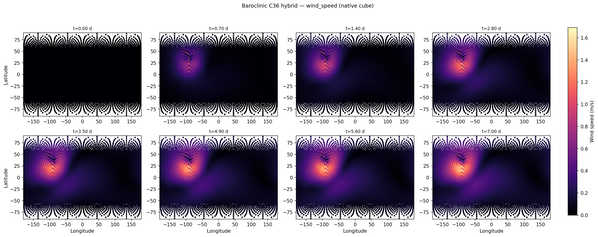}
\caption{Wind speed}\end{subfigure}\\[3pt]
\begin{subfigure}{0.495\textwidth}\includegraphics[width=\textwidth]{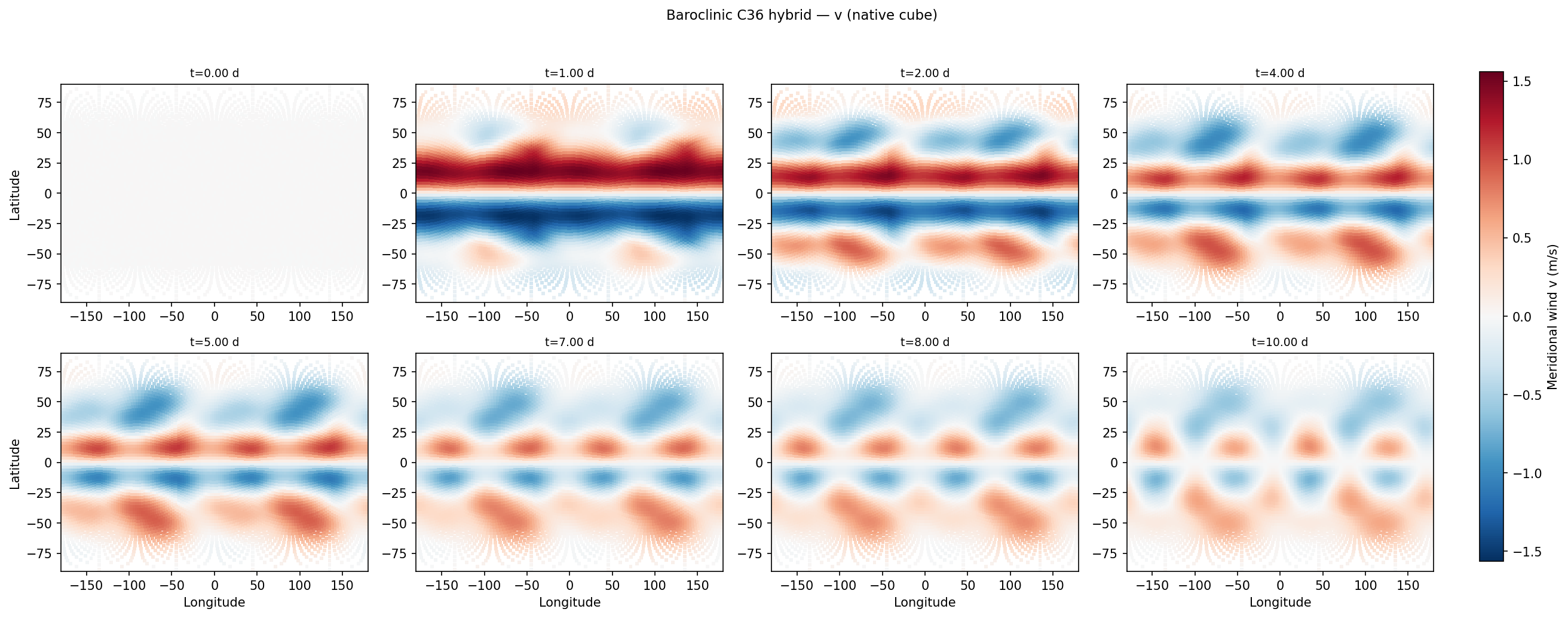}
\caption{Meridional wind on the native cubed-sphere mesh}\end{subfigure}
\caption{Jablonowski--Williamson baroclinic-wave test, hydrostatic cubed-sphere C36, days
0--10. The instability grows at the expected rate and in the expected location. The meridional
wind is additionally shown on the native cubed-sphere mesh, where any residual imprint of the
panel edges would be most apparent.}\label{si:baroclinic}
\end{figure}

legoESM is gated on the standard sphere test
set\cite{Williamson1992,JablonowskiWilliamson2006,HeldSuarez1994}. Figure~\ref{si:cosinebell}
shows cosine-bell solid-body advection (Williamson case 1) on the cubed sphere; the bell
traverses panel edges without visible distortion. Figure~\ref{si:w5w6} shows the Williamson 5
(flow over an isolated mountain) and Williamson 6 (Rossby--Haurwitz wave) tests on the
cubed sphere. The Jablonowski--Williamson baroclinic wave is shown for surface pressure and
near-surface wind speed in Fig.~\ref{si:baroclinic}. Because the equation set and the
numerical discretization are independent choices in legoESM, the same shallow-water problem can
be solved on every grid: in Fig.~\ref{si:gridcompare}, the cubed-sphere, lat--lon,
icosahedral and spectral cores show close visual agreement for cosine-bell
advection, steady zonal geostrophic flow (Williamson~2), flow over a mountain (Williamson~5) and
the Rossby--Haurwitz wave (Williamson~6); the only visible difference is the expected Gibbs
ringing of the truncated spectral (T21) transform around the sharp-edged cosine bell. Global mass is conserved to floating-point round-off, as checked by the test suite; energy and
potential-enstrophy budgets close to within scheme-dependent tolerances, which are specified for
each dynamical core in the acceptance tests distributed with the code (main-text Methods).

\begin{figure}[h]\centering
\includegraphics[width=0.9\textwidth]{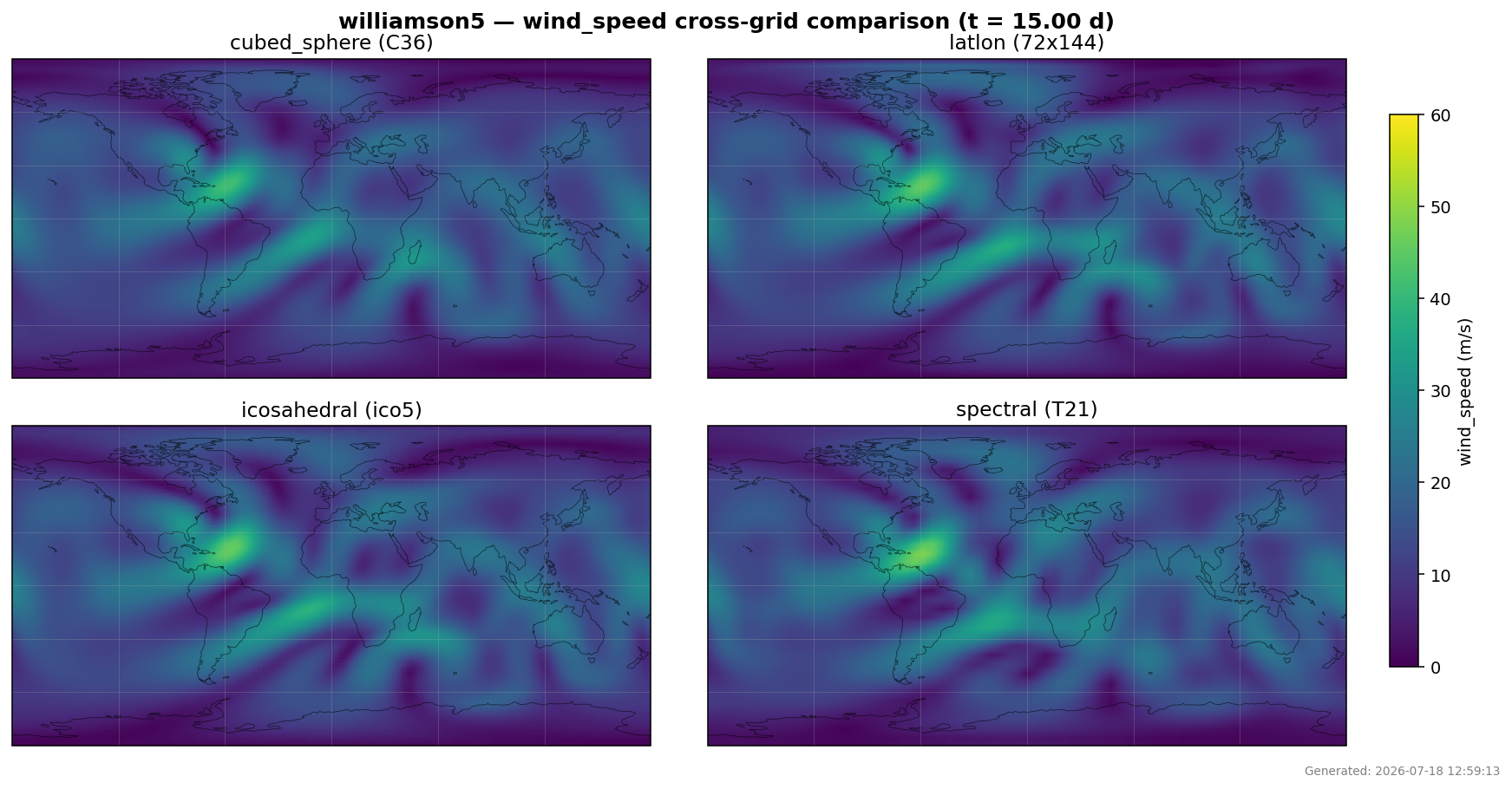}
\caption{Shallow-water Williamson case~5 (flow over an isolated mountain) on the cubed sphere: the balanced height field and its error metrics, with no visible grid imprint at panel edges.}\label{si:williamson5}
\end{figure}

\section{Composability}\label{s3_composability}
Composability operates on three orthogonal axes (main-text Fig.~1): grid, complexity and
physics. Physics composition is additive: because each process tendency is evaluated from the same input state
(process splitting, \S\ref{model_desc}), $\texttt{tendency}(A+B+C)
=\texttt{tendency}(A)+\texttt{tendency}(B)+\texttt{tendency}(C)$ holds to floating-point round-off (up to the order of summation), which is asserted by
tests. This is a property of the parallel process-split discretization, not of numerical models in general: under sequential
(time-split) application of processes, or when a scheme adjusts the state seen by the next one within a time step, the result
depends on the order of operations and exact additivity is lost. A
process is disabled by setting its scheme to \texttt{"none"}, and every factory raises on an
unknown scheme rather than silently defaulting. An entire component can be swapped the same way:
in a short atmospheric integration the same slot accepts (i) the classical physics suite, (ii) a
column neural network, or (iii) a full SFNO dynamical core; and the ocean can be run as
prescribed SST, a slab mixed layer, or a full three-dimensional model from one driver. Table \ref{tab:cores}
lists the dynamical cores; Tables \ref{tab:phys} and \ref{tab:oli} list the physics and ocean/land/ice schemes.

\begin{figure}[!htbp]\centering
\includegraphics[width=\textwidth]{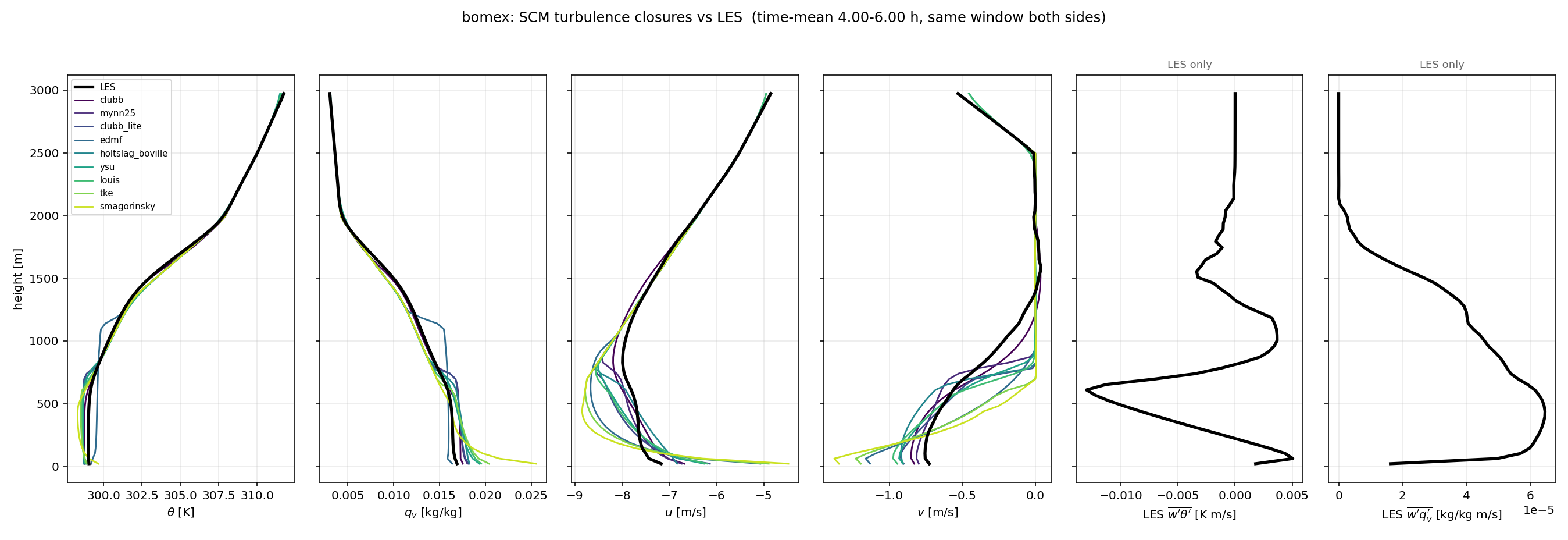}
\caption{\textbf{Boundary-layer turbulence closures against large-eddy simulation (BOMEX).} Mean
profiles of potential temperature $\theta$, water-vapour mixing ratio $q_v$ and the horizontal wind
components $u$ and $v$, averaged over hours 4--6, from nine single-column turbulence closures
(coloured) and from the legoESM large-eddy simulation (black), together with the LES
turbulent fluxes of heat ($\overline{w'\theta'}$) and moisture ($\overline{w'q_v'}$). All closures
reproduce the cloud-layer and inversion structure; they differ mainly in the sub-cloud mixed
layer and in the near-surface momentum and moisture profiles.}\label{si:turbles}
\end{figure}

\begin{figure}[!htbp]\centering
\begin{subfigure}[b]{0.565\textwidth}\includegraphics[width=\textwidth]{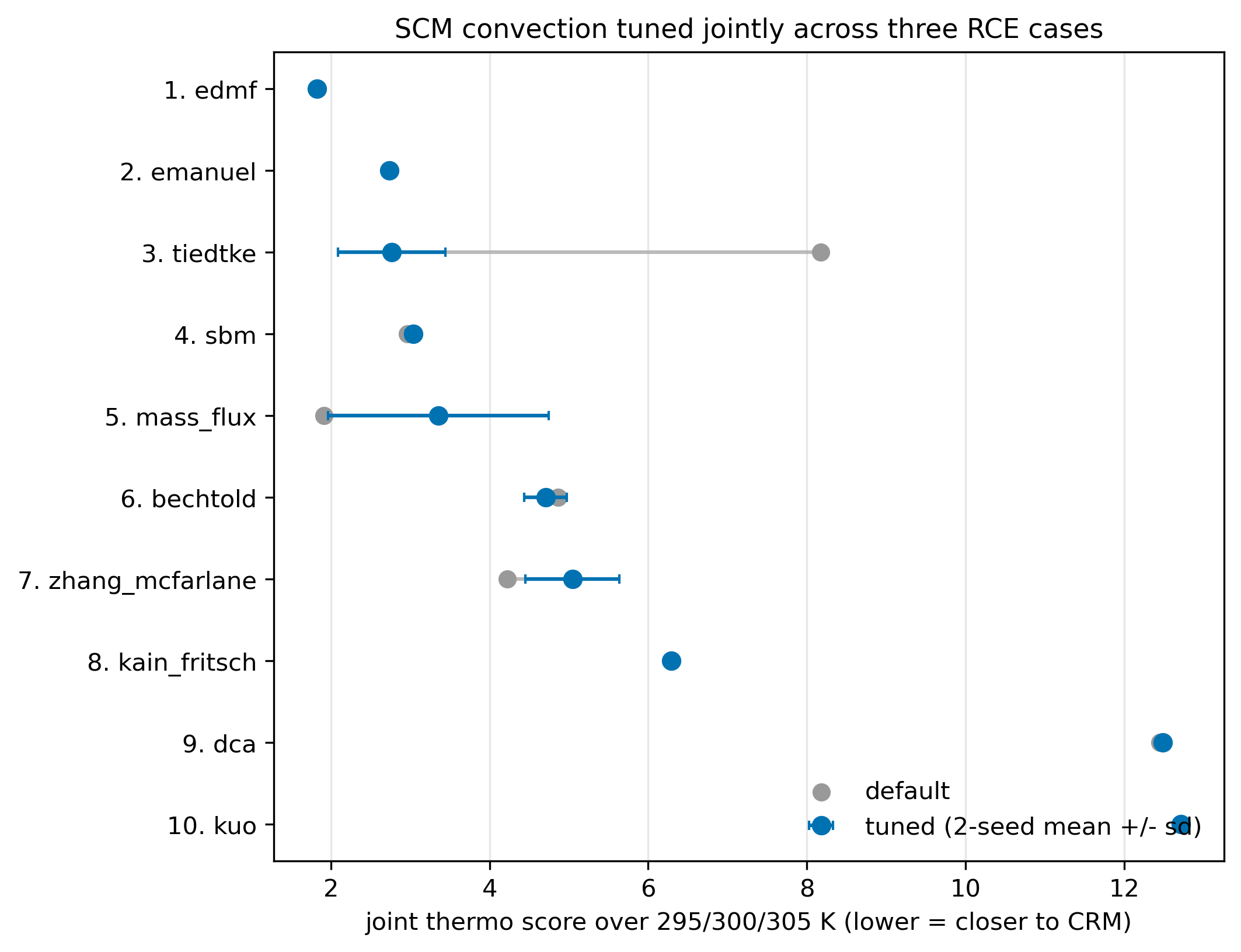}
\caption{Joint score, default and tuned}\end{subfigure}\hfill
\begin{subfigure}[b]{0.425\textwidth}\includegraphics[width=\textwidth]{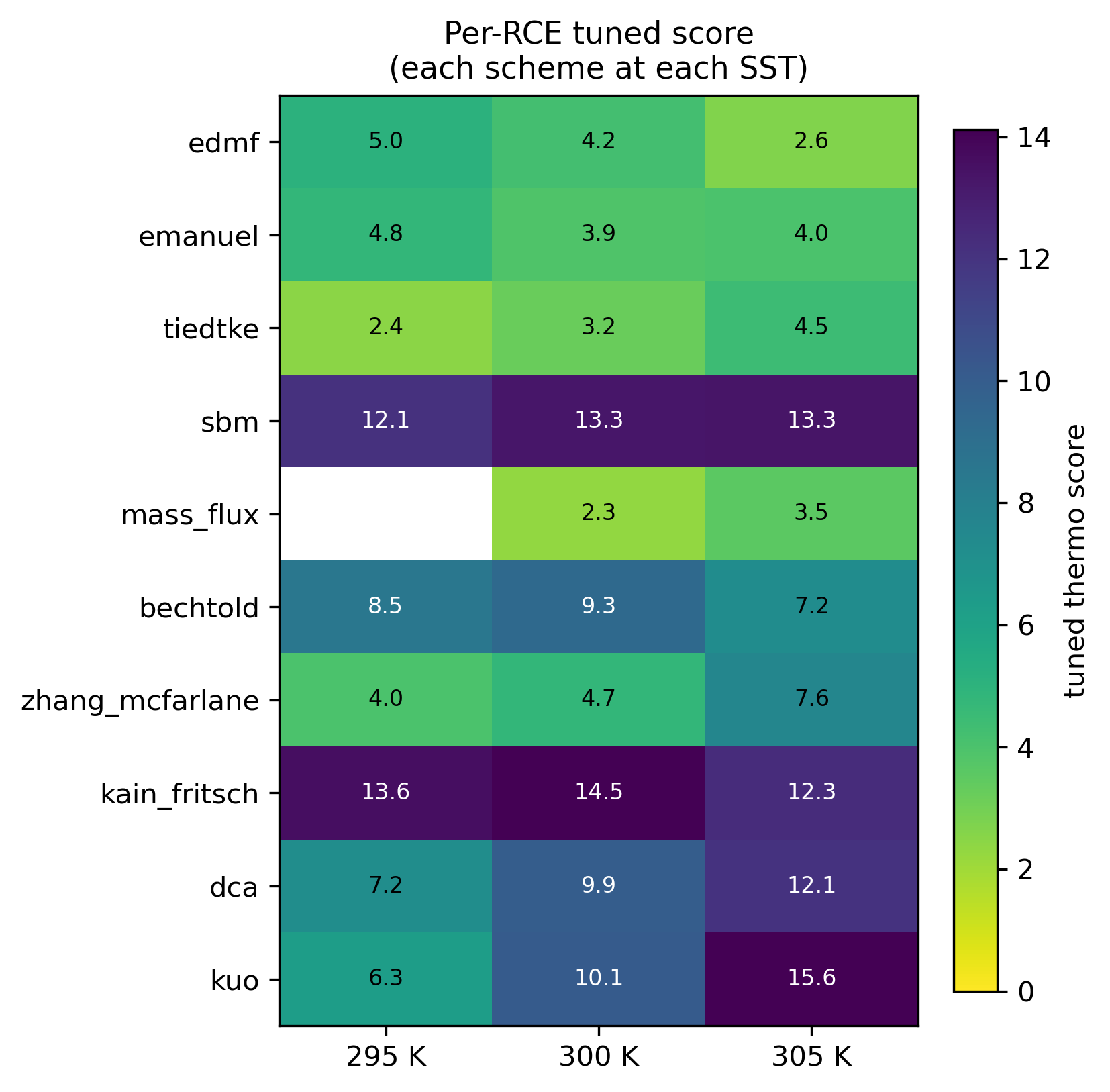}
\caption{Tuned score at each SST}\end{subfigure}
\caption{\textbf{Composability and gradient-based calibration of convection schemes in radiative--convective equilibrium.}
Each of ten convection parameterizations is run, unchanged, in a single atmospheric column to
radiative--convective equilibrium (RCE) at sea-surface temperatures (SSTs) of 295, 300 and 305\,K
and scored against a cloud-resolving RCE simulation of the same case that uses the same
microphysics (Morrison\cite{Morrison2009}) and radiation (RRTMGP\cite{Pincus2019}); lower scores are
closer to the cloud-resolving reference. \textbf{a}, Joint thermodynamic score over the three
cases with default parameters (grey) and after gradient-based calibration, optimized using
Muon\cite{Jordan2024}, of a single parameter set per scheme across all three cases (blue; mean $\pm$ s.d. over two
optimization seeds), ranked by tuned score. All ten schemes reach an equilibrium and are scored.
EDMF is closest to the reference; calibration improves the Tiedtke scheme about threefold, changes
most other schemes little, and slightly degrades the mass-flux and Zhang--McFarlane schemes, whose
tuned parameters also vary between seeds. \textbf{b}, Tuned score of each scheme at each SST: skill
depends on the climate, with Kuo, Zhang--McFarlane and deep convective adjustment degrading with
warming and EDMF improving. The blank cell (mass-flux at 295\,K) is missing. The corresponding profiles are shown in Fig.~\ref{si:convsweep},
and the complete scheme inventory is in Extended Data Table~2.}
\label{si:convrank}
\end{figure}

\begin{figure}[h]\centering
\includegraphics[width=\textwidth]{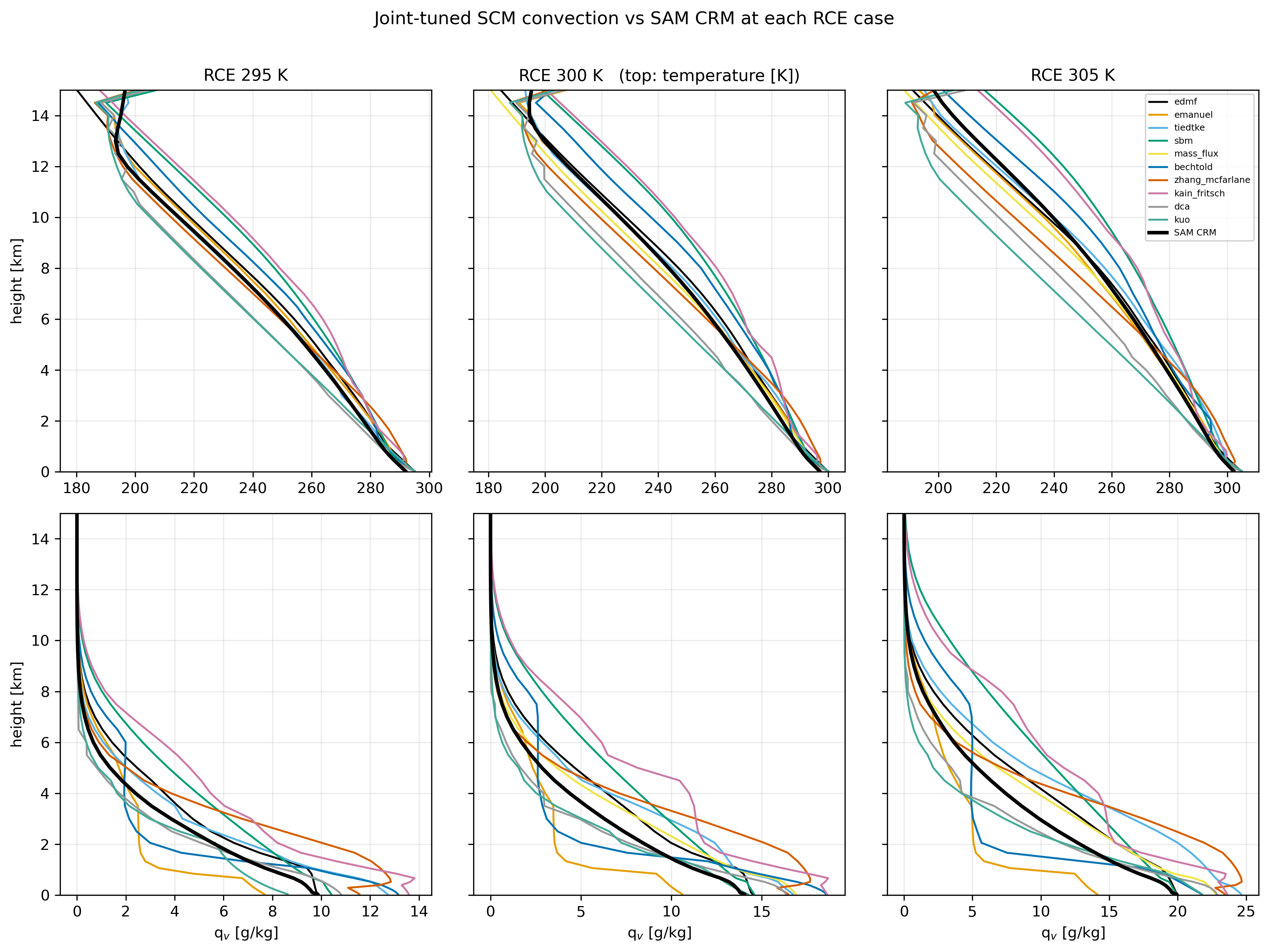}
\caption{\textbf{Jointly calibrated convection schemes against the cloud-resolving reference.}
Equilibrated temperature (top) and water-vapour mixing ratio $q_v$ (bottom) over the lowest 15\,km
in RCE at SSTs of 295, 300 and 305\,K (columns), from the ten single-column convection schemes
after joint calibration (coloured) and from the cloud-resolving reference (SAM configuration;
thick black). Temperature spread is largest in the upper troposphere, where Kain--Fritsch and the
simplified Betts--Miller scheme are too warm and Kuo and deep convective adjustment too cold.
Humidity errors are proportionally larger and concentrated below about 6\,km, where, for example,
Emanuel and Bechtold are too dry and Zhang--McFarlane and Kain--Fritsch too moist, so that schemes
with similar temperature skill can differ markedly in their moisture structure.
Figure~\ref{si:convrank} reduces these profiles to one score per scheme and case.}\label{si:convsweep}
\end{figure}

The same configuration-level swap is used to compare process representations under identical conditions.
Figure~\ref{si:turbles} runs nine boundary-layer turbulence closures, unchanged, in a single
column for the BOMEX trade-wind shallow-cumulus case and compares their mean profiles with a
large-eddy simulation performed with the same model infrastructure. Figures~\ref{si:convrank}
and~\ref{si:convsweep} do the same for the ten convection schemes in radiative--convective
equilibrium at three sea-surface temperatures, before and after gradient-based calibration
(optimized using Muon\cite{Jordan2024}) against cloud-resolving references.

\begin{table}[h]\centering\footnotesize
\caption{Dynamical cores by equation set and discretization family. The discretizations follow
FV3-style cubed-sphere finite volume with PPM transport\cite{Lin2004,PutmanLin2007} and a duo-grid
panel-edge treatment\cite{Mouallem2023}, lat--lon C-grid finite volume\cite{Sadourny1972},
semi-implicit spectral transforms\cite{Bourke1972,HoskinsSimmons1975}, and mimetic TRiSK operators
on Voronoi/MPAS meshes\cite{Thuburn2009,Ringler2010,Skamarock2012}.}\label{tab:cores}
\begin{tabular}{@{}lp{2.7cm}p{2.7cm}p{2.7cm}@{}}
\toprule
Discretization & Shallow water & Hydrostatic PE & Nonhydrostatic \\
\midrule
Cubed-sphere C-D (FV3) & \checkmark & \checkmark & \checkmark \\
Lat--lon C-grid (FV)   & \checkmark & \checkmark & -- \\
Gaussian/spectral      & \checkmark & \checkmark & \checkmark \\
Voronoi/MPAS (TRiSK)   & \checkmark & \checkmark & \checkmark \\
SFNO (learned)         & \checkmark & \checkmark & -- \\
Plane (LES/CRM)        & -- & -- & \checkmark \\
\bottomrule
\end{tabular}
\end{table}

\begin{table}[h]\centering\footnotesize
\caption{Atmospheric physics processes and representative schemes (each composable; set to
\texttt{none} to disable).}\label{tab:phys}
\begin{tabular}{@{}lp{10.5cm}@{}}
\toprule
Process & Schemes \\
\midrule
Radiation & gray (Frierson), RRTMGP correlated-$k$ (LW+SW), ML emulator \\
Convection & Bechtold, Tiedtke, Kain--Fritsch, Zhang--McFarlane, Emanuel, Kuo, unified mass-flux, EDMF, simplified Betts--Miller (SBM), deep convective adjustment (DCA) \\
Microphysics & Kessler, Sundqvist, Morrison, Seifert--Beheng, Thompson, P3, super-droplet, ML \\
Cloud fraction & Sundqvist, Xu--Randall \\
Boundary layer & Louis, Holtslag--Boville, YSU, EDMF, TKE 1.5, MYNN 2.5, CLUBB, CLUBB-lite, Smagorinsky \\
Gravity-wave drag & Lindzen, Hines, McFarlane, Rayleigh, prognostic spectral, ML \\
\bottomrule
\end{tabular}
\end{table}

\begin{table}[h]\centering\footnotesize
\caption{Ocean, land and sea-ice schemes.}\label{tab:oli}
\begin{tabular}{@{}lp{10.5cm}@{}}
\toprule
Component & Schemes \\
\midrule
Ocean vertical mixing & KPP, Richardson, constant; Jayne--St-Laurent tidal \\
Ocean lateral mixing  & harmonic, biharmonic, GM--Redi, Visbeck, Leith, backscatter \\
Ocean grid families (meshes) & C grid (lat--lon, Mercator, tripolar eORCA); cubed-sphere C-D; MPAS Voronoi; FESOM-type triangular \\
Land                  & slab (bucket); multi-layer Richards + Johansen thermal; DALEC carbon; snow \\
Sea ice               & thermodynamic slab; EVP/mEVP dynamic; multi-category ITD; ridging, ponds, brine \\
\bottomrule
\end{tabular}
\end{table}

\section{Multiscale: large-eddy and cloud-resolving suites}\label{s4_multiscale}
\begin{figure}[h]\centering
\begin{subfigure}{0.495\textwidth}\includegraphics[width=\textwidth]{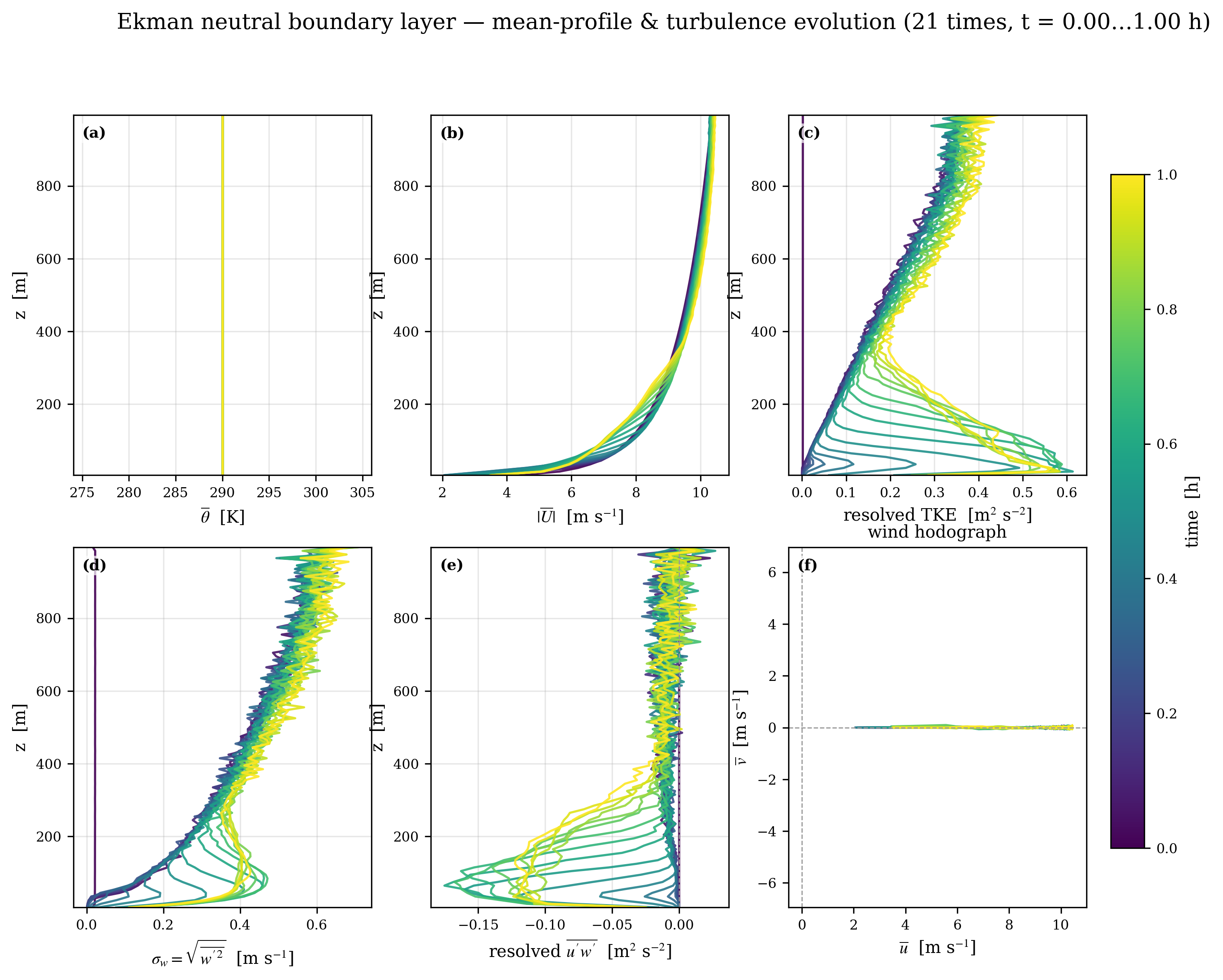}\caption{Ekman layer}\end{subfigure}\hfill
\begin{subfigure}{0.495\textwidth}\includegraphics[width=\textwidth]{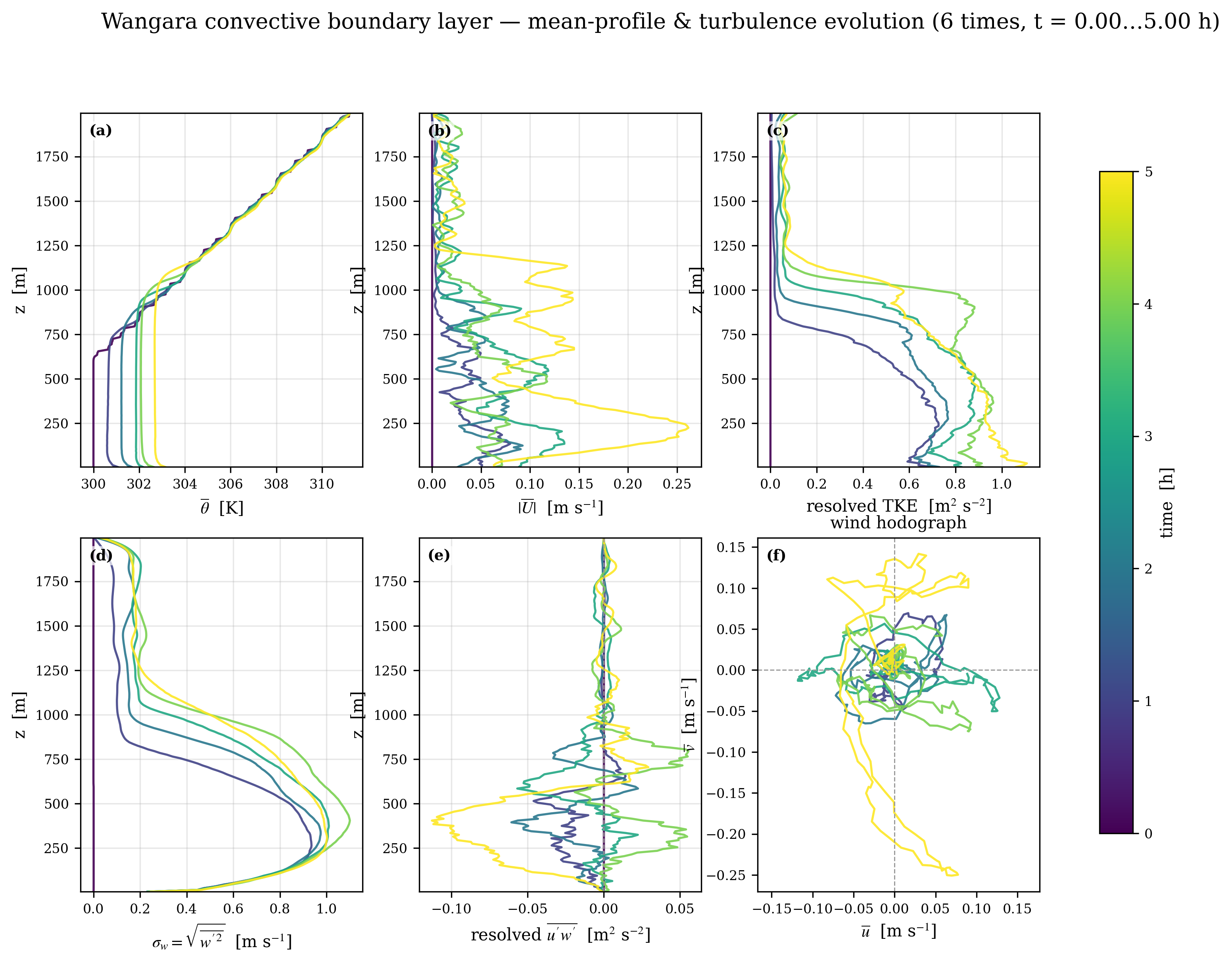}\caption{Wangara diurnal cycle}\end{subfigure}\\[3pt]
\begin{subfigure}{0.495\textwidth}\includegraphics[width=\textwidth]{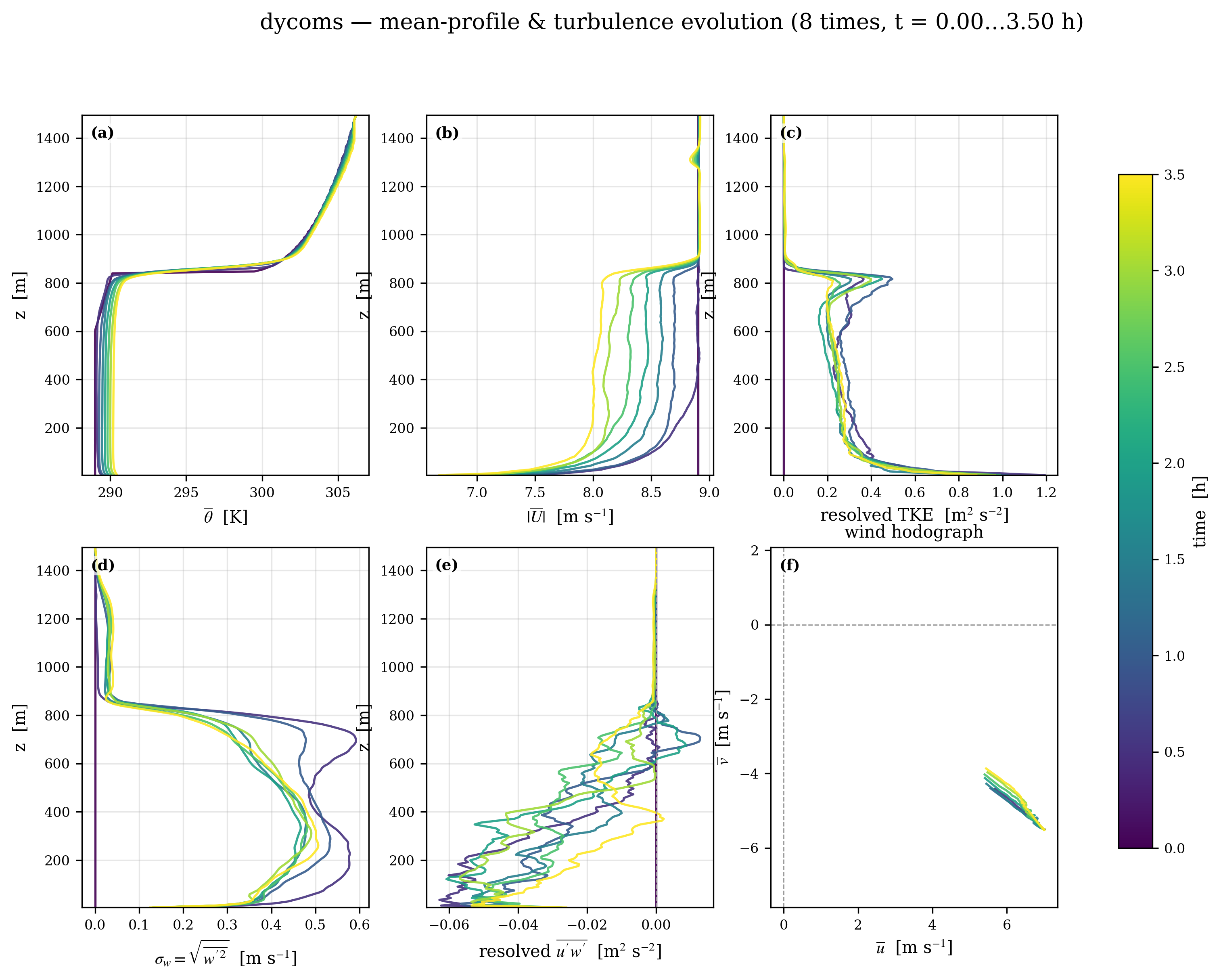}\caption{DYCOMS-II stratocumulus}\end{subfigure}\hfill
\begin{subfigure}{0.495\textwidth}\includegraphics[width=\textwidth]{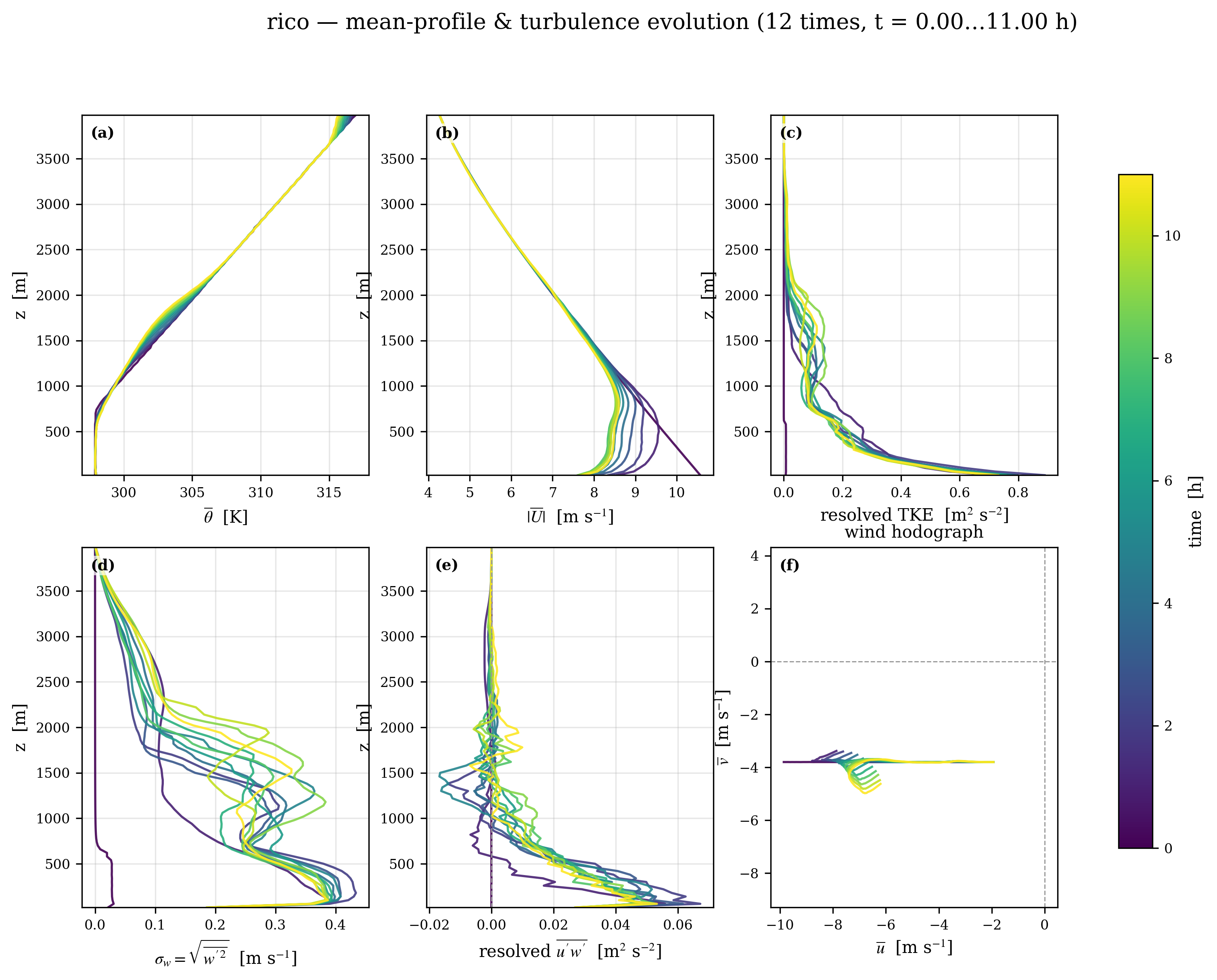}\caption{RICO trade cumulus}\end{subfigure}
\caption{\textbf{Large-eddy simulation of canonical boundary layers.} Evolution of the simulated
profiles for (a) the Ekman layer, (b) the Wangara diurnal cycle, (c) DYCOMS-II stratocumulus and
(d) RICO trade cumulus. Together with the GABLS1 stable boundary layer (Fig.~\ref{si:gabls1}),
these cases span the stable, neutral, diurnal, stratocumulus and shallow-cumulus regimes.}\label{si:les}
\end{figure}

On a doubly periodic plane with the nonhydrostatic core and a scale-dependent dynamic
Smagorinsky closure\cite{Smagorinsky1963,Germano1991,PorteAgel2000}, legoESM reproduces the canonical LES cases: GABLS1\cite{Beare2006}, the neutral Ekman layer\cite{Andren1994}, the Wangara diurnal cycle\cite{Clarke1971}, and the DYCOMS-II\cite{Stevens2005} and
RICO\cite{vanZanten2011} cloud-topped boundary layers (Fig.~\ref{si:les}). Cloud-resolving
RCEMIP\cite{Wing2018} runs (Fig.~\ref{si:crm}) self-aggregate convection; warm-rain
microphysics matches a super-droplet Golovin coalescence benchmark\cite{Shima2009,Bartman2022}
(Fig.~\ref{si:sdm}). The Lagrangian super-droplet method (SDM) is supplied here as a
non-differentiable reference scheme rather than as a trainable component. Its particle-based
formulation is closely connected to what is measured in laboratory and field observations of
individual droplets, quantities that are otherwise very difficult to integrate at the global
modelling scale because of the scale mismatch, which makes it a valuable reference for calibrating
the bulk schemes used in coarser configurations\cite{Lamb2026}.

\begin{figure}[h]\centering
\includegraphics[width=0.75\textwidth]{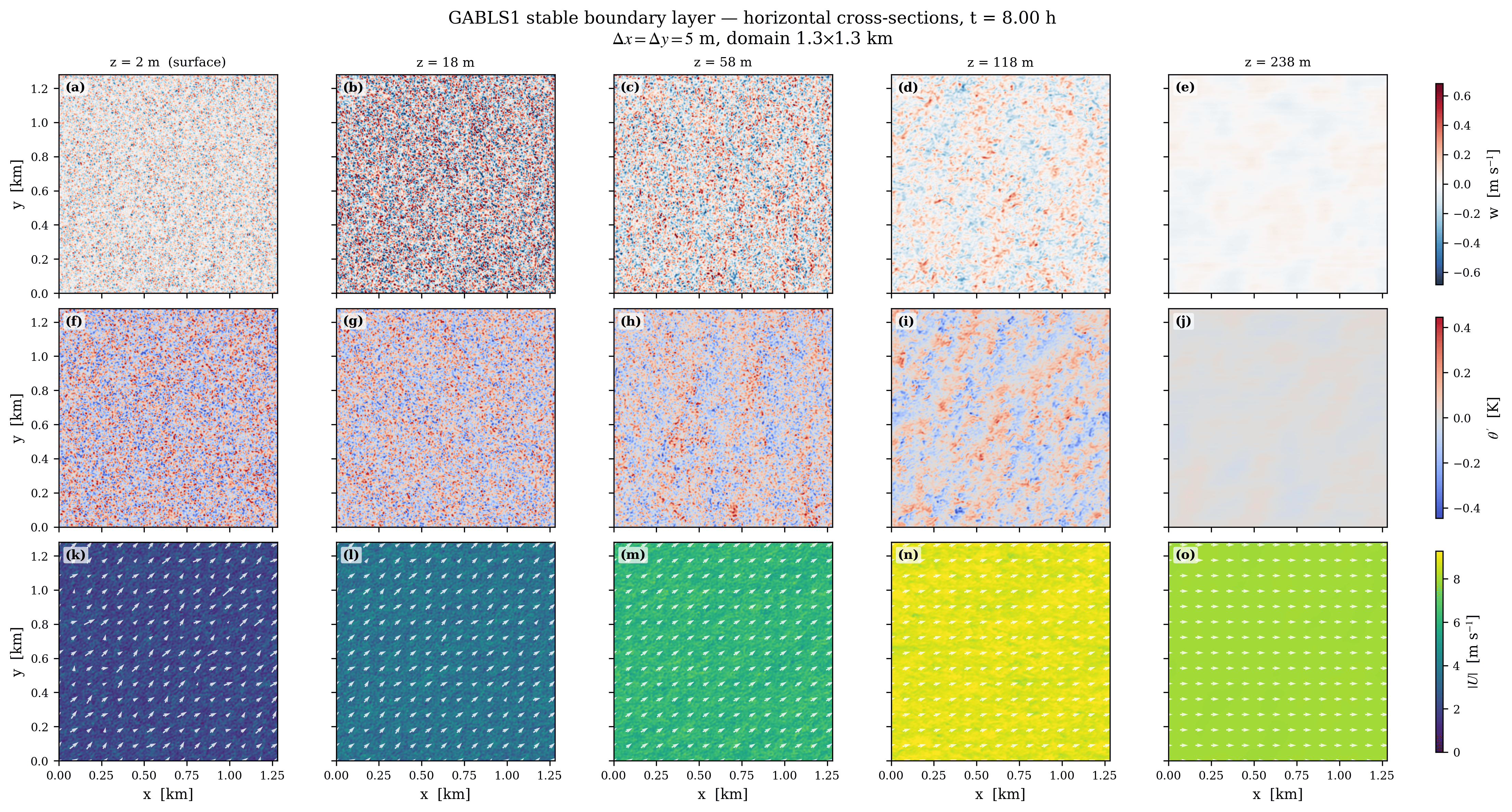}
\caption{\textbf{GABLS1 stable boundary layer.} Instantaneous snapshot from the large-eddy
configuration, showing the shallow, strongly stratified boundary layer that develops under
prescribed surface cooling\cite{Beare2006}.}\label{si:gabls1}
\end{figure}

\begin{figure}[h]\centering
\begin{subfigure}{0.62\textwidth}\includegraphics[width=\textwidth]{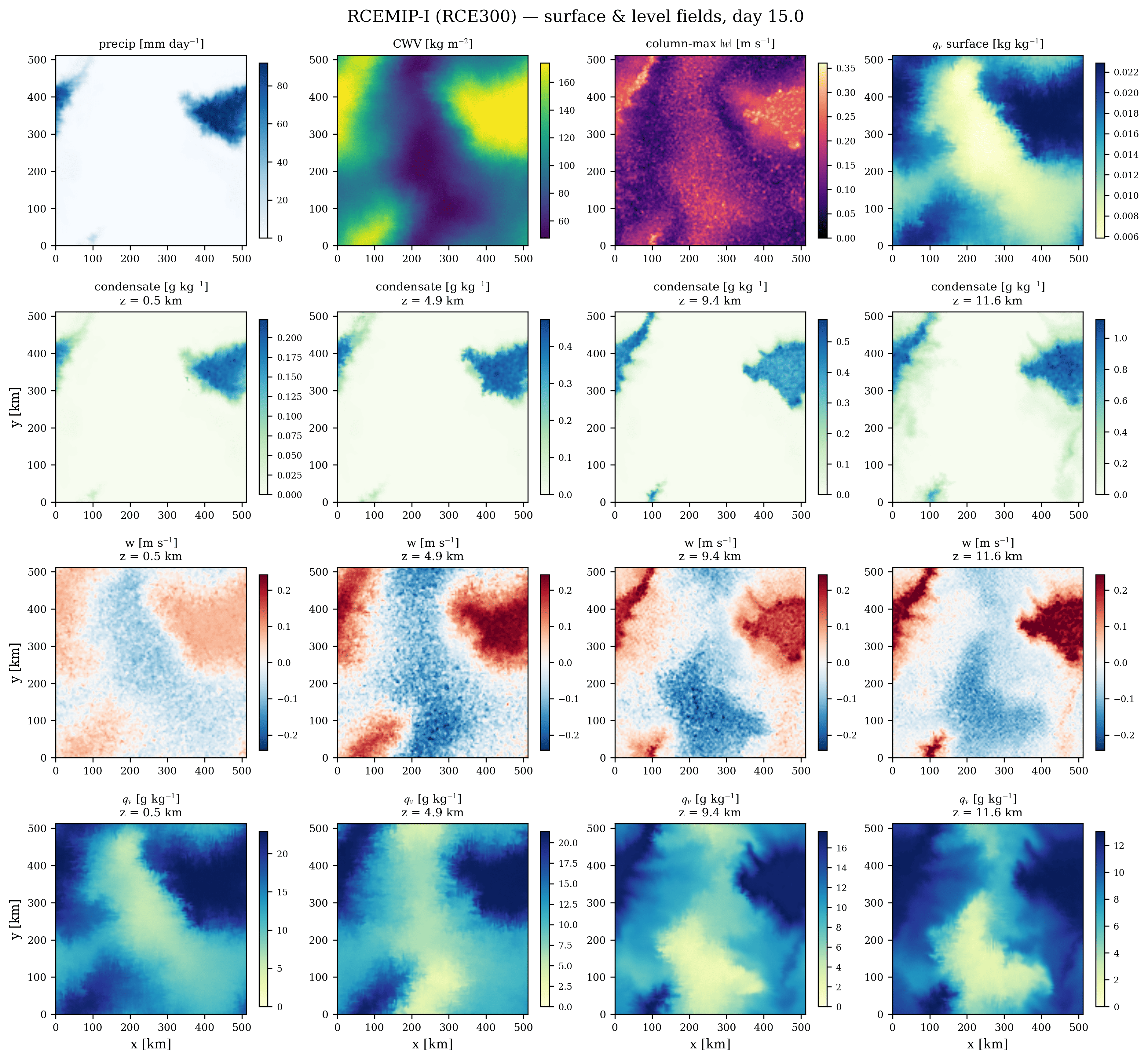}\caption{RCEMIP surface/level fields}\end{subfigure}\hfill
\begin{subfigure}{0.36\textwidth}\includegraphics[width=\textwidth]{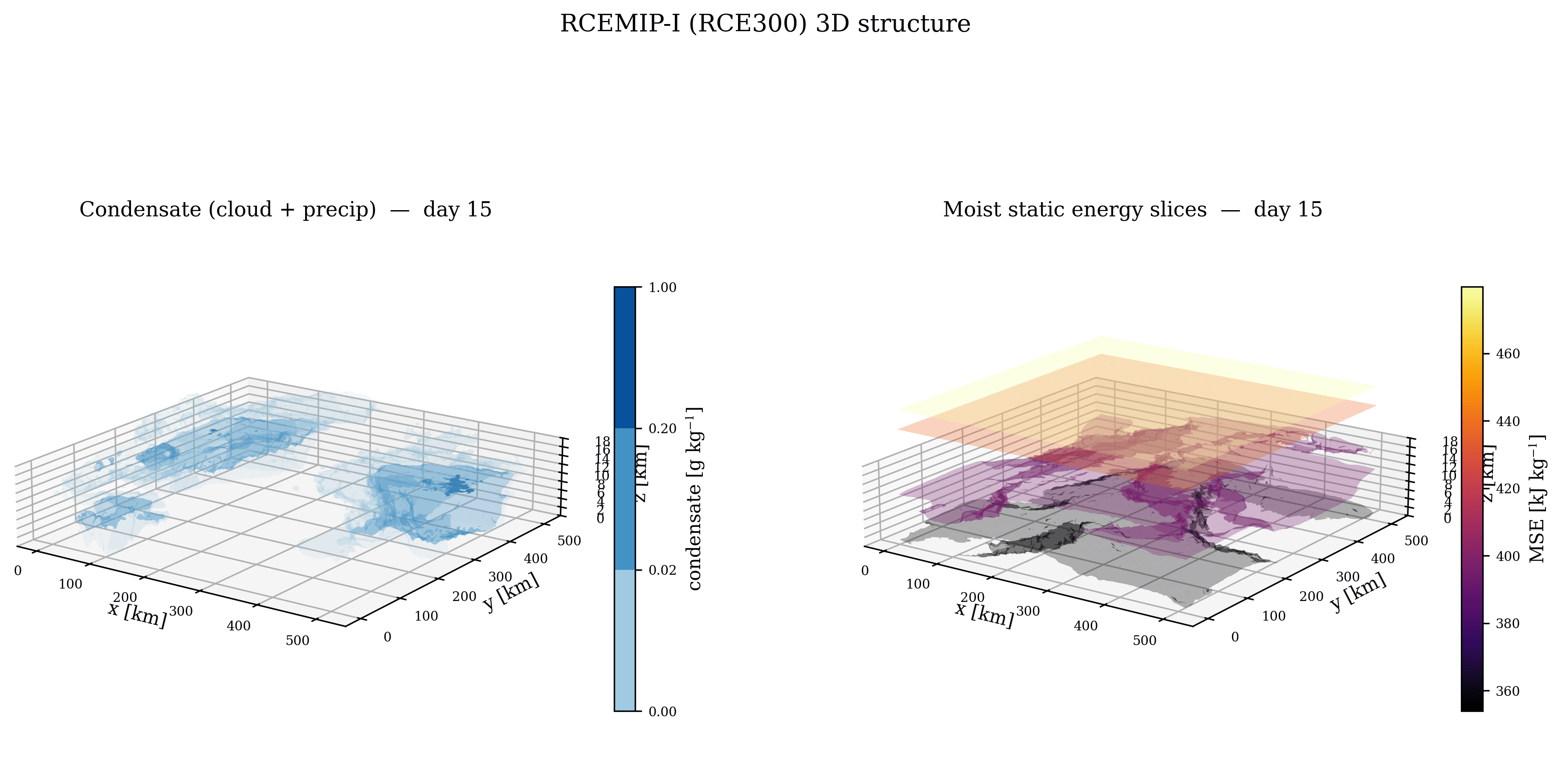}\caption{3D condensate}\end{subfigure}
\caption{\textbf{Cloud-resolving radiative--convective equilibrium (RCEMIP RCE300).}
(a) Surface and level fields, in which convection has spontaneously self-aggregated, and (b) a
three-dimensional snapshot of the condensate field\cite{Wing2018}.}\label{si:crm}
\end{figure}

\begin{figure}[h]\centering
\includegraphics[width=0.7\textwidth]{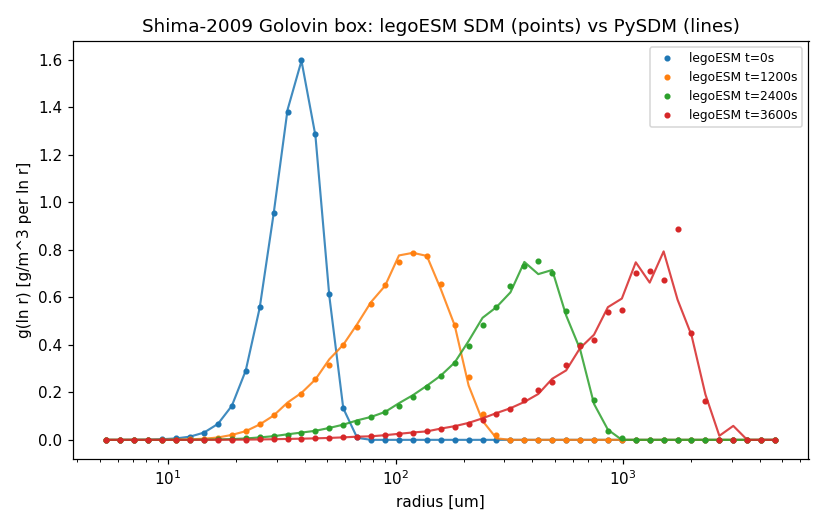}
\caption{Warm-rain collision--coalescence: legoESM super-droplet microphysics vs.\ the PySDM
super-droplet reference on the Golovin analytic benchmark.}\label{si:sdm}
\end{figure}

\section{Atmosphere-only (AMIP) simulations}
\label{si:amip}

The atmosphere-only configuration follows the AMIP protocol of the CMIP6
DECK\cite{Eyring2016}: the ocean surface is prescribed from
observations and the atmosphere and land surface are the only components free
to evolve. Because the sea surface is prescribed, AMIP isolates the behaviour
of the atmosphere and land surface from drift in a prognostic ocean: errors in
the simulated climate arise from the atmosphere, the land, the forcing or the
prescribed boundary conditions rather than from sea-surface temperature drift. It is also the
configuration in which a model can be compared directly with observations over
the satellite era, since the simulated year corresponds to the observed year.
We integrate the standard DECK window, 1~January 1979 to 31~December 2014
(36~years), as a single continuous simulation.

\paragraph{Ocean and sea-ice boundary conditions}

Sea surface temperature (SST) and sea-ice concentration (SIC) are prescribed
from the PCMDI AMIP boundary-condition dataset distributed through
\texttt{input4MIPs}, the same product used by the CMIP models submitting the
DECK \texttt{amip} experiment. These fields are not raw monthly means: they are
the mid-month ``boundary condition'' values of Taylor et
al.\cite{TaylorWilliamsonZwiers2000}, constructed so that \emph{linear
interpolation between adjacent mid-month values recovers the observed monthly
means}. The model therefore uses them as intended, interpolating
linearly in time between the two bracketing mid-month anchors, and the
interpolated value is evaluated once per simulated day and re-applied at every
model step, so that the surface the physics sees is always current rather than
lagged. Because the mid-month anchors deliberately overshoot the physical range
in order to reproduce the monthly means, the physical limits (sea-ice fraction
to $[0,1]$, SST to the seawater freezing point) are imposed \emph{after}
interpolation and never on the anchors themselves; imposing them earlier would
distort the monthly means that the dataset is built to reproduce.
Spatially, the fields are interpolated bilinearly from the observational
$1^\circ$ mesh to the model mesh, and ocean values are extended into land cells
by nearest-neighbour fill, so that no coastal cell can be handed a missing
value.

There is no ocean model and no prognostic sea ice in this configuration. The
sea-ice concentration instead re-partitions each ocean cell: the surface
temperature presented to the atmosphere is the area-weighted blend of the
prescribed SST over open water and an ice-surface temperature over the ice
fraction, and the surface albedo and emissivity are blended the same way, with
the open-water albedo evaluated as a function of solar zenith angle rather than
held constant. Turbulent and radiative surface fluxes are then computed from
this prescribed surface with the same bulk and surface-layer schemes used in
coupled mode. As a consequence, the ocean acts as an infinite heat and moisture
reservoir: surface fluxes are diagnosed from the prescribed surface but cannot
alter it, so atmospheric errors cannot be offset by changes in sea-surface
temperature, and biases that may be partly compensated by SST drift in a coupled
model remain visible.

\paragraph{External forcing}

All external forcings are transient over the full simulation period and are
taken from the CMIP6 forcing datasets, so that the simulation is driven by the
same imposed history as the CMIP6 historical and \texttt{amip} experiments.
Solar forcing uses the CMIP6 solar dataset\cite{Matthes2017} in its spectrally
resolved form: the total solar irradiance carries the 11-year cycle, and the
accompanying band-resolved spectral irradiance is mapped onto the shortwave
$g$-points of the correlated-$k$ radiation scheme\cite{Pincus2019}, so that the
\emph{spectral distribution} of the solar variation reaches the gas optics
rather than only its integral. Insolation is computed for a realistic
orbit\cite{Berger1978}, including the solar declination and the
eccentricity-driven Earth--Sun distance factor, and the diurnal cycle is
resolved explicitly, with radiation called every 15~simulated minutes.
Greenhouse gases (CO$_2$, CH$_4$, N$_2$O, CFC-11, CFC-12) are prescribed as
globally uniform, annually resolved concentrations from the CMIP6 historical
concentration dataset\cite{Meinshausen2017}, interpolated to model time.
Ozone is prescribed from the CMIP6 monthly, zonally averaged,
interannually varying ozone dataset\cite{ChecaGarcia2018}, interpolated in
time, in latitude, and onto each column's own pressure levels at every
radiation call, so that the stratospheric ozone trend and the seasonal Antarctic
ozone hole are represented rather than replaced by a fixed climatology.
Tropospheric aerosols are prescribed as \emph{optical properties} rather than
emissions, using the monthly aerosol optical-property climatology of Kinne et
al.\cite{Kinne2013} resolved across the 14 shortwave bands; this is the
prescribed-aerosol option of the CMIP6 protocol and it represents the aerosol
direct effect only, with no interactive aerosol, chemistry or aerosol--cloud
indirect effect. Stratospheric volcanic aerosol is prescribed from the CMIP6
volcanic dataset, which is height resolved and carries both the extinction and
the single-scattering albedo in 14 shortwave and 16 longwave bands. Both
channels are used: the shortwave extinction and the longwave \emph{absorption}
optical depth are placed at the pressure corresponding to the true altitude of
each aerosol layer, so that El~Chichón (1982) and Pinatubo (1991) produce not
only the surface dimming but also the stratospheric warming that accompanies it.

Because the model runs on a 365-day (no-leap) calendar while several forcing
files are dated on a Gregorian calendar, forcing time is mapped by calendar
date rather than by elapsed days. Over 36~years this matters: mapping linearly
would accumulate a nine-day phase error between the forcing channels and the
model's own seasonal cycle.

\paragraph{Land surface}

The land is fully interactive and is run in the same configuration as in
coupled simulations, rather than being reduced to a simpler surface for the
atmosphere-only case. Soil moisture is advanced by the mass-conservative
mixed-form Richards equation\cite{Richards1931,van1980closed,Celia1990} on a multi-layer
column, soil temperature by heat diffusion with Johansen
conductivity\cite{Johansen1975}, and snow is prognostic with an
albedo feedback. Surface properties (plant functional type, soil texture,
glacier fraction) and the land--sea mask are taken from a single consistent
surface dataset, so that the fraction of a cell treated as land by the flux
calculation is the same fraction the land model integrates; deriving the mask
independently, for example from bathymetry, misclassifies endorheic basins such
as the Caspian and Aral seas as ocean and hands them a prescribed SST with
unlimited moisture supply. Surface exchange uses a Monin--Obukhov surface layer
over land and a bulk-flux algorithm (Extended Data Table~9) over the ocean.
Land surface temperature, soil moisture, soil temperature and snow are
hence prognostic and, together with the atmosphere, form the freely
evolving part of the system; land hydrology may therefore develop biases
relative to observations. Land use and land cover are held fixed at present-day values;
this is a documented simplification relative to the CMIP6 protocol, which asks
for transient land use in \texttt{amip}.

\paragraph{Initialization and spin-up}

The atmosphere is initialized directly from ERA5\cite{Hersbach2020} at
00:00~UTC on 1~January 1979: temperature, horizontal winds, specific humidity
and surface pressure are interpolated onto the model mesh and vertical
coordinate. Because the initial state comes from a reanalysis rather than
from a previous model integration, it is not in balance with the model's own
climate, and the model adjusts rapidly during the first days of the
simulation. The land is cold-started from a uniform soil moisture and
temperature profile with no snow, and no separate spin-up cycle is performed.
The components adjust on different timescales: the atmosphere
loses memory of its initial state within days to weeks, the seasonal snowpack
and the upper soil within a season, whereas the deep soil column equilibrates
only over years. We therefore discard the first simulated year from all
evaluation, and we do not interpret deep-soil temperature or moisture
trends over the early record. The spin-up strategy is stated explicitly here
because, under the CMIP6 protocol, how quasi-equilibrium was established is
part of the experiment definition.

A single realization is integrated. Since AMIP climate is largely, but not
entirely, set by the imposed boundary conditions, unforced atmospheric
variability contributes to any individual year; with one member we therefore
evaluate multi-year statistics and do not attribute year-to-year differences to
the forcing.

\paragraph{Integration protocol}

The 36~years are integrated as one continuous simulation on a single GPU. The
run is chained across scheduler allocations: the model writes a complete restart
checkpoint every 30 simulated days and, on reaching a wall-clock limit,
checkpoints and exits cleanly, after which the next allocation resumes from that
checkpoint. Restarts carry the full prognostic state, including the land column,
the snow and the diagnostic accumulators, so the chain is a single
restart-continuous integration rather than a sequence of independent segments,
and monthly means computed across a restart boundary are not affected by the restart. Diagnostics
(global-mean temperature, water vapour, precipitation and the top-of-atmosphere
budget) are printed once per simulated day, which is fine enough that any
instability can be localized to the day on which it began.

The experiment is defined by one machine-independent configuration file that
fixes the model configuration, the physics selection, the forcing selectors and
the output request; only the data paths are supplied at run time. Reproducing
the experiment on another machine therefore requires changing paths and nothing
else, and every run additionally emits a reproducibility manifest recording the
resolved configuration, the git commit and the library versions.

\paragraph{Output}

Output is written as CF-1.8/CMOR-compliant NetCDF following the CMIP6 data
reference syntax, one variable per file, regridded from the model's native mesh
to a regular latitude--longitude grid. The monthly atmospheric stream follows
the CMIP6 \texttt{Amon} table and includes the near-surface and surface state
(\texttt{tas}, \texttt{ts}, \texttt{ps}, \texttt{psl}, \texttt{hurs}), the
hydrological cycle (\texttt{pr}, \texttt{prw}, \texttt{evspsbl}), the cloud
fields (\texttt{clt}, \texttt{clwvi}, \texttt{clivi}), the full top-of-atmosphere
and surface radiation budget, the turbulent heat fluxes and surface stresses,
and the three-dimensional fields (\texttt{ta}, \texttt{ua}, \texttt{va},
\texttt{hus}, \texttt{hur}, \texttt{zg}, \texttt{wap}, \texttt{clw},
\texttt{cli}) interpolated onto the 19 standard CMIP6 pressure levels. A land
monthly stream (\texttt{Lmon}) carries gross primary production, net ecosystem
exchange, leaf area index, total and surface soil moisture and soil temperature,
and a daily stream carries \texttt{tas}, \texttt{tasmin}, \texttt{tasmax},
\texttt{pr}, \texttt{psl}, \texttt{rsut}, \texttt{rlut} and 850\,hPa winds for
variability and extremes diagnostics. Clear-sky radiation is diagnosed
alongside all-sky radiation at both the top of the atmosphere and the surface,
so that the cloud radiative effect is obtained by differencing and a radiation
error can be separated into a clear-sky (gas, ozone, aerosol, surface albedo)
and a cloud contribution.

\paragraph{Evaluation}

\begin{figure}[!htbp]\centering
\includegraphics[width=0.86\textwidth,height=0.78\textheight,keepaspectratio]{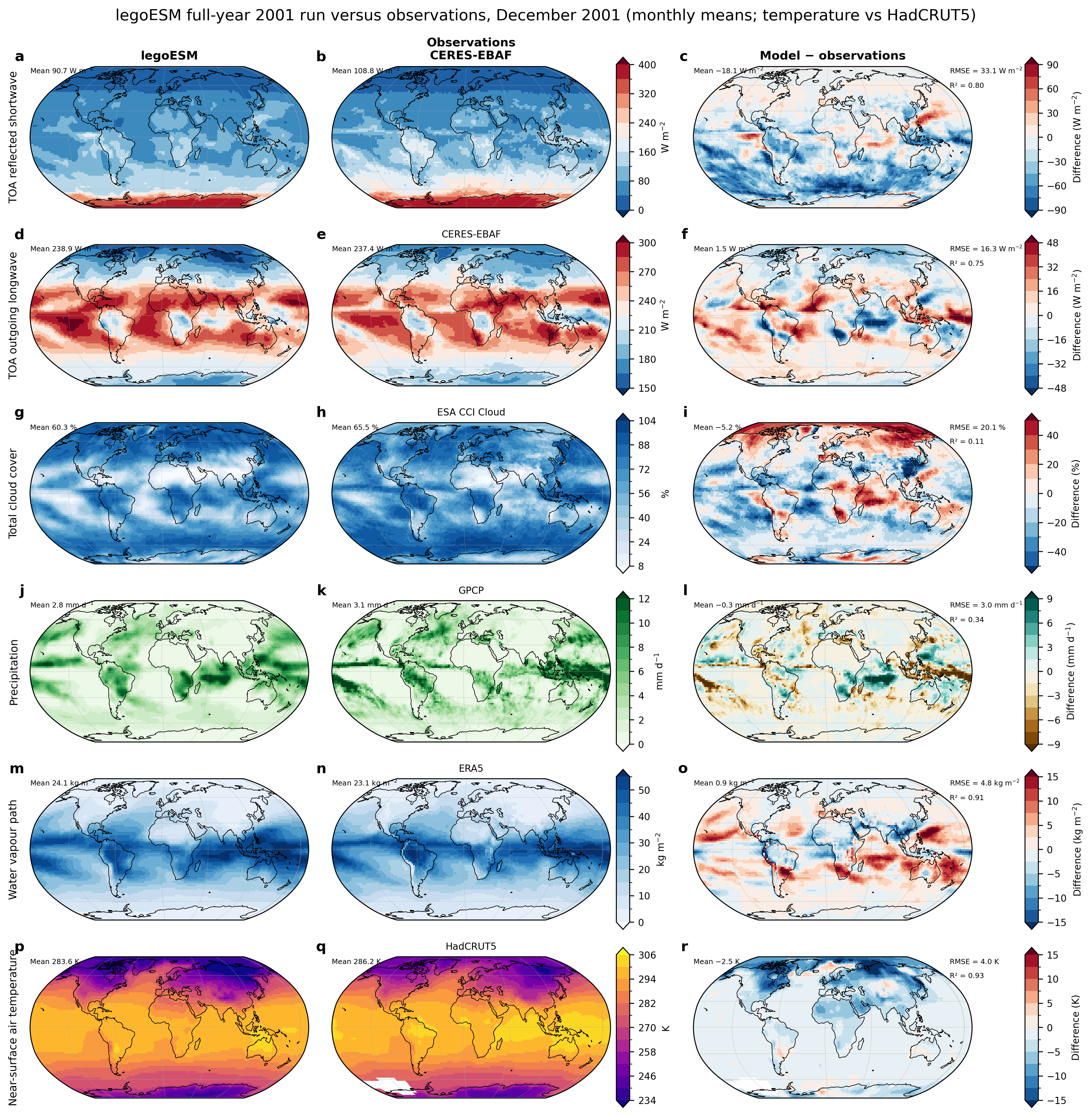}
\caption{\textbf{AMIP simulation against observations.} December 2001 monthly means of
top-of-atmosphere reflected shortwave and outgoing longwave radiation (CERES-EBAF), total cloud
cover (ESA-CCI Cloud), precipitation (GPCP), water-vapour path (ERA5) and near-surface air
temperature (HadCRUT5) from the legoESM atmosphere-only simulation (left; CAM6-like physics
configuration), the observational reference (centre) and their difference (right). Global means
are printed at the top left of each map; the root-mean-square error and the pattern $R^2$ at the
top right of each difference map.}\label{si:amip_eval}
\end{figure}

Figure~\ref{si:amip_eval} compares simulated monthly-mean fields (December 2001) with observations,
using the ESMValTool-based ClimateEval recipes\cite{Righi2020} applied to CMIP6 models. The
large-scale patterns of near-surface air temperature, water-vapour path, precipitation and cloud
cover are reproduced (pattern $R^2$ of 0.93 for near-surface temperature and 0.91 for water-vapour
path). The main biases are a global-mean cold bias of $-2.5$\,K in near-surface air temperature,
a precipitation deficit of about 10\% ($2.8$ versus $3.1$\,mm\,d$^{-1}$ in GPCP), and too little
reflected shortwave radiation at the top of the atmosphere ($90.7$ versus $108.8$\,W\,m$^{-2}$ in
CERES-EBAF), consistent with too little total cloud cover ($60.3$ versus $65.5$\,\%), particularly
over the subtropical stratocumulus decks and the Southern Ocean. These are typical targets for the
gradient-based calibration described in the main text.

\section{Ocean validation}\label{s5:ocean_val}
\begin{figure}[!htbp]\centering
\includegraphics[width=0.78\textwidth,height=0.74\textheight,keepaspectratio]{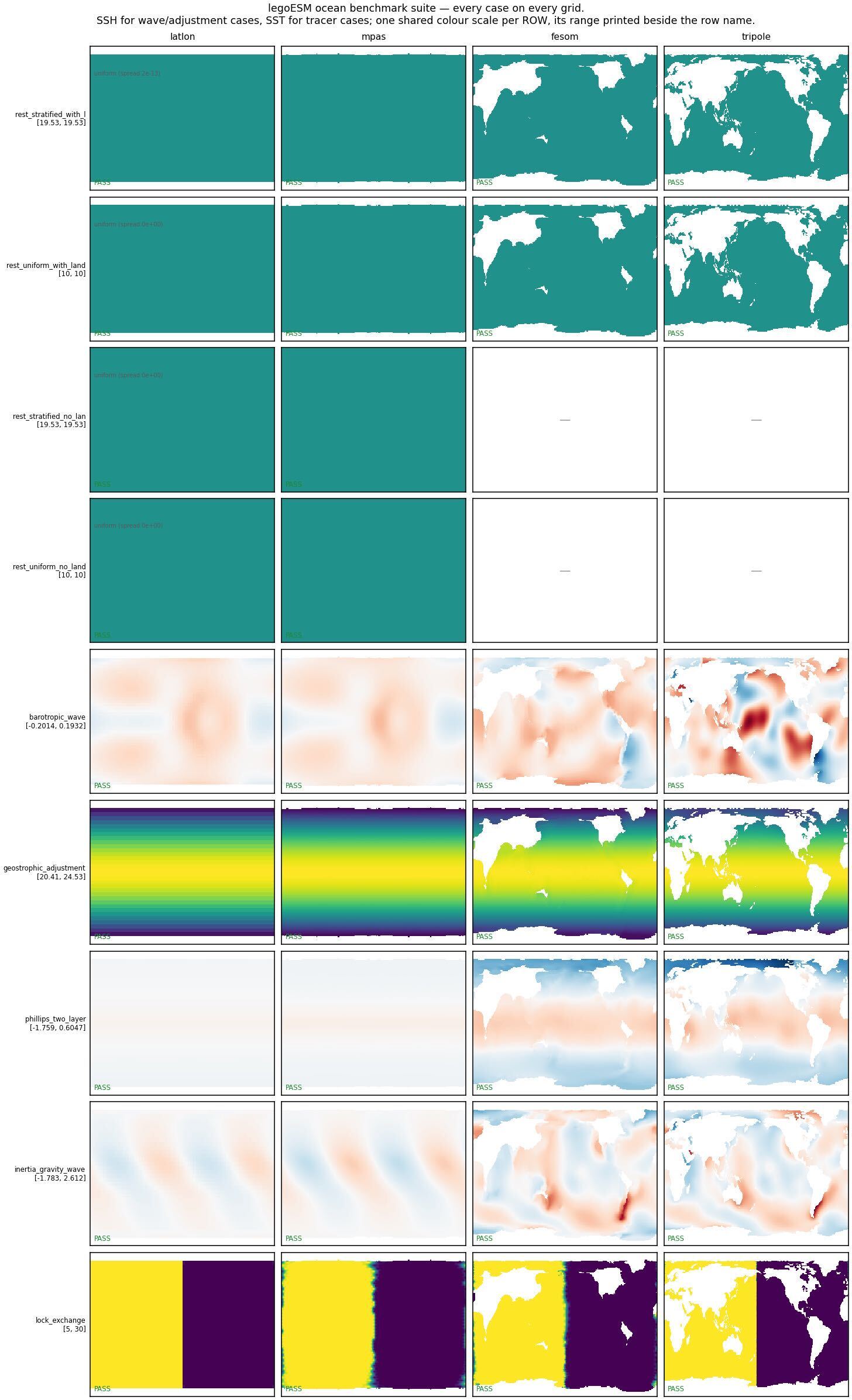}
\caption{\textbf{Ocean benchmark suite on four meshes.} Each row is one idealized case and each
column one mesh (lat--lon, MPAS/Voronoi, FESOM-type unstructured and tripolar); sea-surface
height is shown for the wave and adjustment cases and sea-surface temperature for the tracer
cases, with one shared colour scale per row (range printed beside the case name). Cases without
land are not defined on the realistic-coastline FESOM and tripolar meshes. Every applicable case
passes on each grid on which it is defined.}\label{si:oceanbench}
\end{figure}

\begin{figure}[!htbp]\centering
\includegraphics[width=0.8\textwidth]{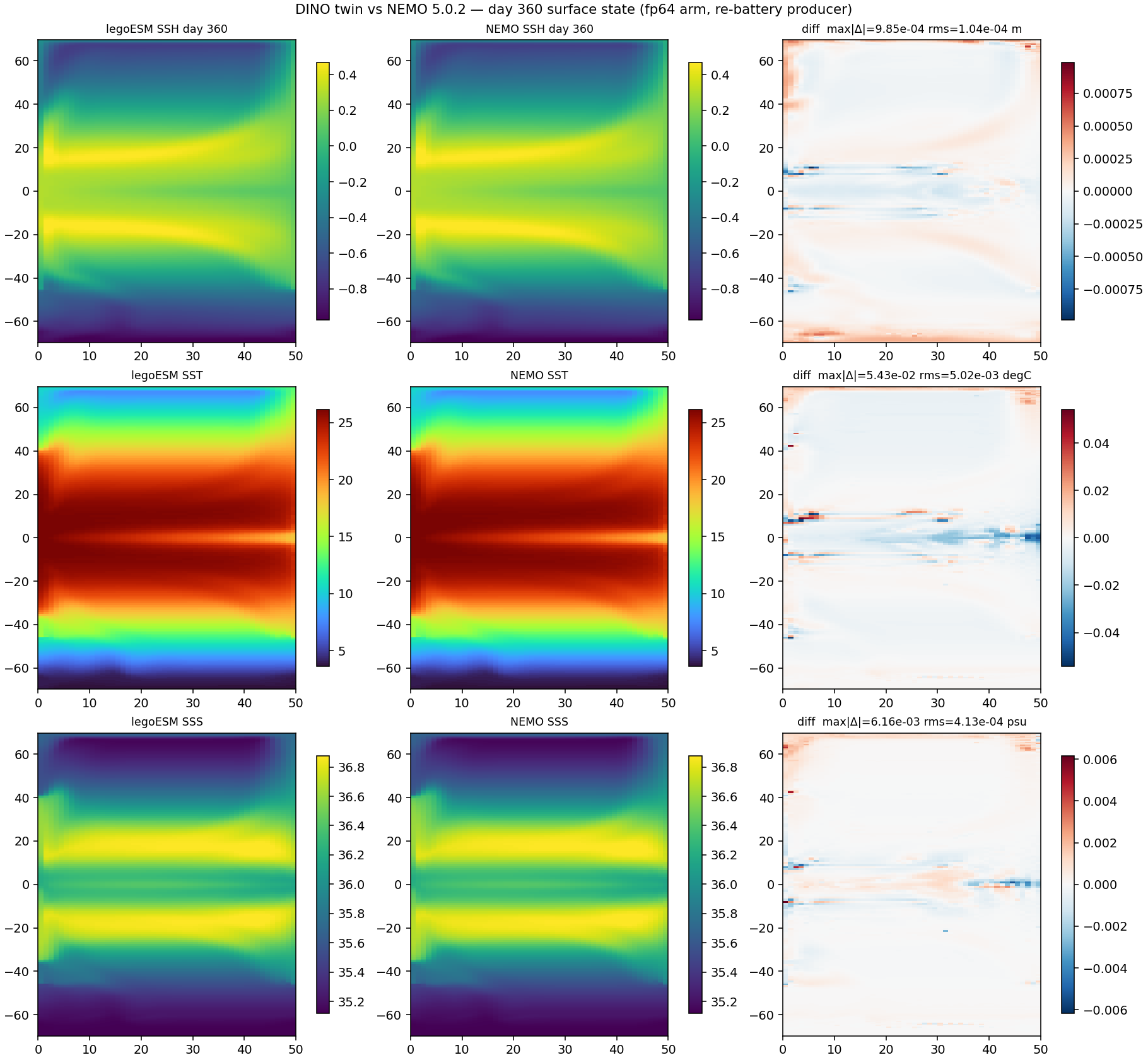}
\caption{\textbf{Twin experiment against NEMO~5.0.2 in the idealized DINO basin.} Sea-surface
height (top), temperature (middle) and salinity (bottom) after 360 days from legoESM (left) and
NEMO (centre), and their difference (right; maximum absolute and r.m.s.\ differences in the panel
titles).}\label{si:dino}
\end{figure}

\begin{figure}[!htbp]\centering
\begin{subfigure}{\textwidth}\includegraphics[width=\textwidth]{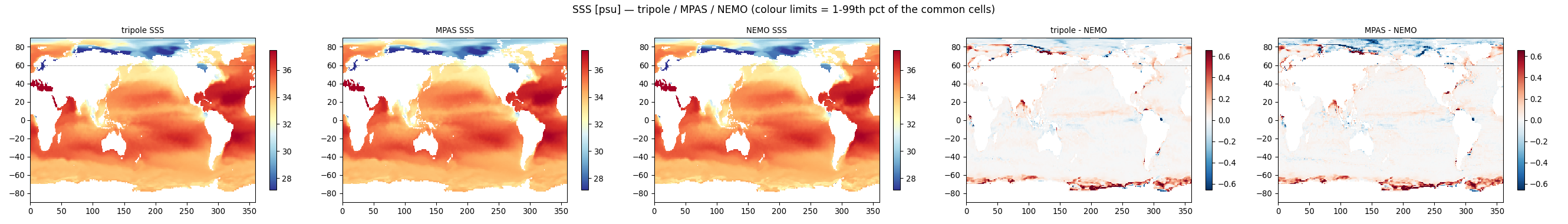}\caption{Sea-surface salinity (psu)}\end{subfigure}\\[3pt]
\begin{subfigure}{\textwidth}\includegraphics[width=\textwidth]{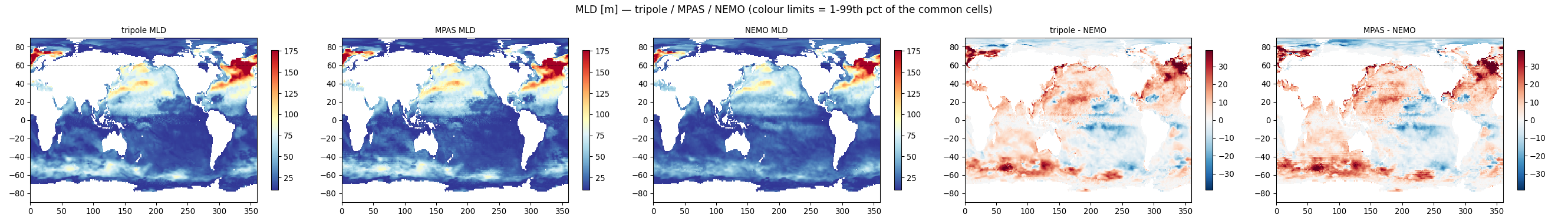}\caption{Mixed-layer depth (m)}\end{subfigure}
\caption{\textbf{Additional global ocean diagnostics against NEMO.} (a) Sea-surface salinity and
(b) mixed-layer depth from the legoESM ocean on the tripolar and MPAS/Voronoi grids and from
NEMO\cite{Madec2017}, with the differences tripolar$-$NEMO and MPAS$-$NEMO (right), on the cells
common to all three grids. Zonal means are shown in main-text Fig.~3b.}\label{si:ocean}
\end{figure}

\begin{figure}[!htbp]\centering
\begin{subfigure}{\textwidth}\includegraphics[width=\textwidth]{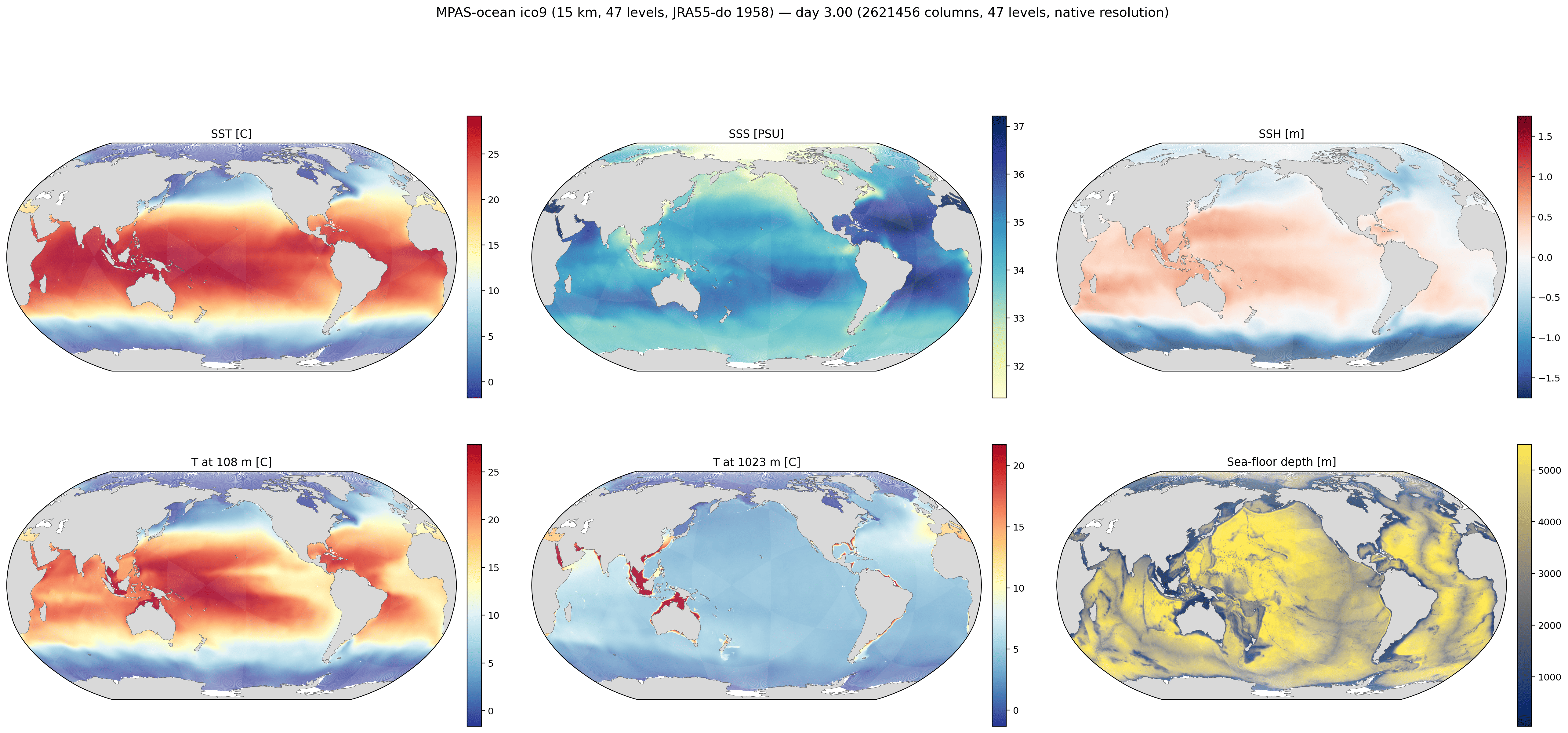}\caption{MPAS-ocean icosahedral mesh at 15\,km (ico9, 47 levels)}\end{subfigure}\\[3pt]
\begin{subfigure}{\textwidth}\includegraphics[width=\textwidth]{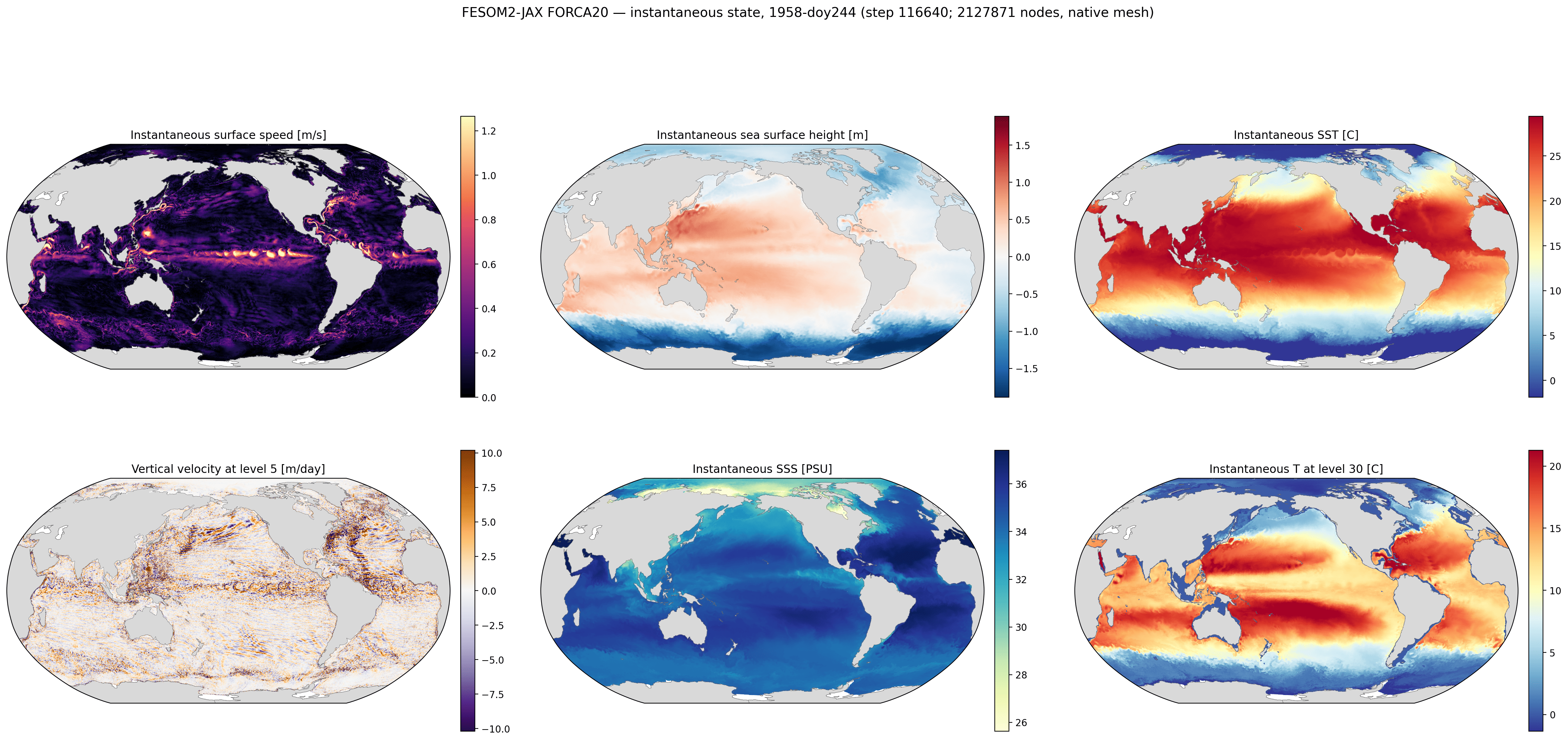}\caption{FESOM2 FORCA20 mesh (2.1 million surface nodes)}\end{subfigure}
\caption{\textbf{Eddy-permitting global ocean configurations on unstructured meshes.} (a) Sea-surface
temperature, salinity and height, temperature at 108 and 1023\,m and sea-floor depth on the native
15-km MPAS icosahedral mesh, three days after initialization with JRA55-do forcing (1958).
(b) Instantaneous surface speed, sea-surface height, temperature and salinity, vertical velocity
at level~5 and temperature at level~30 on the native FESOM2 FORCA20 mesh (1958, day of year
244).}\label{si:oceanhr}
\end{figure}

\begin{figure}[!htbp]\centering
\begin{subfigure}{0.495\textwidth}\includegraphics[width=\textwidth]{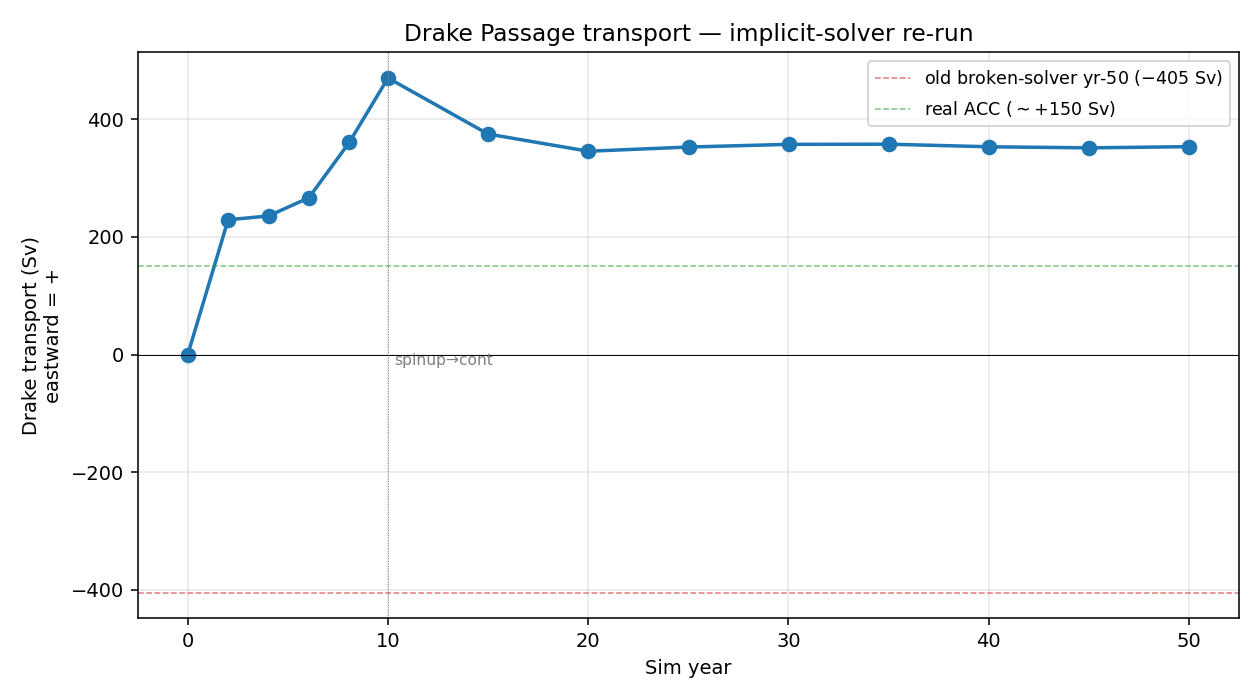}\caption{Drake Passage transport}\end{subfigure}\hfill
\begin{subfigure}{0.495\textwidth}\includegraphics[width=\textwidth]{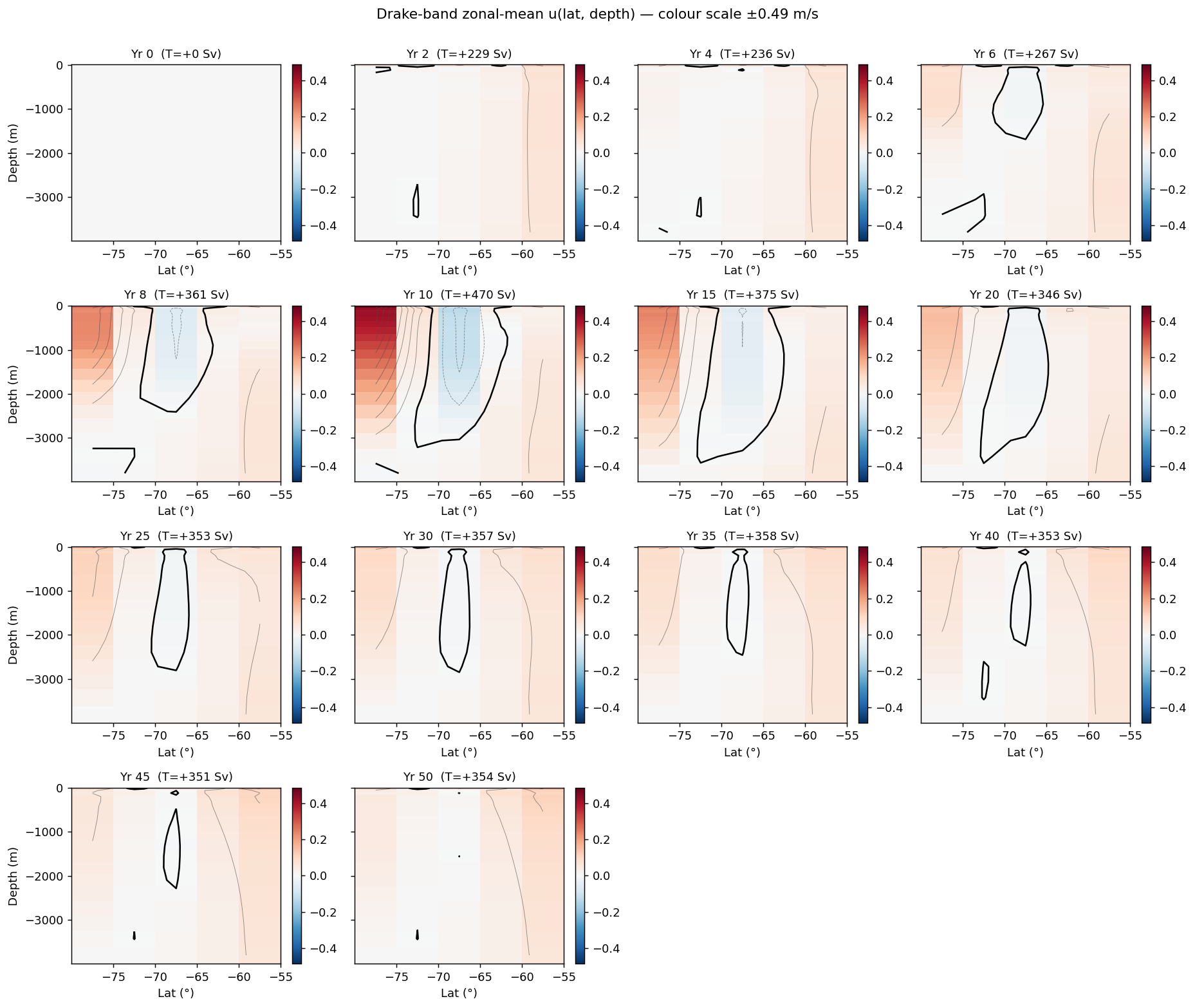}\caption{Zonal-mean zonal velocity}\end{subfigure}
\caption{\textbf{Multi-decadal Drake Passage spin-up.} (a) Zonal transport and (b) zonal-mean
zonal velocity, both of which equilibrate, indicating a stable Antarctic Circumpolar Current under
the implicit solver.}\label{si:drake}
\end{figure}

The ocean is validated in three stages of increasing realism. First, a benchmark suite of
idealized cases (states of rest with and without land and stratification, barotropic and
inertia--gravity waves, geostrophic adjustment, a two-layer Phillips problem and a lock exchange)
is run on four meshes spanning three of the grid families of \S1.1 (the lat--lon and tripolar
C-grid meshes, the MPAS/Voronoi mesh and a FESOM-type triangular mesh). Every applicable case
passes on each grid on which it is defined; cases without land are not defined on the
realistic-coastline FESOM and tripolar meshes (Fig.~\ref{si:oceanbench}).
Second, legoESM is run as a twin of NEMO~5.0.2 in the idealized DINO basin configuration: after
360 days the two models agree to $5\times10^{-3}$\,$^\circ$C (r.m.s.) in sea-surface temperature,
$4\times10^{-4}$\,psu in sea-surface salinity and $10^{-4}$\,m in sea-surface height
(Fig.~\ref{si:dino}). Third, realistic global configurations are compared with
NEMO\cite{Madec2017}. Beyond the sea-surface temperature and zonal means of the main text
(main-text Fig.~3), Fig.~\ref{si:ocean} shows sea-surface salinity and mixed-layer depth on the
tripolar and MPAS/Voronoi grids against NEMO. Figure~\ref{si:oceanhr} shows
eddy-permitting global configurations on an MPAS icosahedral mesh at 15\,km (2.6~million columns)
and on the FESOM2 FORCA20 mesh (2.1~million nodes), the latter being the configuration used in the
scaling experiments of main-text Fig.~4. Finally, Fig.~\ref{si:drake} shows a
multi-decadal Drake Passage spin-up: the zonal transport and zonal-mean zonal-velocity structure
equilibrate, consistent with a stable Antarctic Circumpolar Current under the split-explicit solver.

\section{Land-surface validation at eddy-covariance flux towers}
\label{si:ecsite}

The land component is evaluated offline against point-scale eddy-covariance
measurements, which constrain the coupled water, energy and carbon fluxes at the
sub-daily to seasonal timescales that a coupled simulation cannot isolate. This
section documents the sites, the forcing data, the model configuration, and the
experimental and evaluation protocols.

\paragraph{Sites}

Four FLUXNET towers were selected to span the temperate and Mediterranean
regimes (Table~\ref{tab:ecsites}): two temperate deciduous forests differing in climate
and soil (US-MMS, DE-Hai), a montane evergreen needleleaf forest (DE-Obe), and a
water-limited oak savanna (US-Ton). Peak leaf area index ranges from
2.9\,m$^2$\,m$^{-2}$ at the savanna to 6.2\,m$^2$\,m$^{-2}$ at the closed
forests, and soil texture spans clay to loam, so the four sites exercise
contrasting water-limitation, phenology and surface-roughness regimes.

\begin{table}[h]\centering\footnotesize
\caption{Eddy-covariance sites used for the land-surface validation. $h_c$ is the
canopy height and $z_{\mathrm{ref}}$ the flux-measurement (tower) height, which
sets the reference level of the modelled surface-layer profile.}\label{tab:ecsites}
\begin{tabular}{@{}llccc@{}}
\toprule
Site & Vegetation & Location & $h_c$/$z_{\mathrm{ref}}$ & Period \\
     &            &          & (m) & \\
\midrule
US-MMS\cite{USMMS} & deciduous broadleaf  & 39.3$^\circ$N, 86.4$^\circ$W  & 27/46 & 2015--2018 \\
DE-Obe\cite{DEObe} & evergreen needleleaf & 50.8$^\circ$N, 13.7$^\circ$E  & 20/30 & 2015--2018 \\
US-Ton\cite{USTon} & oak savanna          & 38.4$^\circ$N, 121.0$^\circ$W & 10/24 & 2015--2018 \\
DE-Hai\cite{DEHai} & deciduous broadleaf  & 51.1$^\circ$N, 10.5$^\circ$E  & 34/43.5 & 2010--2013 \\
\bottomrule
\end{tabular}
\end{table}

\paragraph{Forcing data}

Each site is driven by a per-site forcing file combining tower meteorology
(downwelling shortwave and longwave radiation, air temperature, vapour-pressure
deficit, wind speed, surface pressure, precipitation and CO$_2$ mole fraction)
with MODIS leaf area index\cite{Myneni2002} and albedo\cite{Schaaf2002}, the
diffuse fraction of photosynthetically active radiation from the BESS radiation
product\cite{Ryu2018}, and site parameters. The foliage clumping index, which
enters the canopy radiative transfer through the effective leaf area index
$L_{\mathrm{eff}} = \mathrm{LAI}\cdot\mathrm{CI}$, is taken from the global
0.05$^\circ$ monthly CAS-CI product\cite{LiFang2022} derived from the MODIS
BRDF-based clumping index\cite{Wei2019}, and is applied as a per-site monthly
climatology (0.51--0.81 across the four sites). The tower data are the FULLSET
products generated by the ONEFlux processing pipeline\cite{Pastorello2020}: the
AmeriFlux FLUXNET-1F releases at US-MMS (v3-5)\cite{USMMS} and US-Ton
(v4-6)\cite{USTon}, and the corresponding FLUXNET archive products at
DE-Obe\cite{DEObe} and DE-Hai\cite{DEHai}. The forcing is half-hourly except at
US-MMS, which is hourly.

As prognostic soil integration cannot tolerate a missing forcing step, the meteorological variables gap-filled by marginal distribution sampling (MDS)\cite{Reichstein2005} within the ONEFlux
pipeline\cite{Pastorello2020} are used. The remaining interior gaps shorter than 24\,h are filled by linear interpolation, and gaps of at least 24 \, h, together with any gap touching the start or end of the record, are filled with a month-of-year by time-of-day climatology computed from the observed samples alone. Only observed fluxes are used as evaluation targets; filled values do not enter any skill metric.

\paragraph{Model configuration}

The land column is run in its multi-layer configuration. Soil moisture follows
the mass-conservative mixed-form Richards equation\cite{Celia1990} with
van~Genuchten retention\cite{van1980closed} whose parameters are derived from
the site soil texture, solved by Picard iteration on a tridiagonal system. Soil
temperature is advanced by backward-Euler heat diffusion with Johansen thermal
conductivity\cite{Johansen1975}. The column is discretized into eight
geometrically spaced layers reaching approximately 6.4\,m (extended to 10\,m at
the deep-rooted savanna), with free drainage at the base.

Surface exchange uses the two-big-leaf canopy scheme. Shortwave transfer follows
a two-stream partition into sunlit and shaded leaves\cite{Sellers1985,Ryu2011},
each of which carries Farquhar--von~Caemmerer--Berry C$_3$
photosynthesis\cite{Farquhar1980} with Kattge--Knorr thermal
acclimation\cite{KattgeKnorr2007}, or C$_4$ photosynthesis following
Collatz\cite{Collatz1992}, coupled to Ball--Berry stomatal
conductance\cite{Ball1987}. The resulting six-variable system in sunlit and
shaded leaf temperature, the two intercellular CO$_2$ concentrations, and
canopy-air temperature and humidity is closed by Newton--Raphson iteration; an
outer Picard loop reconciles the canopy solution with the soil thermal solver.
Aerodynamic exchange is referenced to the measured tower height at each site,
with roughness length and displacement height scaled from the site canopy height by plant-functional-type ratios, with an LAI dependence. The Ball--Berry cuticular intercept is not down-regulated by soil-moisture
stress, so a live canopy retains a baseline conductance while a dormant canopy
self-limits as its leaf area declines.

Rooting depth and soil-column depth are prescribed modelling choices rather than
measurements: rooting depth is the e-folding depth of an exponential
root-density profile, set between 1 and 5\,m according to the plant functional
type and soil profile, deepest at the phreatophytic savanna.

\paragraph{Experimental protocol}

Each site is integrated as an independent offline single column at the native
forcing timestep over the full analysis window, with soil moisture and soil
temperature carried prognostically. The column is initialized to a vertically
uniform profile: every layer takes the first available prescribed soil
temperature and volumetric water content at the start of the window, the matric
potential is then derived from that water content through the van~Genuchten
retention curve, and the snow, ponding and runoff stores start empty. The
initial water content is clipped to the hydraulic range of the site soil,
which binds only at US-MMS, where the measured water content (0.43\,m$^3$\,m$^{-3}$)
exceeds the porosity implied by its clay texture (0.38\,m$^3$\,m$^{-3}$). No
separate spin-up cycle is performed, so the earliest part of each record reflects
adjustment away from this initial state; all four windows begin on 1 January, so
that adjustment falls in the dormant season.

\paragraph{Evaluation protocol}

A timestep enters the evaluation only if every required forcing field was
observed (not gap-filled), which retains 63, 83, 96 and 60\,\% of
timesteps at US-MMS, DE-Obe, US-Ton and DE-Hai respectively. Model and
observations are additionally co-sampled, that is, compared only on timesteps at
which both are available, so that a model--observation difference cannot arise
from the two being averaged over different samples. Sub-daily values are
aggregated to daily means wherever at least 30\,\% of a day is present, and skill
is summarized by the Nash--Sutcliffe efficiency, the coefficient of determination
and the mean bias.

\begin{figure}[h]\centering
\includegraphics[width=\textwidth]{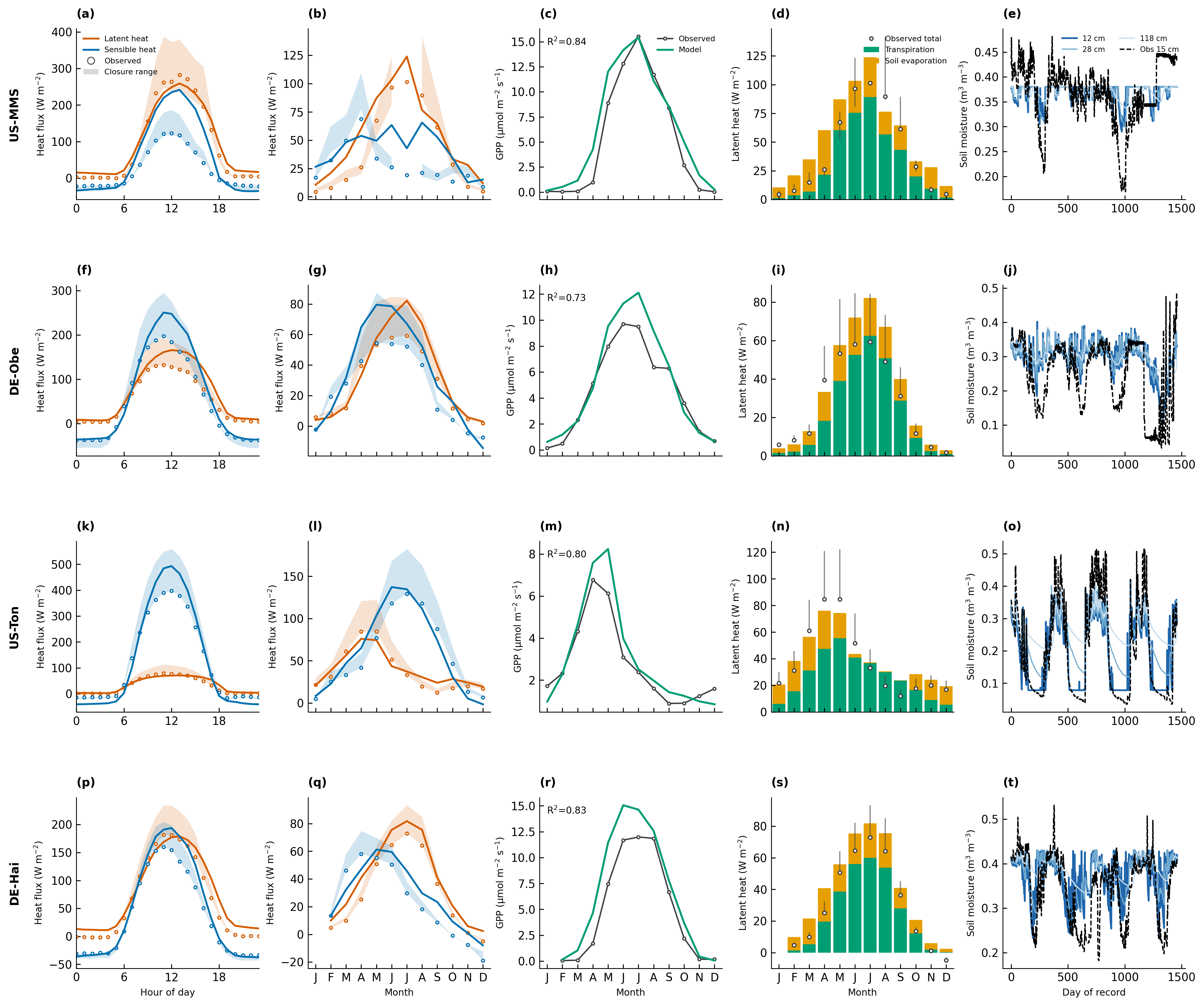}
\caption{\textbf{Land-surface validation against eddy-covariance flux towers.}
The offline land model with the two-big-leaf canopy scheme at four FLUXNET sites:
US-MMS (mesic deciduous forest), DE-Obe (montane needleleaf forest), US-Ton (oak
savanna) and DE-Hai (humid deciduous forest). One row per site (a--t): the mean
summer diurnal cycle and the seasonal cycle of latent (orange) and sensible (blue)
heat; the GPP seasonal cycle; the modelled latent-heat partition into
transpiration (green) and soil evaporation (orange) with the observed total; and
observed soil moisture at the sensor depth with modelled soil moisture at three
depths. Lines are
modelled, markers are raw eddy-covariance observations, and shaded bands span up
to the energy-balance-closure-corrected value; model and observations are compared
only on matched timesteps. Pooled daily Nash--Sutcliffe efficiency against raw
observations is 0.60 (latent heat), 0.57 (sensible heat) and 0.73 (GPP), and 0.69
(latent heat) and 0.68 (sensible heat) against the closure-corrected observations.
Raw fluxes under-close the energy budget by 22--34\,\% here, so the modelled
turbulent fluxes sit near the centre of the closure-uncertainty envelope.}\label{si:ecsite_fig}
\end{figure}

Raw eddy-covariance fluxes under-close the surface energy budget by 22--34\,\% at
these sites, whereas the model closes it by construction. Skill is therefore
reported against both the raw and the energy-balance-closure-corrected
observations, which bracket the true flux. Pooled across sites, the modelled
turbulent fluxes are biased $+17$\,\% against the raw observations and $-17$\,\%
against the corrected observations, and so lie near the centre of the
closure-uncertainty envelope (Fig.~\ref{si:ecsite_fig}). GPP achieved overall $R^2$ values of 0.73--0.84 at the four sites, although there is a positive bias in peak season carbon uptake.

\begin{figure}[h]\centering
\includegraphics[width=0.95\textwidth]{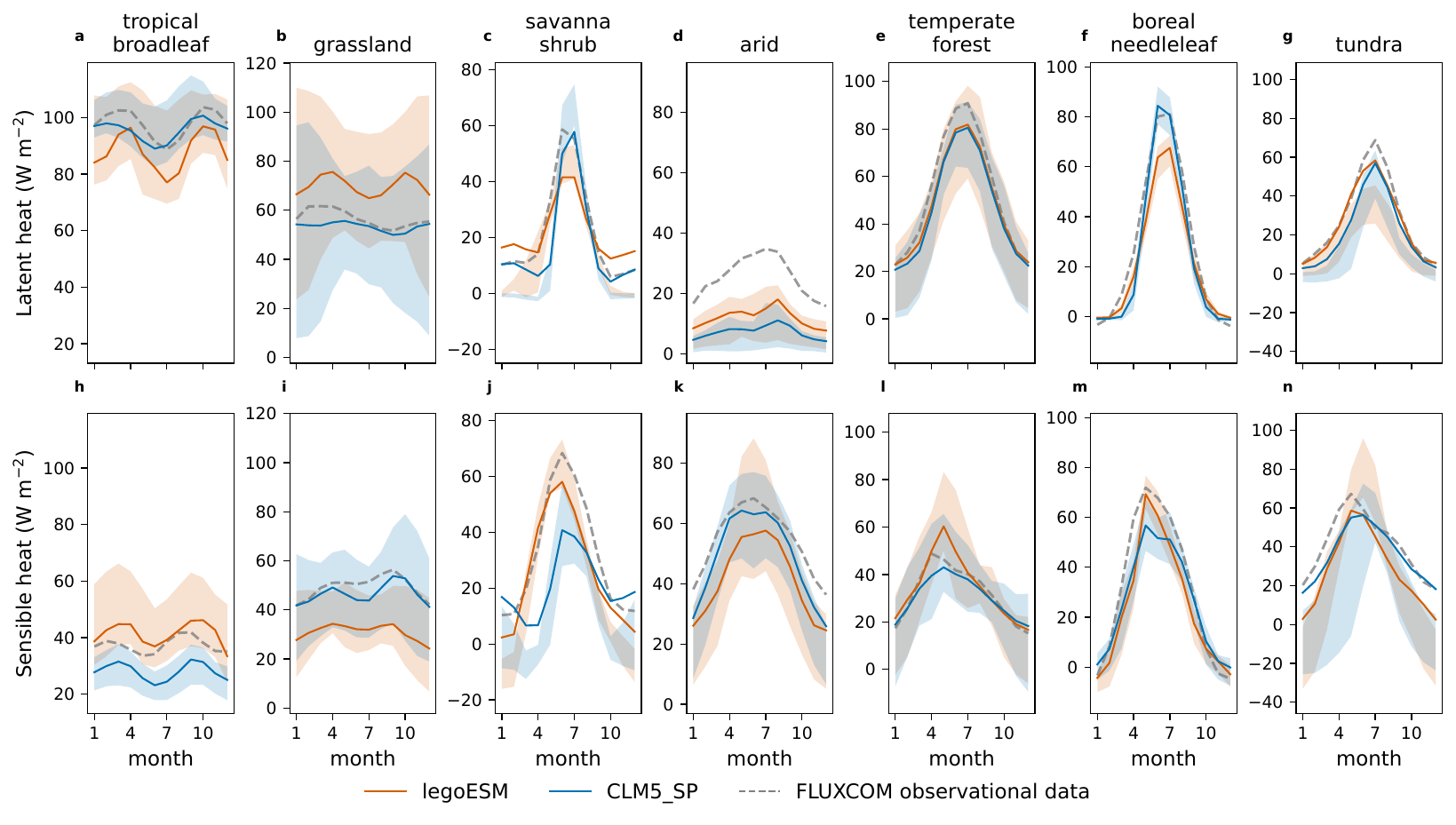}
\caption{\textbf{Seasonality of the turbulent heat fluxes by biome.} Monthly latent (a--g) and
sensible (h--n) heat fluxes from legoESM and from CLM5-SP, against FLUXCOM observational
estimates, aggregated over seven biomes. Shading spans the spread within each biome. The phase of
the seasonal cycle is captured across contrasting vegetation types, with the largest departures in
the arid and tundra classes, where the two models also differ most from each
other.}\label{si:lh_sh_biome}
\end{figure}

\begin{figure}[h]\centering
\includegraphics[width=0.95\textwidth]{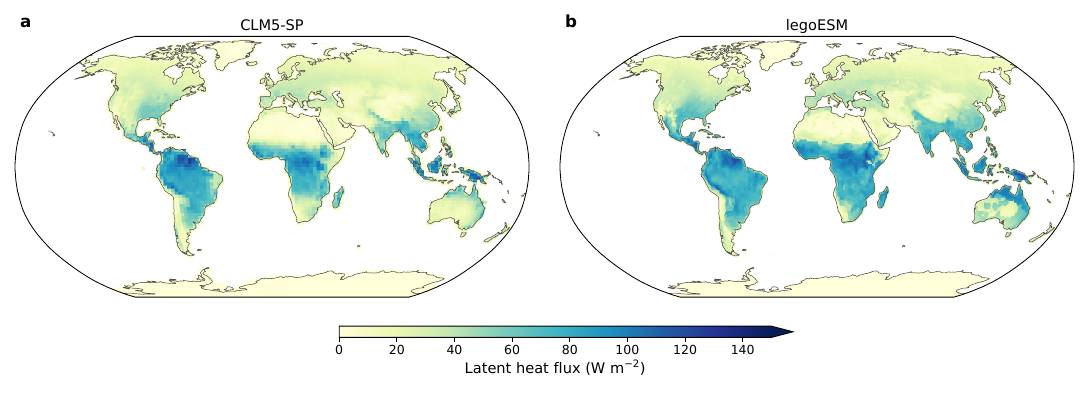}
\caption{\textbf{Global latent heat flux.} Mean annual latent heat flux from (a) CLM5-SP and
(b) legoESM. The large-scale pattern of evapotranspiration, with maxima over the wet tropics and
minima over the arid subtropics and high latitudes, is reproduced.}\label{si:lhflx_maps}
\end{figure}

\begin{figure}[h]\centering
\includegraphics[width=0.8\textwidth]{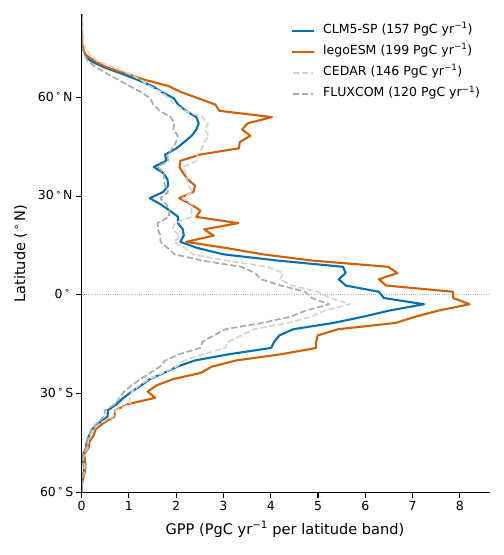}
\caption{\textbf{Zonal distribution of gross primary productivity.} Zonal-mean gross primary
productivity from legoESM and CLM5-SP against the CEDAR and FLUXCOM data-driven products, with
global totals given in the legend. The latitudinal structure, in particular the tropical maximum
and the northern mid-latitude peak, is reproduced, but legoESM overestimates the global total
(199\,PgC\,yr$^{-1}$) relative to both CLM5-SP (157\,PgC\,yr$^{-1}$) and the data-driven
estimates (120--146\,PgC\,yr$^{-1}$), a bias that remains to be
addressed.}\label{si:gpp_zonal}
\end{figure}

Beyond the tower sites, the land component is evaluated globally against CLM5-SP and against
data-driven products (Figs.~\ref{si:lh_sh_biome}, \ref{si:lhflx_maps} and~\ref{si:gpp_zonal}): the seasonal cycle of the
turbulent heat fluxes by biome, the spatial pattern of the mean latent heat flux, and the zonal
distribution of gross primary productivity. The energy fluxes and the latitudinal structure of
productivity are well reproduced, whereas the global productivity total is overestimated.

\begin{figure}[!htbp]\centering
\includegraphics[width=0.85\textwidth]{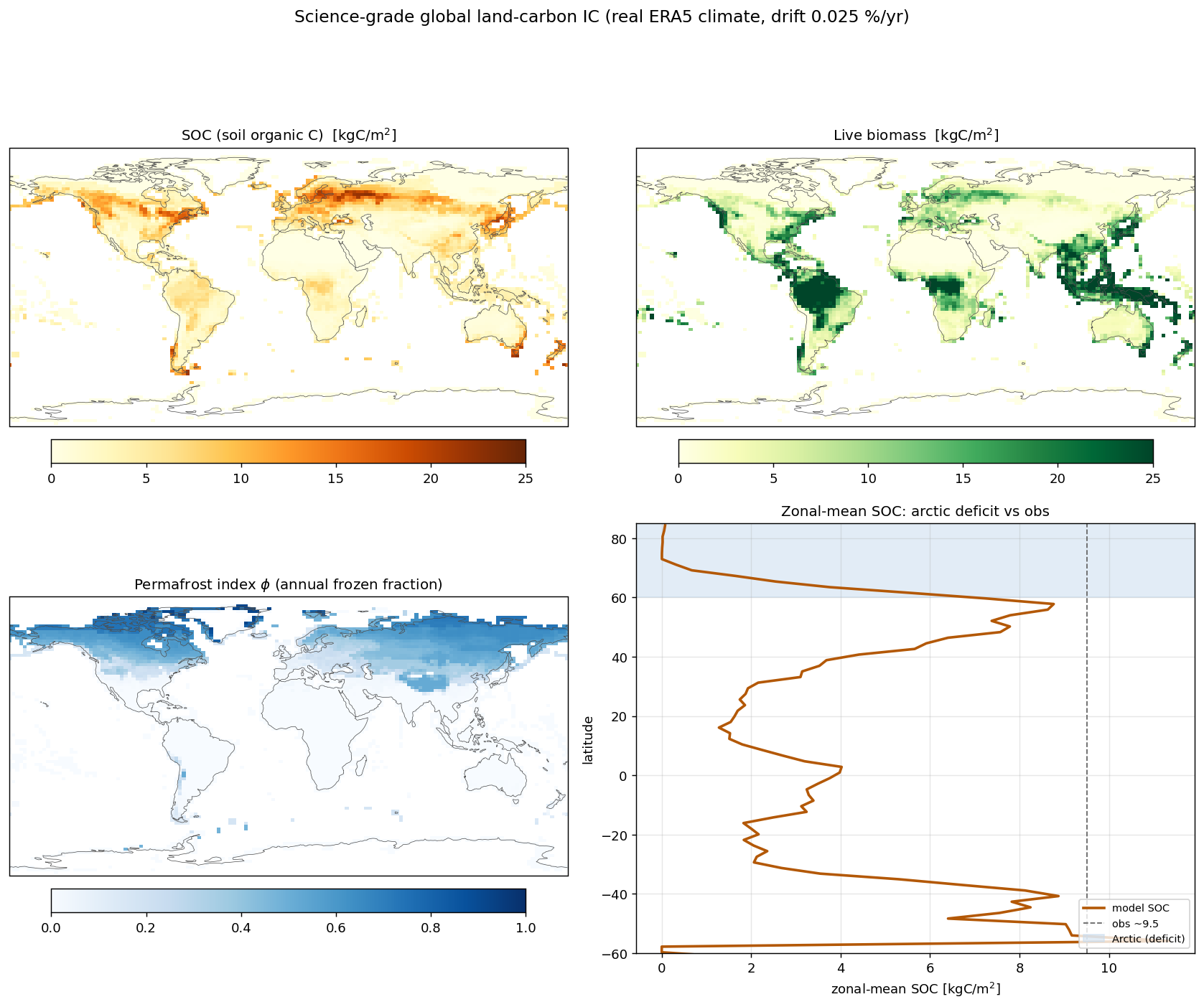}
\caption{\textbf{Initial land-carbon state.} Soil organic carbon, live biomass and permafrost index
(annual frozen fraction) of the equilibrated land-carbon initial condition under ERA5 climate,
and the zonal-mean soil organic carbon against an observational estimate
($\sim$9.5\,kgC\,m$^{-2}$, dashed); the shaded band marks the Arctic deficit north of
$60^\circ$N.}\label{si:carbon_ic}
\end{figure}

\paragraph{Initial carbon pools}
The initial carbon pools of the DALEC-based carbon model\cite{Fang2026} are obtained under ERA5
climate either from the analytical linearized steady state or by differentiable optimization,
which give very similar results, with a residual drift of 0.025\,\%\,yr$^{-1}$
(Fig.~\ref{si:carbon_ic}). Live biomass peaks in the tropical forests and soil organic carbon in
the boreal zone. In this configuration soil carbon north of $60^\circ$N is strongly
underestimated relative to observations; this comparison is what exposed the northern
high-latitude deficit that motivated the refinements of the soil-carbon structure described in
the main text.

\section{Performance and scaling}\label{s7}
The single-GPU acceleration numbers below were measured on one consumer-class laptop GPU (NVIDIA
RTX, thermally limited) plus a 24-core CPU. Single-GPU throughput (Mcells\,s$^{-1}$, fp32)
is 678 (lat--lon atmosphere), 404 (cubed-sphere), 298 (icosahedral), 400 (lat--lon ocean), 313
(tripolar) and 252 (MPAS ocean); GPU acceleration over the CPU is $16$--$21\times$ (atmosphere) and
$33$--$99\times$ (ocean) at fp32.

Strong scaling at production resolution (main-text Fig.~4) was measured on the DKRZ Levante
system, whose GPU nodes carry four NVIDIA A100-80GB GPUs (NCCL communication) and whose CPU nodes
carry two 64-core AMD EPYC 7763 (Milan) processors (gloo communication). For the $\sim$5-km
($4096\times8192$, 26 levels) lat--lon atmosphere, the time per step falls from about 220\,ms on
4 GPUs to about 9\,ms on 128 GPUs at single precision, a parallel efficiency of roughly 75\%, and
at equal node count (four GPUs versus 128 CPU cores) the GPU partition is roughly an order of
magnitude faster. The ORCA12-class tripolar ocean ($3072\times4352$, 75 levels) scales close to
ideal up to 64 GPUs. The unstructured MPAS atmosphere and ocean depart from ideal scaling once
the per-device subdomain becomes small, and the FESOM2 FORCA20
ocean saturates beyond 32 GPUs, where halo exchange on the unstructured mesh dominates.
Throughput keeps rising with device count until the per-device tile becomes communication-bound.

Reverse-mode differentiability is preserved across distributed execution: gradients through MPI
halo exchange (\texttt{sendrecv} with a custom adjoint) and \texttt{allreduce(SUM)} match
single-rank references to $\sim$$10^{-9}$. Documented limits: spectral transforms are
communication-bound on multi-GPU, and cross-node cubed-sphere panel exchange anti-scales on a
commodity TCP fabric (these are capability rather than speed-up results).

\section{Accessibility: a worked example}\label{s8}
Because the model is idiomatic Python, configuring an experiment (here a single column with
gray radiation and a Louis boundary-layer scheme) is a few readable lines, and the same physics
object can be wrapped by a neural network or differentiated. This low barrier is a core design
goal for teaching and rapid prototyping.

\begin{lstlisting}
from legoesm.atmosphere.scm import SingleColumnModel
from legoesm.atmosphere.physics import PhysicsConfig, RadiationConfig, TurbulenceConfig
import jax.numpy as jnp

scm = SingleColumnModel.create(
    physics_config=PhysicsConfig(
        radiation=RadiationConfig(scheme="gray"),
        turbulence=TurbulenceConfig(scheme="louis")),
    nlev=40, dt=300.0,
    T_profile=jnp.linspace(220.0, 295.0, 40), latitude_deg=0.0)

final_state, history = scm.run(nsteps=288, save_every=12)   # 24 h at dt=300 s
\end{lstlisting}

\noindent Swapping \texttt{scheme="gray"} for \texttt{"rrtmgp"}, or \texttt{turbulence} for an
\texttt{"edmf"} or ML scheme, requires changing only the configuration; differentiating the run
with respect to any parameter requires wrapping \texttt{scm.run} in \texttt{jax.grad}.

\section{Physical consistency: operator-level tests}\label{s9}
\begin{figure}[h]\centering
\begin{subfigure}{0.495\textwidth}\includegraphics[width=\textwidth]{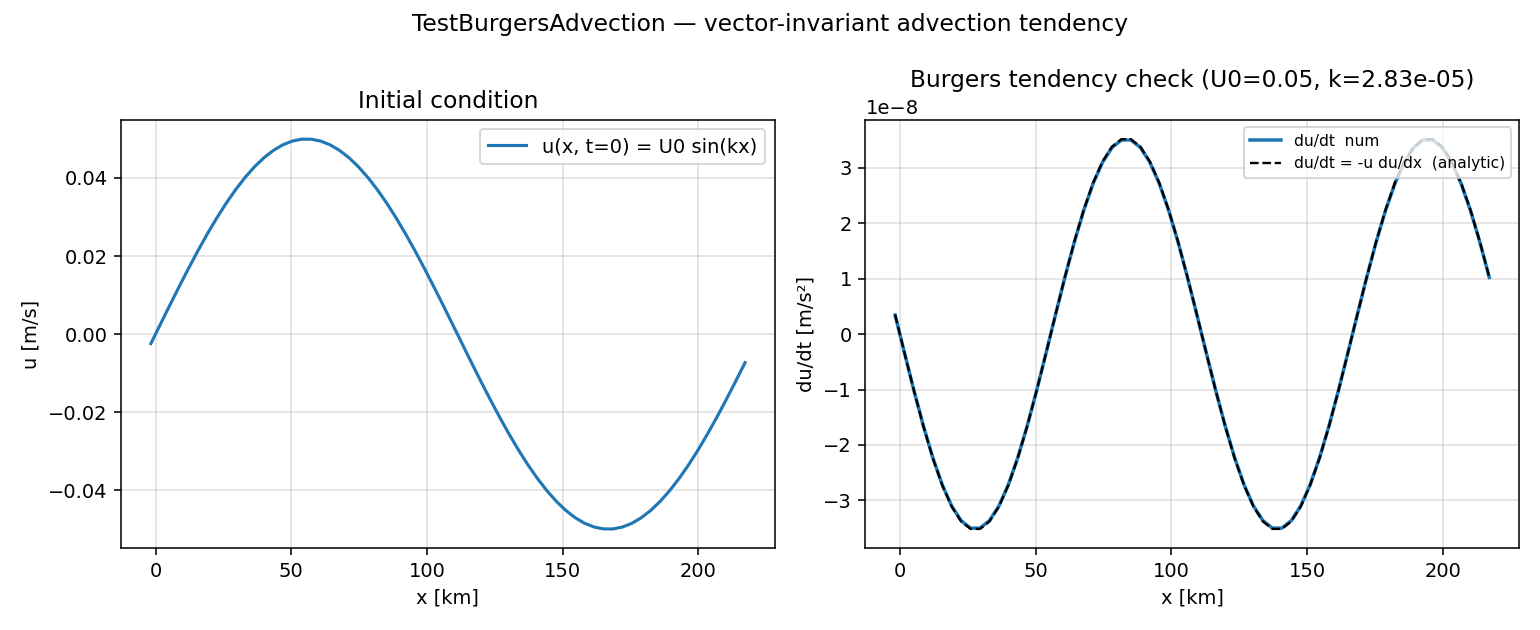}\caption{Vector-invariant (Burgers) advection}\end{subfigure}\hfill
\begin{subfigure}{0.495\textwidth}\includegraphics[width=\textwidth]{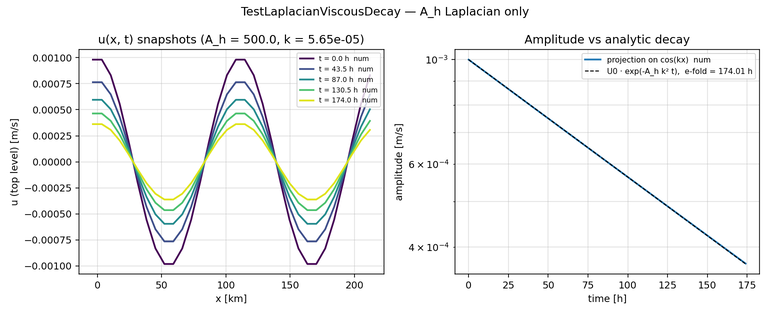}\caption{Laplacian viscous, $k^2$ scaling}\end{subfigure}\\[3pt]
\begin{subfigure}{0.495\textwidth}\includegraphics[width=\textwidth]{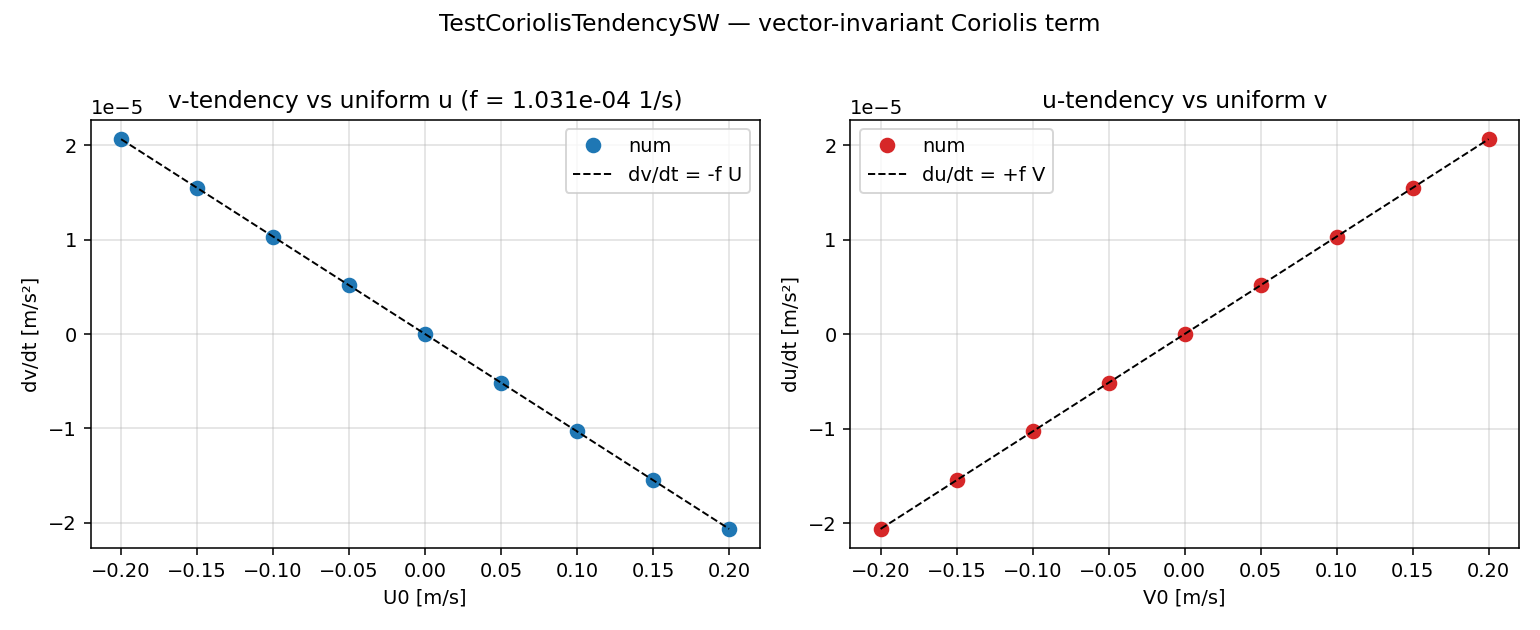}\caption{Coriolis tendency}\end{subfigure}\hfill
\begin{subfigure}{0.495\textwidth}\includegraphics[width=\textwidth]{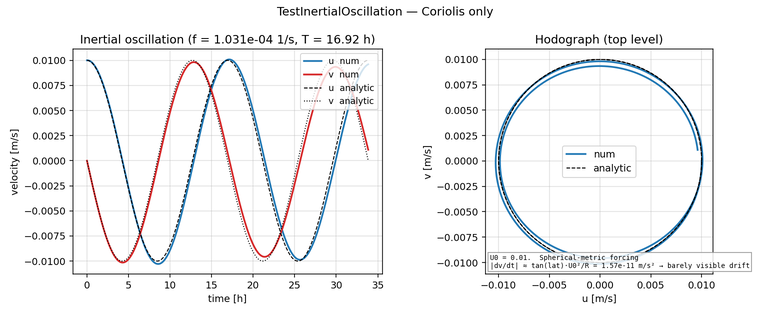}\caption{Inertial oscillation hodograph}\end{subfigure}\\[3pt]
\begin{subfigure}{0.495\textwidth}\includegraphics[width=\textwidth]{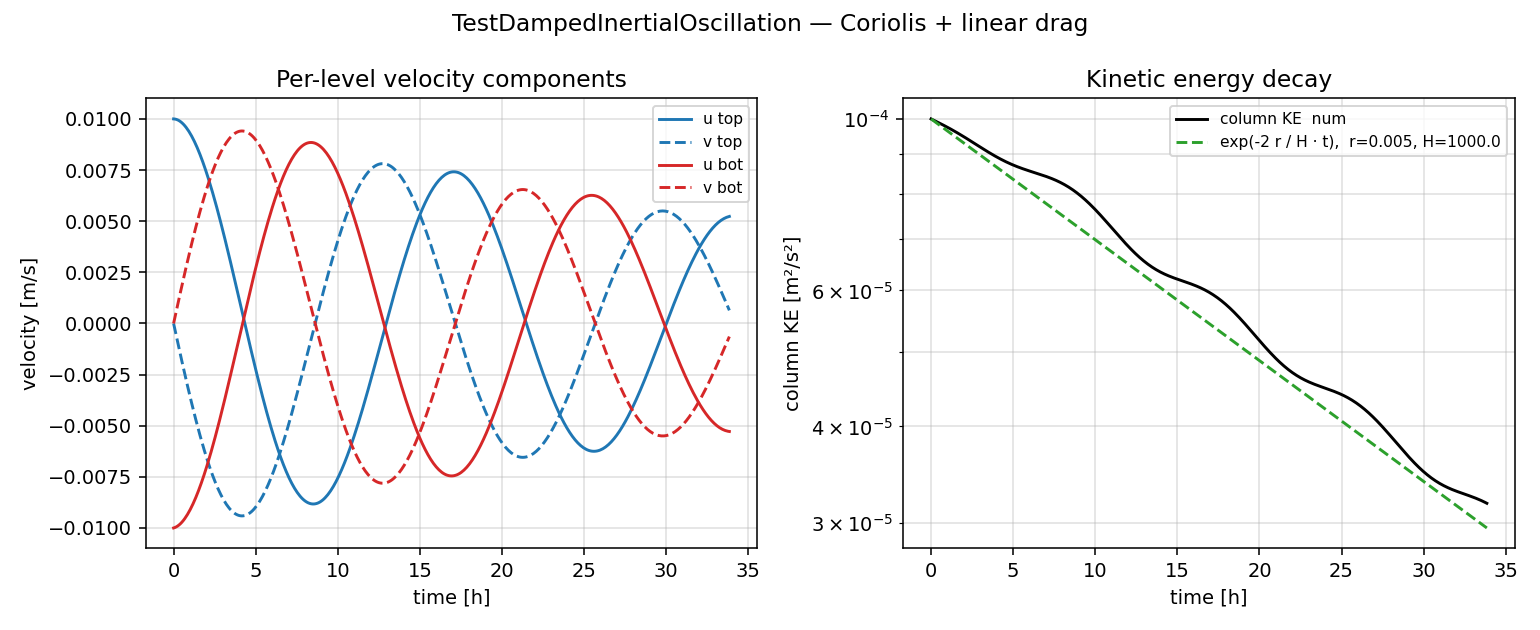}\caption{Damped inertial oscillation}\end{subfigure}
\caption{\textbf{Operator-level verification against analytic solutions.} Simulated tendencies
(solid) are compared with the corresponding closed-form solutions (dashed) for the advective,
viscous, Coriolis and inertial terms.}\label{si:phys}
\end{figure}

\begin{figure}[h]\centering
\begin{subfigure}{0.495\textwidth}\includegraphics[width=\textwidth]{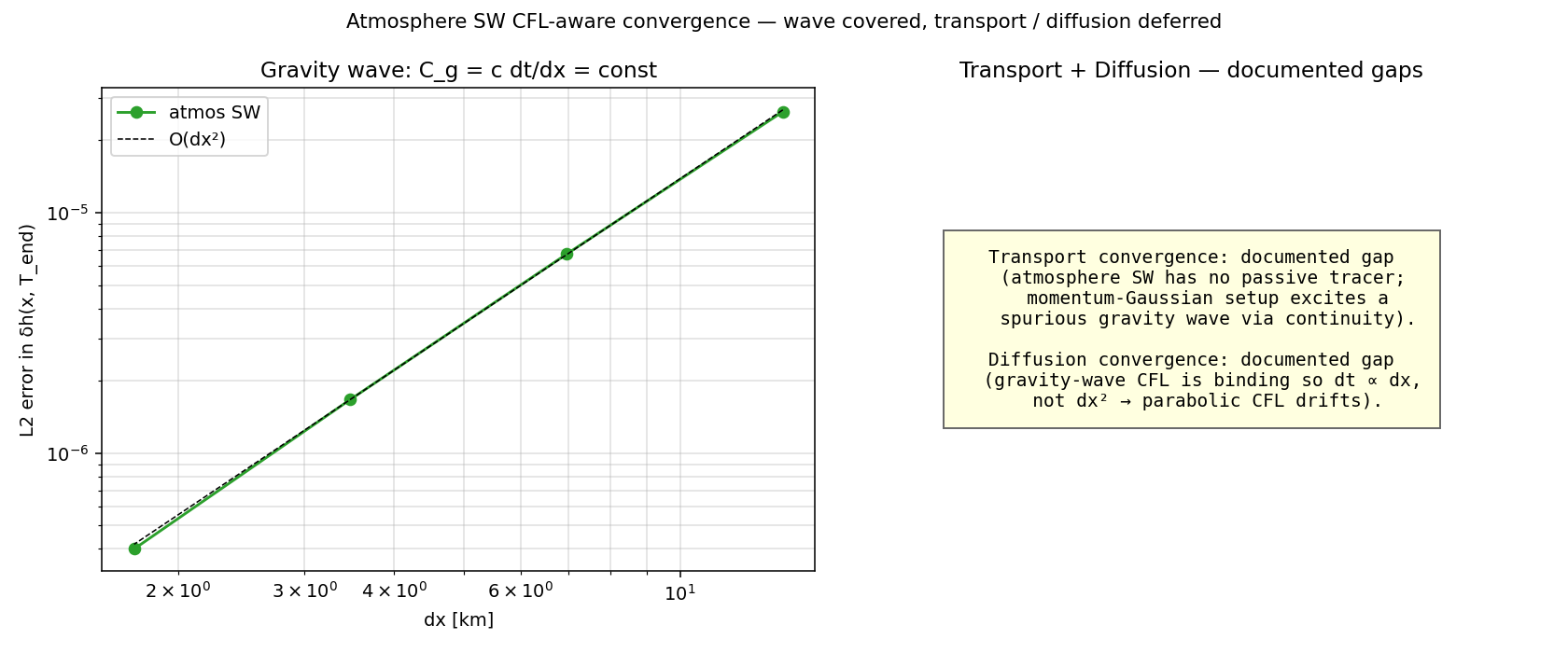}\caption{Atmosphere SW}\end{subfigure}\hfill
\begin{subfigure}{0.495\textwidth}\includegraphics[width=\textwidth]{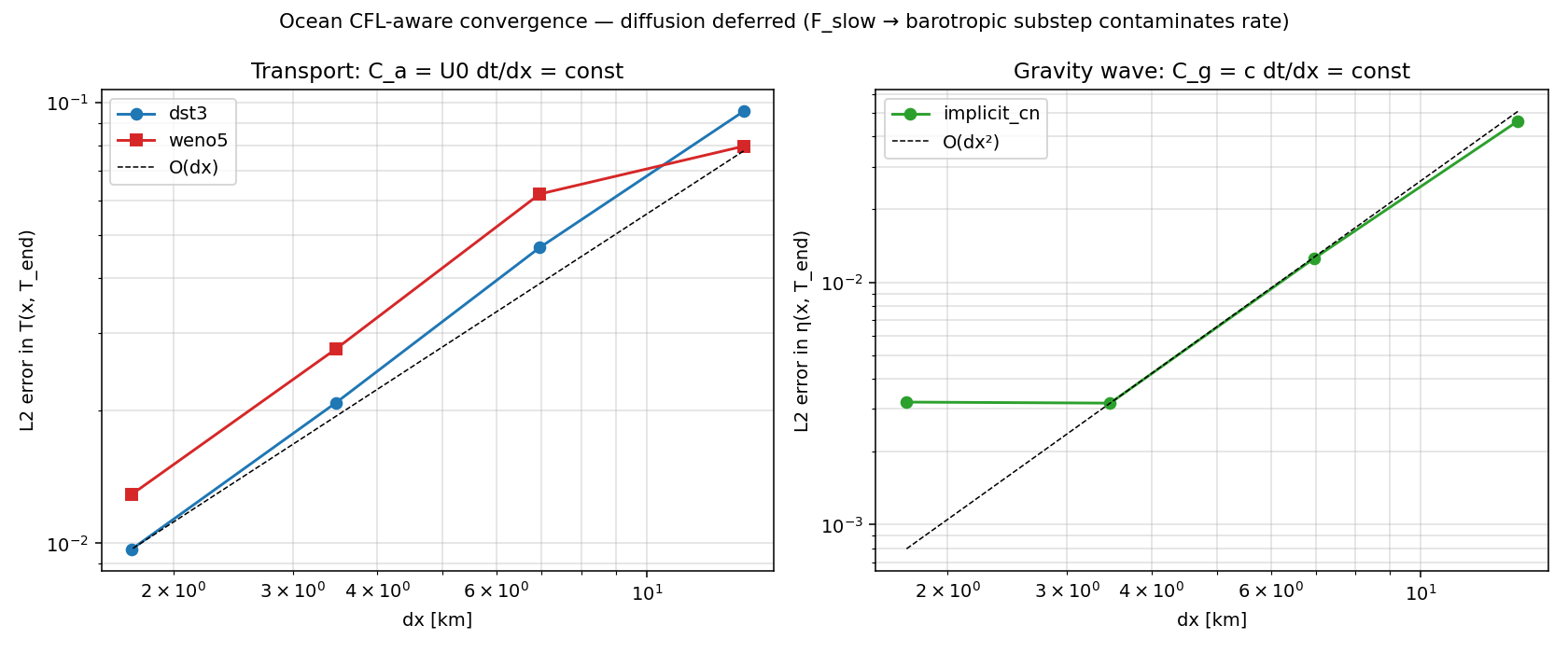}\caption{Ocean}\end{subfigure}
\caption{\textbf{Convergence under CFL-controlled refinement.} Gravity-wave and transport errors
as a function of grid spacing, shown together with the reference slopes expected from the design
order of each scheme, for (a) the atmospheric shallow-water core and (b) the ocean.}\label{si:cfl}
\end{figure}

Individual operators are verified against closed-form solutions before being combined
(Fig.~\ref{si:phys}). The vector-invariant advection reproduces the Burgers tendency $-u\,\pp_x u$; the Laplacian viscous term reproduces
$A_h\,\pp_{xx}u$ and its $k^2$ spectral scaling and exponential decay rate; the Coriolis term
gives $\dd v/\dd t=-fU$, $\dd u/\dd t=+fV$; and free inertial oscillations trace the analytic
hodograph. Under CFL-controlled refinement, finite-volume transport converges at its design order
and the semi-implicit gravity-wave solver at second order (Fig.~\ref{si:cfl}); documented gaps
(e.g.\ a parabolic-CFL regime where $\Delta t\propto\Delta x$) are reported rather than masked.

\section{Variational data assimilation}\label{s10}
Modern weather prediction depends on data assimilation to integrate observations with prior forecasts and provide accurate initial conditions.
We developed a variational data assimilation system for legoESM and evaluated its effectiveness using single-observation experiments. 
We first used a three-dimensional variational data assimilation (3DVar)~\cite{Courtier1998ECMWF3DVar} experiment to test whether a static background-error covariance matrix, $\mathbf{B}$, estimated from time-lagged legoESM forecast differences using the NMC method~\cite{ParrishDerber1992NMC}, captures the spatial and cross-variable covariances needed to spread observational information realistically.

\begin{figure}[h]
\centering
\includegraphics[width=0.98\textwidth]{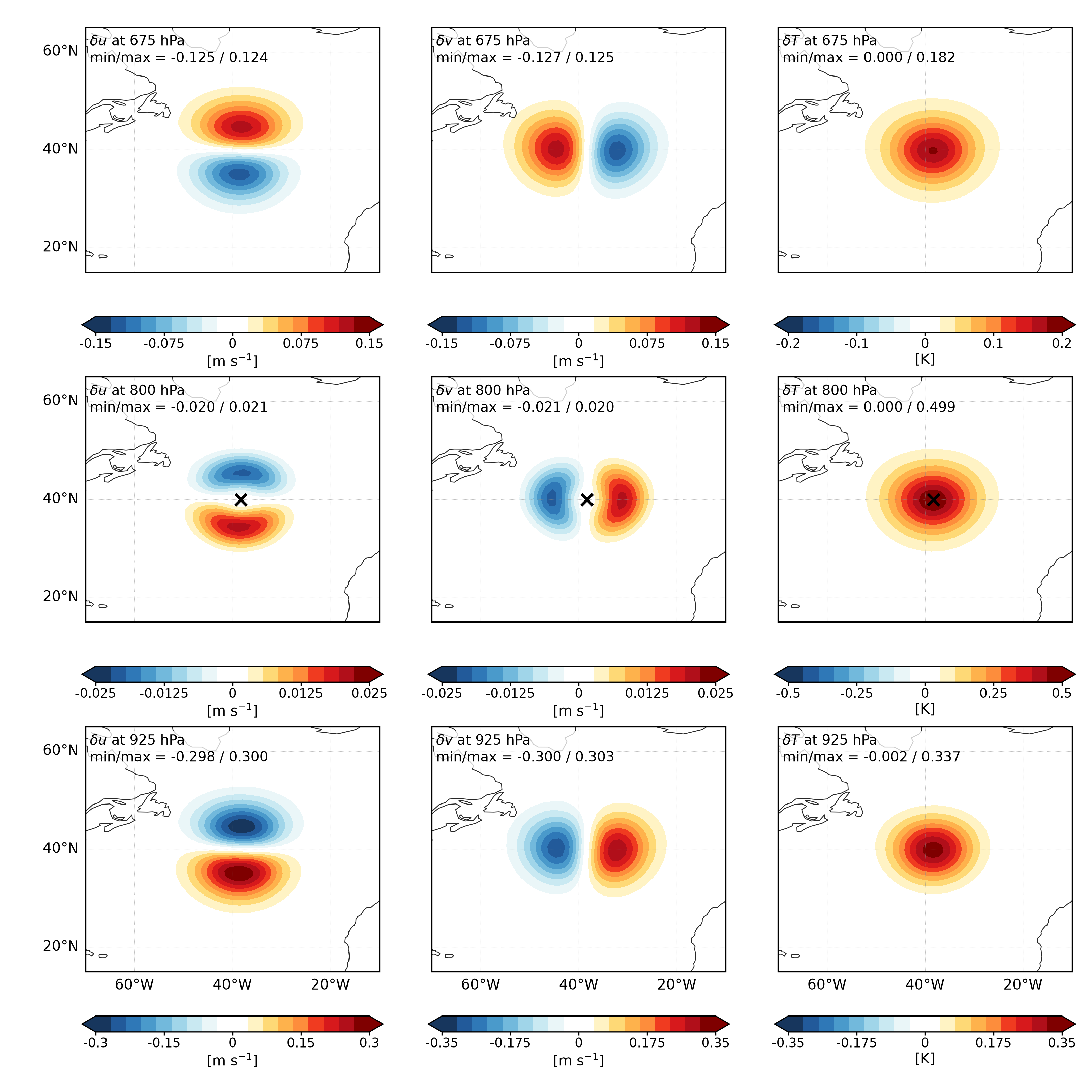}
\caption{\textbf{3DVar single-observation experiment.}
Panels show the temperature and horizontal wind increments at native model levels near 675, 800 and 925~hPa after assimilating a single ($+1~\mathrm{K}$) temperature innovation near 800~hPa.
The background state at ($t_0$) was taken from a 24-h MPAS forecast initialized from ERA5 at 00 UTC on 14 January 2017.
The observation was assigned to the MPAS grid cell near $40^{\circ}$N, $38^{\circ}$W, at approximately 800~hPa.
The cross marks the observation location. The static background-error covariance matrix, $\mathbf{B}$, was estimated from 80 pairs of 48-h and 24-h legoESM forecasts valid at the same times using the NMC method.}
\label{si:single_obs_3dvar}
\end{figure}

Following Jung et al.~\cite{jung2023three}, we assimilated a single (+1~$\mathrm{K}$) temperature innovation near 800~hPa and examined the surrounding temperature and wind increments (Fig.~\ref{si:single_obs_3dvar}). The temperature response decayed smoothly with horizontal and vertical distance from the observation, consistent with the correlation structure encoded in $\mathbf{B}$. The wind response was anticyclonic above and cyclonic below the warm anomaly, as expected from balanced mass--wind coupling. These results indicate that the static $\mathbf{B}$, derived from legoESM forecasts, captures realistic spatial correlations and multivariate balance.

Extending 3DVar through time, four-dimensional variational data assimilation (4DVar)~\cite{Courtier1994Incremental4DVar} optimizes the initial state against observations throughout the assimilation window under the constraint of model dynamics, and underpins many of the world’s leading operational weather prediction systems.
In conventional systems, however, 4DVar requires dedicated tangent-linear and adjoint models that are difficult to develop and maintain~\cite{Solvik2025Backprop4DVar}. By contrast, in the fully differentiable legoESM, both operators can be readily obtained through JAX automatic differentiation, making 4DVar comparatively straightforward to implement.

\begin{figure}[h]
\centering
\includegraphics[width=0.98\textwidth]{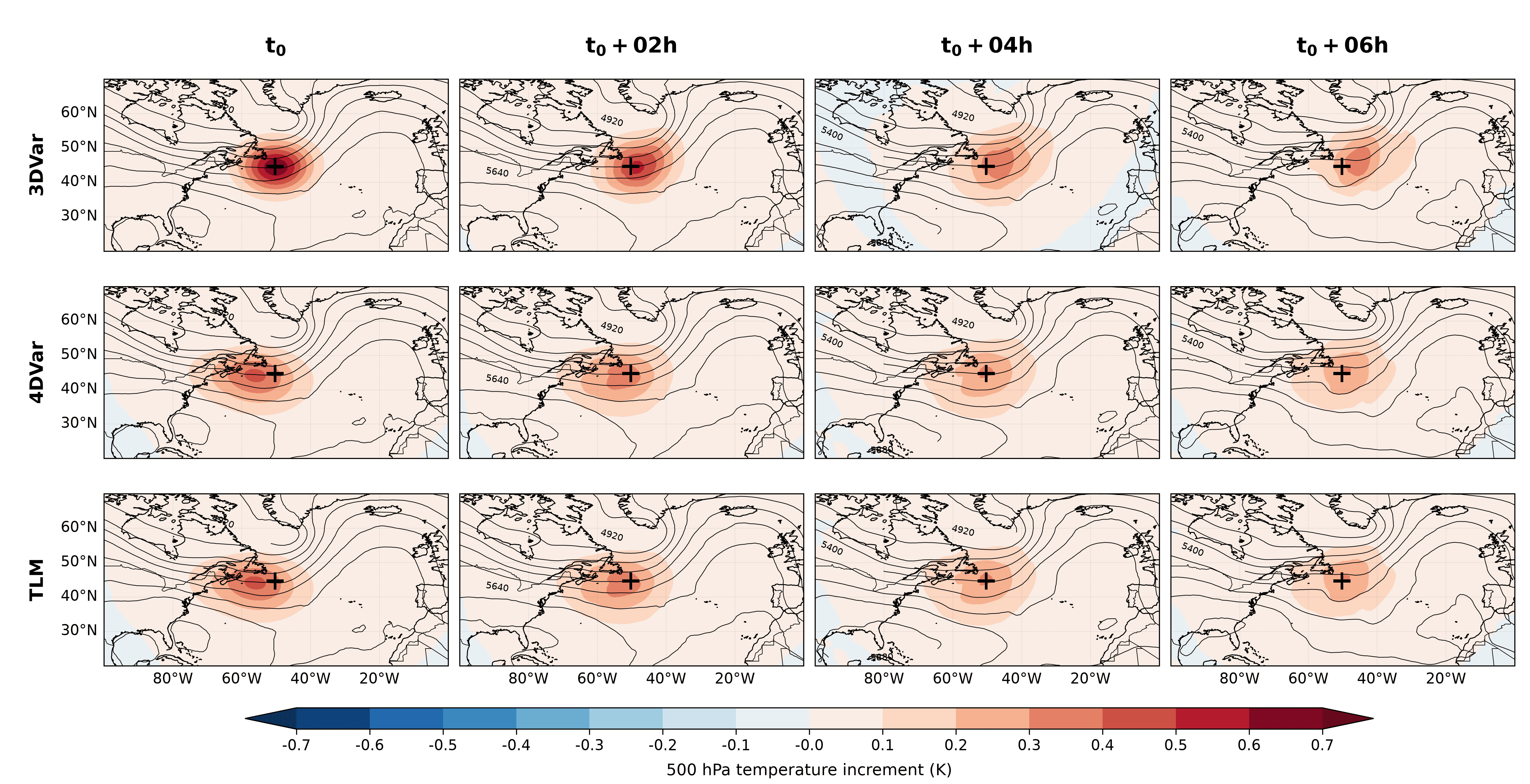}
\caption{\textbf{Comparison of 3DVar and 4DVar for assimilating a future single observation.}
Columns show the evolution of the 500~hPa temperature increment at \(t_0\), \(t_0+2\,\mathrm{h}\), \(t_0+4\,\mathrm{h}\), and \(t_0+6\,\mathrm{h}\).
The synthetic observation has a \(+2\,\mathrm{K}\) innovation and is valid at \(t_0+6\,\mathrm{h}\) near \(45^{\circ}\mathrm{N}\), \(50^{\circ}\mathrm{W}\).
Shading denotes the temperature increment, black contours show the background 500~hPa geopotential height, and the cross marks the observation location.
The top row shows the evolution of the 3DVar increment, for which the observation is treated as valid at \(t_0\).
The middle row shows the 4DVar increment, and the bottom row shows the same increment propagated by the tangent-linear model generated automatically with JAX.
Both 3DVar and 4DVar use the same static $\mathbf{B}$ and background state as in Fig.~\ref{si:single_obs_3dvar}.}
\label{si:single_obs_4dvar}
\end{figure}

To illustrate this capability, we performed a standard 4DVar single-observation experiment~\cite{Huang2009WRF4DVar} with the MPAS dynamical core (Fig.~\ref{si:single_obs_4dvar}). We introduced a synthetic 500~hPa temperature innovation in the Northern Hemisphere midlatitudes at \(t_0+6\,\mathrm{h}\) and used both 3DVar and 4DVar to update the atmospheric state at \(t_0\). In 3DVar, observations within the assimilation window are treated as if they were made at the analysis time, which may misplace their influence in the evolving flow~\cite{Massart2010ThreeDFGAT}. Here, this causes the increment centre to shift east of the observation by \(t_0+6\,\mathrm{h}\). In contrast, 4DVar uses the model dynamics to propagate the influence of the observation at \(t_0+6\,\mathrm{h}\) back to \(t_0\), producing an upstream increment that is carried eastward by the midlatitude westerlies to the observation location by \(t_0+6\,\mathrm{h}\). We also propagated the analysis increment using the tangent-linear model generated automatically by JAX. The result agrees closely with the nonlinear forecast, with a difference of less than 1\%.

\clearpage
\bibliographystyle{unsrtnat}
\bibliography{references}